\documentclass[11pt,letterpaper]{article}

\usepackage{imports}

\newcommand{\poly}{\operatorname{poly}}
\newcommand{\polylog}{\operatorname{polylog}}
\newcommand{\tO}{\ensuremath{\widetilde O}}

\newcommand{\tOmega}{\widetilde\Omega}
\newcommand{\ttheta}{\widetilde\theta}
\newcommand{\tTheta}{\ttheta}
\newcommand{\E}{\operatorname{E}}

\newcommand{\sgn}{\operatorname{sgn}}

\def\set#1{\{ #1 \}}

\def\inprod#1{\langle #1 \rangle}

\def\Bigbar#1{\mathrel{\left|\vphantom{#1}\right.\n@space}}

\newcommand{\eps}{\varepsilon}
\newcommand{\R}{\mathbb{R}}

\newcommand{\N}{\mathbb{N}}
\newcommand{\indic}{\mathds{1}}

\newenvironment{tightemize}{\setlist{nosep,topsep= 0.3em-\parskip}\begin{itemize}}{\end{itemize}}

\newcommand{\universe}{\mathcal U}
\newcommand{\innerprodq}{\textsc{InnerProduct}\xspace}
\newcommand{\owm}[1]{\textsc{1-Way-Marginals}(#1)\xspace}
\newcommand{\owmm}{\textsc{1-Way-Marginals}\xspace}

\newcommand{\ed}[1]{#1_{\eps, \delta}}

\newcommand{\inc}{\textsc{Inc}}

\newcommand{\topk}{\textsc{TopK}\xspace}
\newcommand{\matching}{\textsc{Matching}\xspace}
\newcommand{\kcore}{\textsc{KCore}\xspace}
\newcommand{\cc}{\textsc{CC}\xspace}
\newcommand{\deghist}{\textsc{DegHist}\xspace}

\newcommand{\mincut}{\textsc{MinCut}\xspace}
\newcommand{\stmincut}{\ensuremath{s\text{-}t}-\mincut}
\newcommand{\edgecount}{\textsc{EdgeCount}\xspace}
\newcommand{\SNE}{\textsc{SNE}\xspace}
\NewDocumentCommand{\kmed}{o}{
  \IfNoValueTF{#1}
    {\textsc{\ensuremath{k}-Medioids}}
    {\textsc{\ensuremath{#1}-Medioids}}
}

\renewcommand{\top}[1]{\textsc{Top-}\ensuremath{#1}\xspace}

\newcommand{\alg}{\textsc{Alg}\xspace}

\newcommand{\calA}{\mathcal{A}}
\newcommand{\calB}{\mathcal{B}}
\newcommand{\calD}{\mathcal{D}}
\newcommand{\calE}{\mathcal{E}}
\newcommand{\calH}{\mathcal{H}}
\newcommand{\calI}{\mathcal{I}}
\newcommand{\calM}{\mathcal{M}}
\newcommand{\calR}{\mathcal{R}}

\newcommand{\stream}[1]{\mathbf{#1}}
\newcommand{\init}{\mathrm{init}}

\definecolor{eqcolor}{HTML}{f4bbff}

\newcommand{\apx}{\widetilde}

\newcommand{\lpar}{\left(}
\newcommand{\rpar}{\right)}
\newcommand{\lnor}{\left\|}
\newcommand{\rnor}{\right\|}

\newcommand{\auxV}{W}
\newcommand{\estV}{E}

\newcommand{\ourResult}{\cellcolor{vsoftlimegreen!30}}

\newcommand{\nomulterrmechs}{mechanisms in the exact setting\xspace}
\newcommand{\nomulterrmech}{mechanism in the exact setting\xspace}
\newcommand{\funcnomulterr}{solving $g$ in the exact setting\xspace}
\newcommand{\dec}{\mathrm{dec}}

\usepackage{authblk}

\newcommand{\fundingmark}{\textsuperscript{\(\dagger\)}}

\begin{document}
\title{Improved Lower Bounds for Privacy under Continual Release}

\author[1]{Bardiya Aryanfard\fundingmark}

\author[2]{Monika Henzinger\fundingmark}
\affil[1,2]{Institute of Science and Technology Austria (ISTA), Klosterneuburg, Austria}
\author[3]{David Saulpic}
\affil[3]{CNRS \& Université Paris Cité, IRIF, Paris, France}
\author[4]{A. R. Sricharan}
\affil[4]{Faculty of Computer Science, UniVie Doctoral School of Computer Science DoCS, University of Vienna, Vienna, Austria}

\date{}

\maketitle

\begingroup
\renewcommand\thefootnote{}
\footnotetext{\fundingmark {This project has received funding from the European Research Council (ERC) under the European Union's Horizon 2020
research and innovation programme (MoDynStruct, No.~101019564).\\ \includegraphics[width=1.6cm]{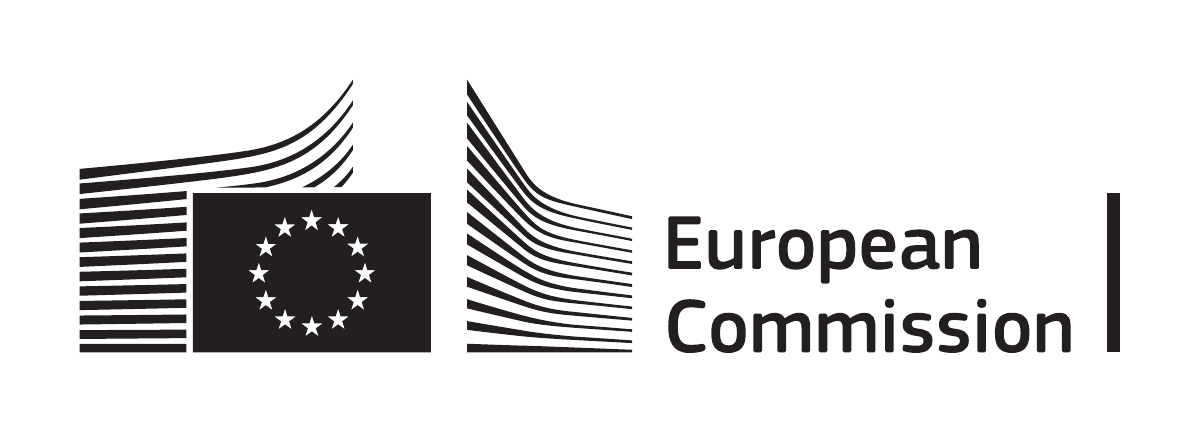}}}
\addtocounter{footnote}{-1}
\endgroup

\begin{abstract}
We study the problem of continually releasing statistics of an evolving dataset under differential privacy. In the event-level setting, we show the first \emph{polynomial} lower bounds on the additive error for \emph{insertions-only} graph problems such as maximum matching, degree histogram and $k$-core number computation. These results represent an exponential improvement on the polylogarithmic lower bounds of Fichtenberger, Henzinger, and Ost~\cite{FHO21graphdp} for the former two problems, and are the first  lower bounds in the continual release setting for the latter problem. Our results run counter to the intuition that the difference between insertions-only vs fully dynamic updates (ie with insertions and deletions) causes the gap between polylogarithmic and polynomial additive error. Indeed, we show that for estimating the size of the maximum matching or $k$-core number of a vertex, allowing small multiplicative approximations is what brings the additive error down to polylogarithmic.
We complement these results with improved upper bounds on the additive error when no multiplicative approximation is allowed.

Beyond graph problems, our techniques also show that polynomial additive error is unavoidable for the Simultaneous Norm Estimation problem in the insertions-only setting. When multiplicative approximations are allowed, we circumvent this lower bound by giving the first continual mechanism with polylogarithmic additive error under $(1+\zeta)$ multiplicative approximations, for any $\zeta >0$, for estimating \emph{all} monotone symmetric norms simultaneously.

In the item-level setting, we show \emph{polynomial} lower bounds on the product of the multiplicative and the additive error of continual mechanisms for a large range of graph problems. To the best of our knowledge, these are the first lower bounds shown for any differentially private mechanism under continual release with multiplicative error.
To obtain these results, we prove a new lower bound on the product of multiplicative and additive error for the \textsc{1-Way-Marginals}\xspace problem, and give reductions from \textsc{1-Way-Marginals}\xspace to our desired graph problems. This generalizes the prior results of Hardt and Talwar~\cite{HT10dpgeometry} and Bun, Ullman, and Vadhan~\cite{bun2018fpc}, who gave lower bounds on the additive error for the special case of mechanisms with no multiplicative error.
 \end{abstract}

\section{Introduction}
\label{sec:intro}
Differential privacy~\cite{dwork2006calibrating} has emerged as the standard for privacy-preserving data analysis in recent years. Research on releasing updated statistics with each update to an evolving dataset under differential privacy, called the \emph{continual release (or observation) setting,} was initiated by the seminal work by Dwork et al.~\cite{dwork2010differentially} and Chan et al.~\cite{chan2011private} on computing prefix sums of an integer-valued stream. It has since seen usage in a wide variety of topics, including subroutines for first-order methods in machine learning~\cite{DBLP:conf/nips/DenisovMRST22, fichtenberger2023differentially}, dynamic maintenance of graph properties~\cite{SLM18graphdp, FHO21graphdp}, and estimating norms of frequency vectors of data streams~\cite{chan2012differentially,epasto2023differentially}.

The general continual release model is commonly defined as follows: Given an initially empty data set, a stream of $T$ update operations arrive, one per time step. After each arrival a new answer has to be output, leading to the disclosure of $T$ outputs.
The goal is to have a small additive error; ideally it should be at most polylogarithmic in the problem parameters, which is typically the length  $T$ of the stream and, e.g.~in the case of graphs, the number of vertices $n$.  Usually the  \emph{incremental} setting (where each operation is an insertion) and the \emph{fully dynamic} setting (where an operation is an insertion or a deletion) have been studied, since the deletions-only setting is often symmetric to the insertions-only setting.

Initial edge-private graph mechanisms~\cite{SLM18graphdp,FHO21graphdp}
focused mostly on the \emph{incremental} setting. Fichtenberger, Henzinger and Ost~\cite{FHO21graphdp} gave mechanisms for various graph problems with polylogarithmic additive error, albeit with small ($1+\zeta$) multiplicative approximations, and complemented these with logarithmic lower bounds on the additive error of any
mechanisms in the \emph{exact setting}
(i.e., mechanisms with no multiplicative error).
The situation in the \emph{fully dynamic} setting, on the other hand, was more dire, where no non-trivial mechanisms were known for most problems of interest.

Towards explaining the lack of such mechanisms in the fully dynamic setting, a recent line of results~\cite{ELM24sublinear,RS25graphdp}
shows that polynomial additive error is unavoidable for
\nomulterrmechs for graph problems such as maximum matching and degree histogram under both insertions and deletions.
In the non-private setting, going from the incremental to the fully dynamic setting usually leads to a decrease in performance (as measured by asymptotic running time). Thus, it is natural to conjecture that the decrease in performance in the private setting (measured by additive error) is due to the change in the set of permissible operations. This leads to the following question:

\begin{center}
\begin{boxwithtitle}{Question 1}
\begin{center}
\it
Does the transition from the incremental to the fully dynamic setting
cause the increase from polylogarithmic to polynomial additive error for differentially private mechanisms?
\end{center}
\end{boxwithtitle}
\end{center}

Prior work~\cite{ELM24sublinear,RS25graphdp} do not answer this question:
The mechanisms with polylogarithmic additive error in the \emph{incremental} setting for many of these problems have a \emph{small multiplicative error}, while their polynomial fully dynamic lower bounds hold only for \nomulterrmechs.

Surprisingly, we are able to refute the intuition that the inherent hardness of obtaining mechanisms with polylogarithmic additive error is caused by the fully dynamic setting for many of these problems: We give polynomial lower bounds for them even for the incremental version.
We study both graph problems -- such as matching or $k$-core -- but also use our techniques for a standard data streaming problem, that of simultaneous norm estimation.
This answer motivates us to refine our understanding of incremental \nomulterrmechs, for which we present improved algorithms -- reducing the gap between the upper and lower bounds.

Secondly,
note that in the incremental model, any object appears only once in the stream -- namely when it is added. However, for the fully-dynamic setting, objects can appear multiple times, and this can lead to two different notions of \emph{neighboring inputs} and, thus, of differential privacy, namely \textit{item-level} neighboring and \textit{event-level} neighboring.
In graphs, for the \textit{item-level} setting of privacy, privacy is guaranteed against \emph{all} updates corresponding to the same edge -- which can be inserted and deleted multiple times over the stream history. This is opposed to the \textit{event-level} setting, where privacy is guaranteed against a single update operation, i.e., one update of one edge.

Question~1 applies to the event-level setting. However, depending on the application, the
 item-level setting  with its stronger privacy guarantees is more relevant. Therefore we also investigate the accuracy guarantees that are possible for differentially private mechanisms in the item-level fully dynamic setting.
Previous item-level lower bounds for graphs by Raskhodnikova and Steiner~\cite{RS25graphdp} were slightly higher than the corresponding event-level bounds, but were still only for the \emph{exact} setting, i.e., without multiplicative error.
Our dichotomy between exact mechanisms and mechanisms with multiplicative error  for incremental results raises the question whether there is a similar dichotomy in the item-level fully dynamic setting. Said differently, is it possible that the item-level lower bounds of~\cite{RS25graphdp} can be circumvented by giving mechanisms with small multiplicative approximations and polylogarithmic additive error? Or are there polynomial lower bounds even with small multiplicative error? This leads us to the following question:

\begin{center}
\begin{boxwithtitle}{Question 2}
\begin{center}
\it
Does there exist a dichotomy between the additive error of exact mechanisms and mechanisms with multiplicative error for the item-level setting as in the
event-level setting?
\end{center}
\end{boxwithtitle}
\end{center}
We answer this question negatively by giving suitable polynomial lower bounds.

\subsection{Our Results}

\begin{table}[t]
\caption{Results on private event-level problems. A tuple $\eta$, $O(\alpha)$ says there exists an $\eta$-multiplicative approximation mechanism with $O(\alpha)$ additive error, while $\eta$, $\Omega(\alpha)$ shows a lower bound of $\Omega(\alpha)$ on any $\eta$-approximation. The superscript $\min$ on a multivariate asymptotic notation means that the notation holds for the $\min$ of all operands, e.g., $O^{\min}(A, B) = O(\min \{A , B\})$. Results that only hold for pure differential privacy are subscripted with $\eps$. We hide the dependence on $\eps$ and $\log (1/\delta)$ for simplicity. The lower bound on \SNE in \cref{thm:inctopklb} holds even for a subset of monotone symmetric norms, namely \topk norms.}
\label{tab:inc}
\renewcommand{\arraystretch}{1.2}
\centering
\small
\begin{tabular}{ccccccc}
\toprule
\textbf{Problem}
& \multicolumn{3}{c}{\textbf{Incremental}}
& \multicolumn{3}{c}{\textbf{Fully Dynamic}} \\
\cmidrule(lr){2-4} \cmidrule(lr){5-7}
& \textbf{Mult. Err.} & \textbf{Add. Err.} & \textbf{Reference}
& \textbf{Mult. Err.} & \textbf{Add. Err.} & \textbf{Reference} \\
\midrule

\multirow{4}{*}{\matching}
& $1$ & $\Omega_\eps(\log T)$ & \cite{FHO21graphdp}
& & & \\
& \ourResult $1$ & \ourResult $\Omega^{\min}(\sqrt[4]{T}, \sqrt[4]{n})$ & \ourResult \cref{lem:incmatching}
& $1$ & $\Omega^{\min}(\sqrt[4]{T}, \sqrt{n})$ & \cite{RS25graphdp, ELM24sublinear} \\
& \ourResult $1$ & \ourResult $\tO^{\min}(\sqrt[3]{T}, \sqrt[3]{n})$ & \ourResult \cref{thm:incGraphMonotone}
& $1$ & $O^{\min}(\sqrt[3]{T}, n)$ & \cite{RS25graphdp} \\
& $1+\zeta$ & $O_\eps(\log^2 n/\zeta)$ & \cite{FHO21graphdp}
& & & \\
\cmidrule(lr){1-1}

\multirow{4}{*}{$\kcore(v)$}
& \ourResult $1$ & \ourResult $\Omega^{\min}(\sqrt[8]{T}, \sqrt[4]{n})$ & \ourResult \cref{lem:inckcore}
& & & \\
& \ourResult $1$ & \ourResult $\tO^{\min}(\sqrt[6]{T}, \sqrt[3]{n})$ & \ourResult \cref{thm:incGraphMonotone}
& & & \\
& $1+\zeta$ & $O_\eps(\log^2 n/\zeta)$ & \cite{FHO21graphdp}
& & & \\
& $2+\zeta$ & $O_\eps(\log^3 n / \zeta^2)$ & \cite{ELM24sublinear}
& & & \\
\cmidrule(lr){1-1}

\multirow{3}{*}{\deghist}
& $1$ & $\Omega_\eps(\log T)$ & \cite{FHO21graphdp}
& & & \\
& \ourResult $1$ & \ourResult $\Omega^{\min}(\sqrt[4]{T}, \sqrt{n})$ & \ourResult \cref{lem:incdeghist}
& $1$ & $\Omega^{\min}(\sqrt[4]{T}, \sqrt{n})$ & \cite{RS25graphdp} \\
& \ourResult $1$ & \ourResult $\tO^{\min}(\sqrt[3]{T}, \sqrt{n})$ & \ourResult \cref{thm:incdeghistub}
& $1$ & $O^{\min}(\sqrt[3]{T}, n)$ & \cite{RS25graphdp} \\
\cmidrule(lr){1-7}
\multirow{2}{*}{\SNE}
& \ourResult $1$ & \ourResult $\Omega^{\min}(\sqrt[4]{T},\sqrt{n})$ & \ourResult \cref{thm:inctopklb} & & &  \\
& \ourResult $1+\zeta$ & \ourResult $O_\eps(\log^6(nT)/\zeta^5)$ & \ourResult \cref{thm:probaBoosted} & & &  \\
\bottomrule
\end{tabular}
\end{table}

\paragraph*{Question 1.}
We refute the common intuition that the hardness of obtaining mechanisms with polylogarithmic additive error is caused by the ability to remove elements from the stream, by giving polynomial lower bounds for the incremental version of many of these problems.
Our results show that for problems such as maximum matching and estimating the $k$-core number of a vertex, the hardness arises not from allowing deletions, but rather from the choice of not allowing small multiplicative approximations. Specifically, we show the following result:

\begin{restatable}{theorem}{incgraphlb}
\label{thm:incgraphlb}
Let $\eps \in [0,1]$ and $\delta \in [0,1/3]$.
Any $(\eps, \delta)$-DP incremental \nomulterrmech for
maintaining the size of the maximum matching, degree histogram, or core number of a vertex\footnote{These problems are formally defined in \cref{sec:prelim}.} must have additive error $\alpha = \Omega( \min\{T^a, n^b\})$, for problem-specific constants $a,b \in [\nicefrac18,\nicefrac12]$ given in \cref{tab:inc}.
\end{restatable}

Our polynomial lower bounds for the real-valued graph problems are tight in the sense that relaxing to either the static setting
or to small multiplicative approximations admit private mechanisms with polylogarithmic additive error.
Our bounds exponentially improve the previous lower bounds of Fichtenberger, Henzinger and Ost~\cite{FHO21graphdp} for maximum matching and degree histogram, and are the first lower bounds for $k$-core estimation under continual release.

We also give improved upper bounds for these problem, significantly reducing the gap between lower and upper bounds in the item-level setting.
More specifically, when $T$ was much larger than $n$, the best \nomulterrmechs given by Raskhodnikova and Steiner~\cite{RS25graphdp} in the fully dynamic case
(for the problems whose lower bound reduction was from inner product queries)
was the trivial mechanism that returns $0$ as the answer at all time steps.
This led to a $\sqrt{n}$ gap between the lower and upper bounds for the problems in \cref{tab:inc}. We present new \emph{incremental} upper bounds that match our lower bounds up to a $\sqrt[12]{n}$ factor, drastically reducing the gap.

\begin{restatable}{theorem}{incgraphub}
\label{thm:incgraphub}
There exist $(\eps, \delta)$-DP incremental \nomulterrmechs for maintaining the size of the maximum matching, degree histogram, or core number of a vertex with additive error\footnote{Here $\ed O$ hides $\poly(1/\eps, \log(1/\delta))$ factors, and $\tO(X) = O( X \polylog(X) )$.}
$\alpha = \ed\tO( \min\{T^c, n^d\})$ for problem-specific constants $c, d \in [\nicefrac16, \nicefrac12]$ which are given in \cref{tab:inc}.
\end{restatable}

\paragraph*{Norm estimation problem.} We show that our lower bound techniques also extend from graph mechanisms to a standard problem for data streams, namely the incremental \emph{norm estimation problem}. This problem requires maintaining a data structure whose output estimates various norms of the frequency vector under continual insertions of elements.
It is especially important in the private setting to estimate multiple norms simultaneously, since querying for each norm separately using composition theorems requires privacy budget proportional to the square root of the number of norms queried, which is prohibitive as the number of queries grow.
This problem has been studied extensively under differential privacy in the streaming setting~\cite{MirMNW11, BlockiBDS12, Smith0T20, WangPS22, BravermanMWZ23}, and there exist mechanisms with polylogarithmic additive error at the cost of a small ($1+ \zeta$) multiplicative error. Since the goal of these mechanisms is to minimize space usage in addition to the additive error, most works use a variant of a bucketing technique which inherently loses some multiplicative error. Towards discovering the limits of the problem in the continual setting even without space constraints, we show that such multiplicative approximations are necessary for simultaneous norm estimation problems (regardless of space usage) by showing polynomial lower bounds on the additive error of any continual mechanism that solves a special case, namely estimating all the \top{k} norms (where the \top{k} norm is the sum of the $k$ largest frequencies).

\begin{restatable}{theorem}{inctopklb}
\label{thm:inctopklb}
Let $\eps \in [0,1]$ and $\delta \in [0,1/3]$.
Any incremental $(\eps, \delta)$-DP \nomulterrmech for \topk that runs with $n$ elements for $T$ timesteps must have additive error $\alpha = \Omega ( \min \{ \sqrt[4]{T} , \sqrt{n} \} )$.
\end{restatable}

This is the first lower-bound for this problem, even in the fully-dynamic setting.
We complement this result by showing that the problem admits polylogarithmic additive error both in the static setting and when allowing small multiplicative error. The static problem can be reduced to computing prefix sums of the sorted frequency vector. For the incremental setting, \emph{we show a strong upper bound on estimating all monotone symmetric norms} simultaneously (SNE) with small multiplicative error and polylogarithmic additive error (See Definition~\ref{def:sne} for a formal definition).
As a special case, this problem consists of estimating all the $\ell_p$-norms and the \top{k} norms of the frequency vector simultaneously. As point of comparison, there is no known way of getting a mechanism for estimating \emph{all} $\ell_p$ norms from a mechanism that just estimates one, since you cannot use composition theorems over all possible $\ell_p$ norms. For \top{k} norms, using a composition theorem would have an accuracy loss that is at least $\sqrt{n}$, while our result has additive error $\polylog(n)$.

\begin{restatable}{theorem}{probaBoosted}
\label{thm:probaBoosted}
Let $\eps>0$ and $\zeta \in (0, 1/2]$.
There exists an $\eps$-DP $(1+\zeta)$-approximate mechanism for \inc\SNE
with additive error
$\tO \lpar \frac{1}{\eps^2 {\zeta}^5} \cdot \log^6(nT)\rpar
\cdot L(e_1)$ for any monotone symmetric norm $L$, where $L(e_1)$ is the norm of the first standard basis vector.\footnote{$L(e_1)$ is allowed to not be $O(1)$ here, since this holds for \emph{any} monotone symmetric norm $L$, not only a fixed one.}
\end{restatable}

\paragraph*{Question 2.}
As our final result, we show that the item-level lower bounds of~\cite{RS25graphdp} cannot be circumvented by allowing multiplicative error: we give a polynomial lower bound on the product of the additive and multiplicative errors. This implies that the item-level fully dynamic setting is too strong to admit mechanisms with small additive error, even with polylog multiplicative error.

\begin{restatable}{theorem}{fdgrapheps}
\label{thm:fdgrapheps}
Let $\eps,\delta \in [0,1]$.
Any $(\eps, \delta)$-DP mechanism in the item-level fully dynamic setting with multiplicative approximation $\eta$ and additive error $\alpha$ for maintaining the size of the maximum matching, subgraph counts, degree histogram, mincut, $s$-$t$ mincut, or core number of a vertex on $n$-node graphs for $T \ge \Omega(1/\delta)$ time steps must have
\[
\eta \cdot \alpha =
\begin{cases}
\Omega( \min\{\sqrt{T}, n\}) & \text{when $\delta = 0$} \\
\tOmega( \min \{ \sqrt[3]{T}, n \} ) & \text{when $0 < \delta < o(1/T)$}
\end{cases}
\]
for edge count, the lower bounds are $\Omega( \min\{\sqrt{T}, n^2\})$  for $\delta = 0$ and $\Omega( \min\{\sqrt[3]{T}, n^2\})$ for $\delta \in (0,o(1/T))$.
\end{restatable}

Even in the exact setting, these constitute the first lower bounds for mincut, $s$-$t$-mincut and $k$-core.

Our results show that even relaxing to prohibitive $o(n/\polylog(n))$-multiplicative approximations still does not result in polylogarithmic additive error, and any $O(\sqrt{n})$-approximate mechanism still needs to have $\Omega(\sqrt{n})$ additive error for long streams (with $T \gg n$).
For all the
fully dynamic graph problems discussed above, while the item-level setting offers high privacy guarantees,
we show that no private mechanism can offer competitive accuracy guarantees.
To achieve these bounds we extend the fingerprinting lower bound technique of~\cite{HT10dpgeometry,bun2018fpc} to work for mechanisms for \owmm that have both a multiplicative \emph{and} an additive error.
\owmm has served as a central hard problem in differential privacy and reductions from this problem have been used repeatedly to show lower bounds~\cite{bassily2014private,jain2021price,jain23countdistinct,RS25graphdp}.
Thus, our strengthened lower bound is likely to be of independent interest.
A different extension of mechanisms with only multiplicative error was given in \cite{DBLP:conf/colt/PeterTU24}. We compare with their work in the related work section (\Cref{sec:relatedwork}).

\subsection{Technical Overview}
\label{sec:techoverview}

\subsubsection{Event-level incremental graph problems}
Our lower bounds in this setting reduce from the inner product query problem, which consists of a secret dataset $x \in \{0,1\}^d$, and a sequence of $m = \Theta(d)$ query vectors $q^1, \ldots, q^m$ with $q^j \in \{0,1\}^d$. It has been shown~\cite{DN03pods, De12dplb} that answering all the inner product queries $\inprod{x,q^1}, \ldots, \inprod{x,q^m}$ privately must incur an additive error of $\Omega(\sqrt{d})$ on at least one of the answers.

Previous dynamic lower bound reductions from this problem were restricted to functions that were \emph{additive across connected components}. These reductions proceeded by constructing a gadget where each component corresponded to one bit of $x$ and one bit of $q^j$, where the function value of the $i\textsuperscript{th}$ component was $1$
whenever the bits $x_i$ and $q_i^j$ were simultaneously equal to $1$, and was $0$
otherwise. Since the function was additive, one could sum up the function values across all
components to obtain $\inprod{x, q^j}$. The edges corresponding to the current query were then deleted, and edges for the next query were inserted. A generalization of these gadgets to the weighted setting, called  $2$-edge distinguishing gadgets, were the main tool used by  the event-level reductions of~\cite{RS25graphdp}.

We extend these lower bound reductions in two ways. (1) First, by taking all bits of the vectors $x$ and $q^j$ into account simultaneously, we can obtain lower bounds even for \emph{non-additive} functions such as estimating the core number of a vertex. { (2) Prior constructions insert suitable edges when a new query vector $q^j$ arrives, next ask a query to determine the suitable output and finally ``reset the state of the construction'' by deleting the previously inserted edges. We show how to construct gadgets for ``incremental erasure'', i.e., the state is reset using edge insertions instead of deletions.}

We give the general lower bound framework in \cref{sec:incgraph}, and describe here only the lower bound for computing the core number of one fixed vertex $v$ (i.e.,~the largest $k$ such that $v$ belongs to a subgraph with minimum degree $\ge k$), which implies a lower bound for computing the core number of all vertices. The general version of this lower bound is formally stated in \cref{thm:incgraphlb}. Our reduction places $v$ into a subgraph where its core number is the sum of two quantities: a value independent of the current query, and the desired inner product value.

The input to our reduction is a private dataset $x \in \{0,1\}^d$ and (non-private) vectors $q^1, ..., q^m \in \{0,1\}^d$.
Our goal is to build a graph and a sequence of edge insertions such that the core number of vertices of the graph after specific updates allow us to compute all the inner-products $\inprod{x, q^i}$.

\WrapLR{\includegraphics[height=1.1in]{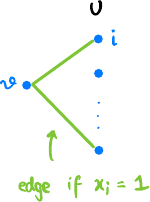}}
{}
{
Our graph consists of a dedicated vertex $v$ whose core-number at various times in the input sequence will be the value of inner products; a set $U = \{u_1, \ldots, u_d\}$ that will encode the private dataset $x$; and sets of vertices $V^j, W^j$ (each with $d$ vertices) that will be used for the $j\textsuperscript{th}$ query. This creates a graph with $\theta(md)$ vertices.
To encode the dataset $x$, we add an edge $(v, u_i)$ if the $i\textsuperscript{th}$ bit of the secret dataset $x$ is $1$.
The answer to each query could be as high as $d$, and we want to increase the core number of $v$ by one for each query bit. The core number of $v$ at the end of the reduction could thus be as high as $\Omega(md)$. We want the query vertices to have high core numbers, since the core number of $v$ should increase by $1$ whenever a query bit and dataset bit are set to $1$, which requires $v$ to belong to progressively larger cores.
Towards this, we insert a complete graph between all the $V$ and $W$ vertices, giving a ${\theta(md)}$-complete
graph with $\theta(m^2d^2)$ edges.
}

\WrapLR{}{\includegraphics[height=1.2in]{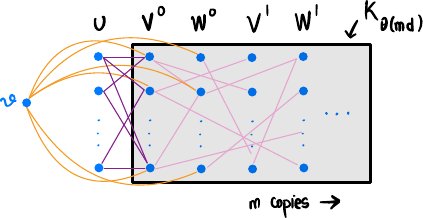}}
{
Answering the first query goes as follows.
We start with $v$ being connected to all of $V^0 \cup W^0$. There is a complete bipartite graph between $U$ and $V^0$, and \emph{no edges} between $U$ and $W^0$.
Before the first query is inserted, it holds that the degree of $v$ is $2d + \|x\|_1$, while its core number is $2d$ since all of its neighbors in $U$ do not belong to any subgraph with degree $\ge 2d$.
Since the core number is at most the degree of a vertex, this gives a slack of $\|x\|_1$ for possible increase in the core number of $v$. Whenever a query bit \emph{and} dataset bit are simultaneously set to $1$, we use this slack to increase the core number of $v$ by $1$.
}

\WrapLR{\includegraphics[height=1.1in]{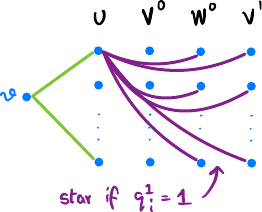}}
{}
{More specifically, to answer the first query $\inprod{x, q^1}$, we insert specific ``query edges''. For each $i$ with $q^1_i = 1$, we add all the edges from $u_i$ to $W^0$ and $V^1$.
If $q^1_i = 0$, then we do not insert any edges.
We show that after all the query edges are inserted, the core number of $v$ is exactly $k = 2d+\inprod{x,q^1}$.
Firstly, any vertex $u_i$ with $q^1_i = 0$ cannot belong to a subgraph with induced degree $\ge k$, since the degree of $u_i$ in the whole graph is $\le d + 1$. This means that all neighbors of $v$ in $U$ whose query bit is set to $0$ do not contribute to its core number.
This upper bounds the core number of $v$ by $k$.}

To show that the core number of $v$ is at least $k$ as well, we show that there exists a subgraph with minimum degree $k$ as certificate: this is the subgraph induced by the set of vertices $S = \{v\} \cup \{u_i: q^1_i = 1\} \cup V^0 \cup W^0 \cup V^1 \cup W^1$.
First, all vertices in $V^0, V^1,W^0$ and $W^1$ have degree more than $k$: indeed, they induce a complete graph on $4d$ vertices, hence have degree at least $4d-1 \geq k$. Second, the vertices in $\{u_i: q^1_i = 1\}$ are connected to all of $V^0$ (as all the vertices of $U$), and by construction to all vertices in $W^0$ and $V^1$. Hence, they have degree at least $3d \geq k$.
Finally, the degree of $v$ is precisely $k$: it has $2d$ neighbors in $V^0 \cup W^0$, and by construction exactly $\inprod{x, q^1}$ neighbors in $\{u_i: q^1_i = 1\}$. Thus, the core number of $v$ is at least $k$. Together with the previous argument it follows that the core number of $v$ is exactly $k$ and the answer $\inprod{x, q^1}$ can be computed by subtracting $2d$ from the core number of $v$.

\WrapLR{}{\includegraphics[height=1.1in]{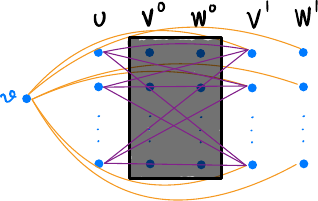}}
{To complete the processing of the first query, we then ``reset'' the gadget by adding all missing edges from $U$ to $W^0$ and $V^1$, and also adding edges from $v$ to all of $V^1 \cup W^1$. This transforms the gadget to the same ``state'' it was before the first query, with $V^1$ and $W^1$ replacing the roles of $V^0$ and $W^0$: $v$ is connected to all of $V^1$ and $W^1$, vertices in $U$ are connected to $V^1$ but not $W^1$. The only change is the core number of $v$: as there is a complete graph between vertices in $V$ and $W$, it increases by exactly $2d$.
Therefore, one can easily repeat this process for each $q^j$ in order to answer all queries. Doing so, the number of edges insertions to the graph is $T = O(m^2 d^2) = O(d^4)$ and the number of vertices $n = O(md) = O(d^2)$. Hence, the lower bound for inner product queries translate to a lower bound of $\Omega( \min \{ \sqrt[8]{T}, \sqrt[4]{n} \})$ for computing core numbers in insertion-only streams.
}

We now discuss mechanisms for incremental graph problems with no multiplicative (but with additive) error. We call this the \emph{exact setting}.
While the earlier upper bounds in the exact setting were achieved by a combination of (i) the mechanism that recomputes every $\sqrt[3]{T}$ time steps and (ii) the mechanism that returns $0$ at all time steps~\cite{RS25graphdp}, we show that in the incremental setting, the natural mechanism is to use the sparse vector technique (SVT)~\cite{Dwork2009complexity} to track the true answer and then to update it every time the true value differs significantly from the currently output value. While earlier use of SVT in the incremental setting by Fichtenberger et al.~\cite{FHO21graphdp} was only for $(1+\zeta)$-approximate mechanisms, we show that it can be directly used to obtain upper bounds
for \nomulterrmechs as well.
We present the details in \cref{sec:incgraphub}.
Our mechanism is similar to the mechanism of Henzinger et al.~\cite{HSS24random} which works in the fully dynamic setting. Applying this approach to estimating the core number of a vertex gives an upper bound of $O( \min \{\sqrt[6]{T}, \sqrt[3]{n} \})$.

\subsubsection{Item-level fully dynamic graph problems}
Previous lower bounds by Raskhodnikova and Steiner~\cite{RS25graphdp} for \nomulterrmechs were reductions from the one-way marginals problem, which is the problem of estimating the column sums of an $n \times d$ binary matrix.
To show multiplicative approximation lower bounds, we white-box the one-way marginals lower bounds on the additive error shown by Hardt and Talwar~\cite{HT10dpgeometry} and Bun et al.~\cite{bun2018fpc}, and use them to give multiplicative approximation lower bounds. While earlier work concentrated only on the additive error, we show that even when the \emph{product of the multiplicative and the additive error} is small, one can decode column sums accurately for the one-way marginals problem.
Our reduction  makes use of the specific structure of the instances that are used in the \owmm lower bound reductions, which we expand on next.

We give a short overview of our modified reduction. In the original reduction of Bun et al.~\cite{bun2018fpc}, each row of the matrix, an element of $\{0,1\}^d$, corresponds to a so-called \textit{fingerprinting code} of a user. The goal is to return the normalized column sums, rounded in some way to $\{0,1\}^d$, such that the following property is true for most columns: If all the users have the same bit in a column, then the returned bit has to match this common bit.
In~\cite{bun2018fpc} it is shown that any $(\eps,\delta)$-differentially private algorithm for the one-way marginals problem when rounded to the closest bit, has to have additive error at least a constant (for the normalized setting, which corresponds to linear additive error in the unnormalized setting).

We show that for the same fingerprinting matrices, the rounding condition can be extended to give a lower bound even for mechanisms with multiplicative and additive error. The key contribution here is to notice that the only columns that are constrained by the above condition are the uniform columns: the all-$1$s column and the all-$0$s column; for all other columns, any answer is acceptable. Since only the uniform columns are constrained, we show that even when the product of the multiplicative and additive error is smaller than some constant, the rounding algorithm can still distinguish the all-$1$s and the all-$0$s columns. Informally, our lower bound is as follows (see \Cref{cor:owmepslb,thm:apxowmGenepsdelta} for formal statements).

\begin{restatable}[Informal]{theorem}{fdowmeps}
\label{thm:fdowmeps}
Any $(\eps,\delta)$-DP mechanism for (normalized) $\owm{n,d}$ with multiplicative error $\eta$ and additive error $\alpha$ satisfies $\eta \cdot \alpha = \Omega(1)$, with $n = O(d)$ for $\delta = 0$ and $\tO(\sqrt{d})$ for $0<\delta=o(1/n)$.
\end{restatable}

Since the range of (normalized) \owm{$n,d$} is $[0,1]^d$, for the unnormalized version of \owm{$n,d$} this lower bound translates into $\eta \cdot \alpha = \Omega(n)$.
We can use this fact to show that mechanisms with constant multiplicative error $\eta$ cannot have additive error linear in the length of the output range. The hardness results shown by \cite{HT10dpgeometry} and \cite{bun2018fpc} follow as special cases of our theorem by setting $\eta=1$, which correspond to \nomulterrmechs.

Armed with our lower bound, we show that a class of reductions that includes those presented earlier by Raskhodnikova and Steiner~\cite{RS25graphdp} using $1$-edge distinguishing gadgets for item-level fully dynamic lower bounds are actually approximation-preserving reductions to the \owmm problem.
While their reductions only work for functions that are additive across connected components (such as maximum matching), our general reductions work for a much larger class of graph functions that includes estimating the (global) mincut, $s$-$t$ mincut, and core number of a vertex.
As did~\cite{RS25graphdp} we define a general framework as follows. We first define a class of graphs, called \emph{marginals solving family} (and informally called \emph{gadgets}), that have to fulfill certain properties.
Our gadgets are different from the gadgets in~\cite{RS25graphdp} in the sense that we do not necessarily have vertex-disjoint gadgets corresponding to each row of the matrix. In fact, this is the reason why our lower bounds are not limited to additive functions.
Then we show that if a graph problem has a gadget that fulfills the properties, then there is a lower bound for the product of the additive and multiplicative error for any continual mechanism maintaining that graph problem. The gadgets have parameters that give varying lower bounds for different graph problems based on these parameters.

If a graph problem has a marginals solving family $\mathcal{F}$, then any all-$0$, resp.~ all-$1$ column in the matrix $M$ of the\owmm problem is mapped by $\mathcal{F}$ into a specific graph whose value for the desired graph problem is $0$, resp.~a large value (which equals the lower bound that we can prove). The transformation of the graph from one column to the next column is done by edge updates and gives rise to a sequence of graphs. The mapping from \owmm to a graph sequence needs to be smooth with respect to the neighboring relation, so that two neighboring databases for \owmm map to two item-level neighboring graph sequences.
For this, our reduction for each graph problem is designed such that all of the updates corresponding to the $i\textsuperscript{th}$ row (for the different columns) of the \owmm instance are mapped to (potentially multiple) updates on the \emph{same} edge, say $e_i$, of the desired graph problem for every $i\in[n]$, and each of the edge updates changes the value of the graph problem of interest by exactly $1$ (or a constant, depending on the problem).

We describe our general lower bound framework for the item-level setting in \cref{sec:fd-item}, and here, in order to show the general idea of our lower bounds, we only explain our lower bound for calculating \stmincut (for fixed vertices $s$ and $t$). As mentioned above, we do so by reducing from \owmm. The input to our reduction is a private matrix $M\in\{0,1\}^{n\times d}$. We introduce a sequence of $T=n+2nd$ updates on a graph $G$ with $n+2$ vertices (including $s$  and $t$) such that the size of its \stmincut at time step $t_i=2ni$ is equal to the $i\textsuperscript{th}$ column sum ($\sum_{j\in[n]}M_{j,i}$) for all $i\in[d]$. Let the vertex set of $G$ be $V=\{s,t,v_1,\ldots,v_n\}$. The initial edge set is empty and the update sequence is as follows:

\textbf{Initialization: } In the first $n$ time steps, we insert the edges $(v_i,t)$ for every $i\in[n]$.

\textbf{Calculating the $i\textsuperscript{th}$ column sum (for every $i\in[d]$): } Time steps $[(2i-1)n+1,(2i+1)n]$ are corresponding to the $i\textsuperscript{th}$ column sum. In the first half of these time steps, in time step $(2i-1)n+j$, we insert the edge $(s,v_j)$ if $M_{j,i}=1$ and have a no-op time step otherwise. Note that  after time step $2in$, for every $j$ with $M_{j,i}=1$, $v_j$ has an edge to both $s$ and $t$ (and only to them), and for every other $j$, $v_j$ only has an edge to $t$. Note that $(\{s\},V\setminus\{s\})$ is a minimum $(s,t)$-cut, and the edge set corresponding to this cut is $\{(s,v_j)|M_{j,i}=1\}$. Therefore, the value of \stmincut at this time step is $\sum_{j\in[n]}M_{j,i}$, which is the $i\textsuperscript{th}$ column sum. Then, in the next $n$ time steps, at time step $2in+j$ (for every $j\in[n]$), we undo the update on time step $(2i-1)n+j$, i.e, remove the edge $(s,v_j)$ if $M_{j,i}=1$ and have a no-op time step otherwise. Note that after time step $(2i+1)n$, the edge set of the graph is, regardless of the entries of the matrix $M$, equal to $\{(v_j,t)|j\in[n]\}$.

Note that in the update sequence above, changing the $j\textsuperscript{th}$ row of $M$, only affects insertions and deletions of the edge $(s,v_j)$. Thus, the sequences corresponding to any two neighboring instances of \owmm, will be item-level neighboring update sequences for the \stmincut problem. Note that the \stmincut instance mentioned above has $N=\theta(n)$ nodes and $T=\theta(nd)$ time steps. Thus, by using an $(\eps,\delta)$-DP mechanism that solves \stmincut with multiplicative error $\eta$ and additive error $\alpha$ for inputs with at most $N$ nodes and at most $T$ updates as a blackbox, one can have an $(\eps,\delta)$-DP mechanism that solves \owm{$n,d$} with multiplicative error $\eta$ and additive error ${\alpha}/{n}$. Where for $\delta>0$, we can find $d=\theta((T\eps)^{2/3})$ and $n=\tTheta(\min(\frac{T^{1/3}}{\eps^{2/3}},N,T))$, and for $\delta=0$, $d=\theta(\sqrt{T\eps})$ and $\theta(\min(\sqrt{\frac{T}{\eps}},N,T))$, such that the number of nodes in the reduction above is less than $N$ and the number of updates is less than $T$, and for both cases the number of samples is less than the sample complexity stated in \cref{thm:fdowmeps}. Thus, from our lower bound on \owmm we have for every $(\eps,\delta)$-DP and $(\eta,\alpha)$-accurate mechanism for \stmincut, if $0<\delta<o(1/T)$, then $\eta \cdot \alpha =\tOmega(\min(\frac{\sqrt[3]{T}}{\eps^{2/3}},{N},T))$, and if $\delta=0$, then $\eta \cdot \alpha =\tOmega(\min(\sqrt{\frac{T}{\eps}},N,T))$.

\subsubsection{Incremental simultaneous norm estimation problems}
We next show our techniques can also be  extended to non-graph problems, namely to the popular streaming problem of estimating norms of the frequency vector of the stream. Specifically, we show polynomial lower bounds for simultaneous norm estimation (SNE), by showing that estimating all \topk norms of the frequency vector in the incremental model requires large additive error. This result was stated more formally in \Cref{thm:inctopklb}.

Our reduction constructs a stream of insertions on $n$ elements such that for each query $q^{j}$, the number of elements with the highest frequency corresponds to the value of the inner product $k^*=\inprod{x, q^j}$ at a certain time step, and the remaining elements have strictly smaller frequency.
In our reduction, this highest frequency will be $j+1$ for the $j\textsuperscript{th}$ query and there will be exactly $k^*$ many items with frequency $j+1$. Thus, $\top{k}$ will equal $[ \top{(k-1)} ]+ (j+1)$ for all $k \le k^*$,
and for all $k > k^*$, $\top{k}$ will be at most $[\top{(k-1)}] + j$. When viewing the plot of $k$ against $\top{k}$, the slope of this plot is $j+1$ until $k^*$,
and it is at most $j$ afterwards.
We use the private output of the mechanism to determine when the slope of the plot changes from $j+1$ to $j$.
See Figure~\ref{fig:inctopk} for an illustration.
The function that maps $k$ to $\top{k}$ is stable in the sense that
\[
\mathrm{privateEstimate}(k^*) - k^* \le O(\text{error of the \top{k} mechanism}).
\]
This ensures that the accuracy guarantees of the $\top{k}$ mechanism translate into accuracy guarantees for estimating this ``change point'' $k^*$, which is then the accuracy guarantee for inner product as well. The known lower bounds for inner product then give a $\Omega(\min\{\sqrt[4]{T},\sqrt{n}\})$ lower bound on the additive error for estimating \topk norms.

We complement this lower-bound by an upper bound that shows that mechanisms with an arbitrarily small constant
multiplicative approximation of $(1+\zeta)$, it is possible to reduce the additive error to polylogarithmic (see \Cref{thm:probaBoosted} for an exact statement). This mirrors our previous graph results, and shows again that the difficulty of the problem lies in avoiding any multiplicative approximation.
Note that the SNE problem comes in two flavors, where the goal of the first is to maintain a vector (called the \textit{vector version}) whose norm approximates the frequency vector for all the relevant norms, while the goal of the second (called the \textit{data structure version}) only requires to maintain a data structure that can be queried to obtain any norm. We give mechanisms for both versions and discuss the vector version first.

The first straightforward mechanism for the incremental monotone symmetric norm estimation problem would be to just compute a private histogram of the input stream, i.e., for each element $i$ maintain a private estimate $\hat f_i$ of its frequency. One can guarantee using a continual histogram mechanism that $\hat f_i - f_i \leq \log^2 T$. While this is good for any individual element -- close to the best one can do, in light of the lower bound of $\log T$ ~\cite{dwork2010differentially} -- this yields a very poor approximation for some norms: for instance, the $\ell_1$ norm of $f$ and $\hat f$ may differ by $n  \log^2 T$.

To improve over this, let $\tau^f  \approx {\log^2 (T)}/{\zeta}$ be a value that we will use as a threshold.
We first note that whenever $f_i$ is large, $\hat f_i$ is a very good multiplicative approximation to $f_i$. More specifically, when $f_i$ is larger than the threshold $\tau^f$, then $\hat f_i \in (1\pm \zeta) f_i$ for some constant $\zeta$. Thus, we are left with dealing with low-frequency elements, for which the error can accumulate.
Following the approach of Blasiok et al.~\cite{Blasiok2017norms} from the non-private setting, we deal with these elements as follows. (1) First, we group them into \textit{levels} of elements with approximately the same frequency: ideally, we would like the $j\textsuperscript{th}$ level to contain all elements with frequency $f_i \in [2^j, 2^{j+1})$. (2) If that were possible, then we could estimate the number of elements in each level, called the \emph{size of the level}.
As earlier, if the size of the level $b_j$ is large enough, then we have a correct private estimate $\hat b_j$. Thus, we could represent the level $j$ by $\hat b_j$ elements with frequency $2^j$ which is a private proxy for those elements.
(3) What remains are elements with low frequency present in levels with small size: the combination of those two conditions ensure that there cannot be too many elements satisfying this condition, and that they also cannot contribute significantly to any norm. We replace their frequencies by $0$, and incorporate the error accrued in the norm estimate into the additive error.

While the above argument almost works, there is one crucial issue with respect to privatizing the frequencies. To privately determine if an element has high frequency, we can check if the private estimate $\hat f_i$ is larger than the threshold $\tau^f$.
However, to determine the level size for the elements with low frequency, i.e.,~$f_i < \tau^f$, we \emph{must} use their true frequency. This is because even small variations in the privatized frequency could cause high errors for the estimated level sizes.
We can easily deal with elements that have $\hat f_i \geq \tau^f$, or that have $f_i < \tau^f$ \emph{and} $\hat f_i < \tau^f$, however, the critical issue comes from the fact that there are elements for which $f_i \geq \tau^f > \hat f_i$.
Indeed, the guarantee we have on $\hat f_i$ is only that $\hat f_i \leq f_i$ and $\hat f_i - f_i \leq {\log^2(T)}$, but $\hat f_i$ is computed from $f_i$ by adding a random noise which cannot be exactly estimated in advance.
Thus, one of our key contributions is to design the threshold such that the dropped elements, namely those with $f_i \geq \tau^f > \hat f_i$, have a negligible contribution to the norm. To show this, we need the threshold $\tau^f$ to be chosen from a carefully designed probability distribution where, with probability $2/3$, only a tiny fraction of elements are dropped; using properties of monotone symmetric norms, we show that any such norm of the frequency vector does not change much even after dropping these elements.

Our mechanism thus first picks $\tau^f$ and then computes the following vector: for every element $i$ with $\hat f_i \geq \tau^f$, the estimated frequency is $\hat f_i$, and for any level of elements with frequency in $[2^j, 2^{j+1})$ and $2^{j+1} < \tau^f$, the mechanism computes the size of the level $\hat b_j$ privately. If $\hat b_j > {\log^2(T)}/{\zeta}$, then, as before, $\hat b_j$ is almost the true size $b_j$, i.e.,~$\hat b_j = (1 \pm \zeta) b_j$. Therefore the mechanism adds $\hat b_j$ many elements with estimated frequency $2^j$ to the output vector. If $\hat b_j \leq {\log^2(T)} / \zeta$, then there are few elements and their frequency is small, so we can simply ignore them. This incurs an additive error $\polylog (T) \cdot L(e_1)$ for any monotone symmetric norm $L$, where $L(e_1)$ is the $L$-norm of the first standard basis vector. The accuracy guarantee holds with constant probability for this mechanism at any fixed time step. For the data structure version of the problem, we boost the accuracy by running a main mechanism which internally runs $O(\log T)$ copies of the mechanism for the vector version. The main mechanism, thus, maintains $O(\log T)$ frequency vectors. When given a query, the main mechanism calculates the queried norm on all the frequency vectors, determines the median frequency vector based on the norm, and outputs its answer. This then gives \Cref{thm:probaBoosted}.

We note that the basic approach of separating high and low frequency elements, and estimating the levels' sizes, was introduced in the non-private setting by Blasiok et al.~\cite{Blasiok2017norms}, and using continual histogram to estimate the relevant quantity is quite a natural idea. Our key contribution of this result is the randomization of the threshold $\tau^f$, and the accompanying accuracy analysis.
\subsection{Related Work}
\label{sec:relatedwork}
\paragraph{Other work on fingerprinting-based lower bounds.} Peter, Tsfadia, and Ullman~\cite{DBLP:conf/colt/PeterTU24} use a variation of Bun et al.~\cite{bun2018fpc}'s method to show a hardness result for fingerprinting codes with additional constraints on the codewords.
Namely, they consider the case that for $b\in\{0,1\}$, at least a $\frac{1-\alpha}{2}$ fraction of columns are $b$ in all codewords. For the \owmm problem, they use a reduction from their constrained version of fingerprinting codes to show a lower bound on the additive error (where the additive error is measured by the $\ell_2$ error), which is a function of the maximum pairwise Euclidean distance of the rows. They use the assumption that there is only a small fraction of columns with mixed entries (in their reduction using the constrained fingerprinting hardness result), to bound the maximum pairwise Euclidean distance of the rows of the matrix. The idea is that if at least $d'$ out of $d$ columns are uniform in all of their entries, then every pair of rows is different in at most $d-d'$ entries, so their Euclidean distance is at most $\sqrt{d-d'}$.
Since we deal with problems whose accuracy is measured with the $\ell_\infty$ norm, and since their technique does not achieve different bounds for the $\ell_\infty$ error than
that of Bun et al.~\cite{bun2018fpc}, their methods are orthogonal to ours and cannot be used to show our multiplicative approximation results.

They apply their technique to show a lower bound for differentially private mechanisms for $(k,z)$-clustering for all $z\ge 1$ that allow both multiplicative and additive error. Specifically, they show a lower bound on the additive error which has inverse dependency on the square root of the multiplicative error (ignoring polylogarithmic factors). As they do not analyze graph problems and work with a different norm, their results are incomparable with ours.

\paragraph{Privacy for graph problems.}
Private release of graph statistics has been an avenue of fruitful results with the definition of edge differential privacy by Nissim et al.~\cite{NRS07smoothsensitivity}, including results on graph cuts~\cite{gupta2010differentially,gupta2012iterative,eliavs2020differentially,dali2023neurips}, effective resistances~\cite{liu24soda}, All Pair Shortest Paths~\cite{sealfon2016shortest,Fan22neurips,chen2023soda,bodwin2024icalp}, and densest subgraph~\cite{nguyen21icml,farhadi22aistats,DLR22kcore}.
Song et al.~\cite{SLM18graphdp} and Fichtenberger et al.~\cite{FHO21graphdp} initiated the study of graphs in the dynamic setting, and Fichtenberger et al.~gave \nomulterrmechs with polylogarithmic error for minimum spanning tree, number of high-degree nodes, and edge count in the incremental setting. For other problems with low sensitivity, they gave $(1+\zeta)$-approximation mechanisms based on the sparse vector technique with polylogarithmic additive error.
 Epasto et al.~\cite{ELM24sublinear} and Raskhodnikova and Steiner~\cite{RS25graphdp} give the first polynomial lower bounds for fully dynamic mechanisms in the event-level setting, as mentioned in \cref{tab:inc}. \cite{RS25graphdp} also give lower bounds in the item-level setting. Our results extend their framework in both settings.

\paragraph{Simultaneous Norm Estimation problem.}
Blasiok et al.~\cite{Blasiok2017norms} study the \emph{non-private} monotone symmetric norm estimation (SNE) problem in the streaming model where the focus is on having low memory, and the mechanism only provides an output for the last time step.
They give a mechanism with multiplicative error $(1+\zeta)$, and small memory that depends quadratically on a parameter they call \emph{maximum modulus of concentration} $mmc(L)$ ($O(\log n)$ for $\ell_p$ norm with $p\in[1, 2]$, but $\sqrt{n/k}$ for \top{k}).
Braverman et al.~\cite{BravermanMWZ23} give a differentially private mechanism for this problem in the streaming model. Using sublinear memory (that depends on the $mmc$ parameter), they can output at the end of the stream a vector which solves the SNE problem with multiplicative error $(1 + \zeta)$, but their privacy analysis relies on the fact that the mechanism only output once, while our mechanism gives $T$ outputs.

The private estimation of singular $\ell_p$ frequencies has received attention in the streaming model, which is the model that outputs only once. In this model prior work
focuses on  estimating a \emph{single} $\ell_p$ norm with $p$ given as input.
Specifically results are known for the case of $p=0,1,2$~\cite{MirMNW11,dwork2010pan,BlockiBDS12,sheffet17differentially,choi2020differentially,Smith0T20,bu2021differentially}, $p \in (0, 1]$~\cite{WangPS22}, and $p \in [0,\infty)$~\cite{blocki2023differentially}.
This differs from our setting which \emph{simultaneously estimates multiple norms} under \emph{continual observation}. There has been one prior work in the continual observation setting.
For the special case of a \emph{given} $\ell_p$ norm, Epasto et al.~\cite{epasto2023differentially} presented a mechanism to privately estimate the norm, which works for any fixed $p$ given to the mechanism at start. Our work extends this result by allowing \emph{simultaneous} estimation of all monotone symmetric norms -- covering all $\ell_p$ and \top{k} norms. Note that our mechanism does not need to know at the start, which norms will be asked for, only when the user actually wants to estimate a norm, they need to disclose the norm.
\section{Preliminaries}
\label{sec:prelim}

\paragraph*{Notation.}
We denote $\{ 1, 2, \ldots, h \}$ by $[h]$.
All our problems have a universe $\universe = [h] \cup \{\bot\}$ of elements including the empty element $\bot$, and all our streams have a fixed length $T \in \mathbb N$.
The empty element has no influence on the value of the function that is computed on any set $S$ of elements, so the value of the function on $S$, $S \setminus \{\bot\}$, and $S \cup \{\bot\}$ are identical.
For frequency estimation problems, we use $n=h$ to denote the number of elements, and for graph problems, we use $n$ to denote the number of nodes where each element $i \in \universe$ corresponds to an edge in the graph.

\paragraph*{Incremental and fully dynamic problems.}
A \emph{fully-dynamic} stream, which allows both insertion and deletion of elements, is a sequence $\stream u = u^1, \ldots, u^T$ with $u^t \in \{ +,- \} \times \universe$.
An \emph{incremental} stream $\stream u$ is a fully dynamic stream with only insertions, i.e., $u^t \in \{+\} \times \universe$ for all $t \in T$.
An insertions-only problem is prefixed with \textsc{Inc}, while a fully-dynamic problem is prefixed with \textsc{FD}.

\paragraph*{Frequency vector.}
The frequency vector $f^t \in \mathbb N^h$ at time $t$ counts the number of times each item is present in the dataset after taking into account all insertions and deletions up to time $t$. Formally,
\[
f^t = (f^t_1, \ldots, f^t_h),
\ \text{and} \quad
f^t_i = \sum_{t \in [T]} \indic\left(u^t \in \{+,-\} \times \{i\}\right) \cdot \sgn(u^t),
\]
where $\sgn(u^t) = 1$ if the first coordinate of $u^t$ is $+$, and $-1$ otherwise.

We consider the most general setting where we allow the frequency vector to be non-binary and negative. Thus, items can be inserted to the stream even when they are already present, and also deleted when absent. Note that the $\bot$ element is not counted in the frequency vector, so having an update of $\bot$ corresponds to a time step with no changes for all the problems we consider.

\paragraph*{Static problems.}
In the static version of the dynamic problems discussed, all the updates are provided upfront, and only one estimate of the value is required after all the updates have been performed.

\paragraph*{Neighboring streams.}
Two streams are  \emph{event-level} neighboring if they are identical at all time steps except for one. Formally, $\stream u$ and $\stream v$ are event-level neighboring if there exists a $t^* \in [T]$ such that $u^t = v^t$ for all $t \neq t^*$.
Two streams are \emph{item-level} neighboring if the streams are identical in the occurrences of all elements except for those of a single element $x \in \mathcal U$, and one stream can be obtained from the other by replacing some subset of occurrences of $x$ with $\bot$. Formally, $\stream u$ and $\stream v$ are item-level neighboring if there exists an element $x \in \mathcal U$ such that either
\[
v^t \in
\begin{cases}
\{u^t\} & \text{for all $t \in [T]$ such that $u^t \not\in \{+,-\} \times \{x\}$} \\
\{u^t\} \cup (\{+,-\} \times \bot) & \text{otherwise}
\end{cases}
\]
or the above restriction holds with the roles of $u$ and $v$ swapped.

\paragraph*{Differential privacy.}
Let $\eps, \delta \ge 0$.
A randomized mechanism $\calA: \universe^T \to \mathcal O$ is said to be $(\eps,\delta)$-differentially private (DP) if for all neighboring streams $\stream u$ and $\stream v$ and for all subsets of outputs $S \subseteq \mathcal O$, we have $ \Pr[\calA(\stream u) \in S] \le e^{\eps} \cdot \Pr[\calA(\stream v) \in S] + \delta $.
If $\delta=0$, then we say the mechanism is $\eps$-DP.

\paragraph*{Continual observation.}
An algorithm in the continual observation setting gets as input a stream $\stream u$. At each time step $t\in [T]$, it obtains as input $u^t$, and must output some $x^t \in \mathcal X^t$ only using $u^1, \ldots, u^t$ before moving on to the next time step. In this setting, the output space $\mathcal O$ for defining privacy is the product of $T$ many output spaces $\mathcal X$, one at each time step, namely $\mathcal O = \mathcal X^T$.

\paragraph*{Approximations.}
Since we work with randomized mechanisms whose accuracy guarantee holds with some probability, we say that a randomized mechanism $\mathcal A$ for a real-valued problem $g$ has $\eta$ multiplicative error and $\alpha$ additive error if
the output of the mechanism satisfies, on all inputs $x$, with probability at least $2/3$,
$
\eta^{-1} \cdot g(x) - \alpha \le \mathcal A(x) \le \eta \cdot g(x) + \alpha
.$
When $\eta=1$, we call this the \emph{exact setting}, since there is no multiplicative error.
For vector-valued functions, the output of the mechanism must satisfy on all inputs $x$, with probability at least $2/3$,
\[
\| \mathcal A(x) - \eta^{-1} \cdot g(x) \|_\infty \le \alpha
\quad\text{and}\quad
\| \mathcal A(x) - \eta \cdot g(x) \|_\infty \le \alpha.
\]
The output in the continual observation setting is a vector in the output space $\mathcal O = \mathcal X^T$, and thus the accuracy guarantee holds with probability at least $2/3$ \emph{over all time steps} as well.

\paragraph*{\topk}
Let $\mathcal{U}$ be a universe with $n$ elements.
The \topk norms problem is
to output an $n$-dimensional vector containing the estimate of
the norms $\|f\|_{\top{k}} = \sum_{i = 1}^k |f_{\sigma(i)}|$ of the frequency vector $f$ for all $k \in \{ 1, \ldots, n \}$,
where $\sigma$ is the sorting permutation, i.e., $|f_{\sigma(1)}| \ge |f_{\sigma(2)}| \ge \cdots |f_{\sigma(n)}|$. This interpolates between the $\ell_1$ and the $\ell_\infty$ norms, with $\top{1} = \ell_\infty$ and $\top{n} = \ell_1$.
\paragraph*{Monotone symmetric norms.}
A norm $L$ is monotone if for all $f, f' \in \R^n$ with $|f_i| \le |f'_i|$ for all $i$, it holds that $L(f) \le L(f')$. It is symmetric if for all permutations $\sigma$ of $n$ elements, it holds that $L(f) = L(f_{\sigma(1)}, \ldots, f_{\sigma(n)})$. It is known~\cite{bhatia2013matrix} that a norm is monotone if and only if  $L(f) = L(|f_1|, \ldots, |f_n|)$, which is called \emph{gauge-invariance}.

\begin{definition}[The \SNE problem]
\label{def:sne}
The Monotone Symmetric Norm Estimation (SNE) problem is to estimate the frequency vector $f$ for all monotone symmetric norms simultaneously. This comes in two flavours. Given a frequency vector $f \in \N^n$ with $n = h$ elements,
the \emph{vector version} of SNE is to maintain a vector $\hat f \in \R^n$ where $\hat{f}$ is an approximation with $\eta$ multiplicative error and $\alpha$ additive error if the following holds for all symmetric norms $L$ simultaneously:
\[
|L(\hat{f}) - \eta^{-1} \cdot L(f)| \le \alpha
\quad\text{and}\quad
| \mathcal L(\hat{f}) - \eta \cdot L(f) | \le \alpha.
\]
The second version, called the \emph{data structure version} of SNE, is to maintain a data structure $\mathfrak d$ that can be queried to estimate any particular monotone symmetric norm, where $\mathfrak d: \N^n \times \mathcal L \to \R^n$ is an approximation with $\eta$ multiplicative error and $\alpha$ additive error if the following holds for all symmetric norms $L$ simultaneously:
$|\mathfrak{d}(f, L) - \eta^{-1} \cdot L(f)| \le \alpha$ and $
|\mathfrak{d}(f,L) - \eta \cdot L(f)| \le \alpha
$.
\end{definition}
As earlier, privacy is defined with respect to all possible outputs of $\mathfrak{d}$ for any query. For the mechanisms we design, the data structure will maintain a finite set of frequency vectors and return one of them for any queried norm, so it suffices to argue privacy with respect to returning the set of output vectors since the choice of vector to return can be performed as post-processing.

\paragraph{Constructing a graph from a stream.}
Given a stream $\stream u$ where $\mathcal U$ is the set of edges of a graph, we construct the graph $G^t$ at time $t$ with all positive frequency edges, namely
$e \in G^t$ if and only if $f^t_e > 0$,
where $f^t_e$ is the cumulative frequency of edge $e$ in the first $t$ updates of $\stream u$ as defined above.
Thus event-level neighoring streams correspond to the standard edge-neighboring event-level setting, and item-level neighboring streams correspond to the standard edge-neighboring item-level setting, since one can replace every operation that moves $f^t_e$ outside $\{0,1\}$ by $\bot$.

\paragraph*{Graph problems.}
A graph problem $g$ is a vector-valued function $g: \mathcal G \to \R^k$ where $\mathcal G$ is the space of all undirected, unweighted, simple graphs.
We consider the following problems on graphs, and the property to be returned by a mechanism for the problem. Some of the problems have designated special vertices (such as $s$ and $t$ for \stmincut), which are assumed to be encoded in the graph representation (such as being the first two vertices in the representation).

\begin{tightemize}
    \item \matching: maximum number of vertex disjoint edges present,
    \item \deghist: degree histogram $= (d_1, \ldots, d_{n-1})$ where $d_i = $ number of vertices of degree $i$,
    \item $\kcore(v)$: core number of a particular vertex $v$, which is defined as
    \[\max \{k \in [n] \mid v \in \text{a subgraph of $G$ with $\min{}$ degree} \ge k\},\]
    \item \mincut: minimum number of edge removals needed to disconnect the graph,
    \item \stmincut: minimum number of edge removals needed to disconnect vertices $s$ and $t$,
    \item $f_{\geq\tau}$: number of vertices with degree at least $\tau$,
    \item \edgecount: number of edges in the graph.
\end{tightemize}

For our lower bounds, we use the following two problems.

\paragraph*{$\innerprodq(d, m)$.}
Given $d, m \in \N$,
the input is a private dataset $x \in \{0,1\}^d$, and a sequence of $m$ public query vectors $q^1, \ldots, q^m$ with $q^j \in \{0,1\}^d$. The output consists of private estimates of the inner products $\inprod{x,q^1}, \ldots, \inprod{x,q^m}$.
Two neighboring instances differ in one bit of $x$.

Based on the results from \cite{DN03pods,DMT07dplb,MirMNW11,De12dplb},
Jain et al.~\cite{jain23countdistinct} used the following lower bound on a linear number of inner product queries to show lower bounds for fully dynamic count of distinct elements.
\begin{restatable}[\citealp{DN03pods, De12dplb}]{theorem}{innerproductlb}
\label{thm:innerproductlb}
There exists a constant $\zeta \in \N$ such that for any
$d \in \mathbb N$, $\eps \in [0,1]$, and $\delta \in [0,1/3]$,
any $(\eps, \delta)$-DP \nomulterrmech that solves $\innerprodq(d, \zeta d)$ up to additive error $\alpha$ satisfies $\alpha = \Omega(\sqrt{d})$.
\end{restatable}

\paragraph{$\owm{n,d}$.}
Given $n, d \in \N$,
the input is a private matrix $Y\in(\{0,1\}^d)^n$, and the output consists of private estimates of the normalized column sums $\frac{1}{n}\sum_{i=1}^nY_{i,j}$ for each column $j \in [d]$. Two neighboring matrices for this problem differ in a single row.

We extend the following lower bounds on the additive error of \owmm in \cref{sec:owmlb}, to a lower bound on the product of multiplicative and additive error.

\begin{theorem} [\citealp{HT10dpgeometry, bun2018fpc}]
For every $\eps, \delta\in[0,1]$ and $d,n\in\mathbb N$, there exists $\alpha=O(1)$, such that if there exists an $(\eps,\delta)$-differentially private mechanism to solve \owm{$n,d$} with additive error $\alpha$, then if $\delta=0$, $n=\Omega({d}/{\eps})$ and if $\delta>0$ and $\delta=o(1/n)$, $n=\tOmega(\sqrt{d}/\eps)$.
\end{theorem}

Jain et al.~\cite{jain2021price} used a sequential embedding technique from this problem to prove polynomial lower bounds under continual observation, which was then extended by other works to show similar bounds for other problems under continual observation~\cite{jain23countdistinct,HSS24random,RS25graphdp}.

\section{Insertions-only Graph Lower Bounds}
\label{sec:incgraph}
To show our lower bounds for algorithmic problems in graphs, we introduce the notion of
\emph{extendable bitwise AND gadgets}, which is an extension of $2$-edge distinguishing gadgets of Raskhodnikova and Steiner~\cite{RS25graphdp} to the incremental graph setting. Intuitively, our gadgets allow embedding an inner product query using the ``AND part'' of the gadget, then the ``extensibility'' of the gadgets allows ``resetting the impact'' of the previous query for inserting the next query.

Since our gadgets are defined to embed inner product queries into our incremental graph problems, we only consider the case when $m = \zeta d$ for the constant $\zeta$ from \cref{thm:innerproductlb} to simplify notation, even though it could be defined for any sequence of queries in the most general case.

\paragraph*{Extendable bitwise AND gadget$(d)$ for graph problem $g$.}
Given a dimension $d \in \mathbb N$,
an extendable bitwise AND gadget consists of the following:
positive and increasing functions $v(d)$ and $t(d)$, where the former denotes the number of nodes in the gadget, and the latter
an upper bound on the number of edge insertions,
an initial graph $H_{\init}=(V, E_0)$ with $|V| = v(d)$, along with
\begin{enumerate}
    \item edges $e_1, \ldots, e_d \in \binom{V}{2}$ such that for any $x \in \{ 0,1 \}^d$, one can define $H^1_x = H_{\init} \cup \{ e_i \}_{i \in [d]: x_i = 1}$ satisfying the conditions below, and
    \item for any sequence of $m = \zeta d$ query vectors $q^j$ with $ j \in [m]$ as in \cref{thm:innerproductlb}, a sequence of pairs of edge sets $(S^j, T^j)$ and decoding functions $\dec_j: \textrm{Range}(g) \to \R$ such that the following conditions hold:
    \begin{enumerate}
        \item $\dec_j$ is $1$-Lipschitz continuous with respect to the error norm  $\|\cdot \|$ on $g$, i.e., for any pair $x,y \in \textrm{Range}(g)$
        \[
        | \dec_j(x) - \dec_j(y) | \le \|x - y\|.
        \]
        \item
        inserting $S^j$ into $H^j_x$ creates a graph $Q^j_x = H^{j}_x \cup S^j$ which satisfies $$\dec_j(g(Q^j_x)) - \dec_j(g(H^{j}_x)) = \inprod{x,q^j},$$
        \item
        inserting $T^j$ into $Q^j_x$ creates the graph $H^{j+1}_x = Q^j_x \cup T^j$, i.e., it ``resets the query''.
    \end{enumerate}
\end{enumerate}

When the function $g$ is real-valued, we use the identity function $\dec_j = \mathrm{id}$ as the decoding functions.
In general we use the gadgets for all dimensions $d \in \mathbb N$ together to construct our lower bounds, so we use $v(d)$ and $t(d)$ to denote the relevant parameters with their dependence on $d$.
We hide the dependence on $g$ when talking about a single function $g$.
We may insert $O(d)$ many $\bot$s in the stream,
but since it will always hold that $t(d) = \Omega(d)$,\footnote{This is because $t(d)$ is a positive and increasing function.} these $O(d)$ insertions of $\bot$ are asymptotically subsumed by $t(d)$. The lower bound of $\Omega(\sqrt{d})$ from inner product then directly leads to lower bounds for the variables of the graph stream, $T$ and $n$.

\begin{algorithm}[t]
\DontPrintSemicolon
\caption{Reduction from \innerprodq to \funcnomulterr}
\label{alg:incgadgetlb}
\KwInput{A secret dataset $x \in \{ 0,1 \}^d$, and $m = \zeta d$ queries $q_1, \ldots, q_m$ with $q^j \in \{ 0,1 \}^d$.}
\KwOutput{Inner product answers $\inprod{x, q^1}, \ldots, \inprod{x, q^m}$}
Set $d = \min \left\{ {t^-(T)}, {v^-(n)} \right\}, $ $n' = v(d)$, and $T' = t(d)$.\;
Initialize an $(\eps, \delta)$-DP mechanism $\mathcal A$ for \funcnomulterr
with $n'$ vertices for $T'$ timesteps with additive error $\alpha$.\;
Insert edges of $H_\init$ in any arbitrary order.\;
\For{$t = 1, \ldots, d$}{
    \textbf{if} $x_t = 1$ \textbf{then} insert edge $e_t$ \textbf{else} insert $\bot$ \algcomment{insert dataset $x$}
}
\For{$j = 1, \ldots, m$}{
    $\apx h$ $\leftarrow$ $\dec_j$ evaluated on the query value from $\mathcal A$ on the current graph $H^j_x$. \;
    Insert all edges in $S^j$ in any arbitrary order. \algcomment{insert query $j$}
    $\apx q$ $\leftarrow$ $\dec_j$ evaluated on the query value from $\mathcal A$ on the current graph $Q^j_x$.\;
    Set $\textrm{inprod}^j = \apx q - \apx h$.\;
    Insert all edges in $T^j$ in any arbitrary order. \algcomment{neutralize query $j$}
}
\end{algorithm}

\begin{restatable}{lemma}{incgadgetlb}
\label{lem:incgadgetlb}
Let $\eps \in [0,1]$ and $\delta \in [0,1/3]$, and let $g$ be a graph problem that admits an extendable bitwise AND gadget with $v(d)$ vertices and $t(d)$ edges.
For any $(\eps, \delta)$-DP mechanism for the event-level setting that is (i) \funcnomulterr and  (ii) running on $n$-node graphs for $T$ timesteps with additive error $\alpha$, it holds that
\[
    \alpha = \Omega \left( \min \left\{ \sqrt{t^{-}(T)} , \sqrt{v^{-}(n)} \right\} \right),
\]
where $f^-$ is the generalized inverse of a function $f: \N \to \N$ defined as $f^-(x) = \sup \{d \mid f(d) \le x \}$.
\end{restatable}

\begin{proof}
Let $\mathcal{A}$ be a mechanism that is \funcnomulterr with additive error $\alpha$. We denote by $\mathcal A(G)$ the query value from the mechanism on the graph $G$. The accuracy guarantee tells us that with probability $2/3$, for any graph $G$ generated during the execution of the mechanism,
\[
    \| g(G) - \calA(G) \| \le \alpha
.\]

We condition on the event that the accuracy guarantee holds, and
given an instance of \innerprodq of dimension $d$, we show how to privately answer it with an additive error of at most $2\alpha$ using the mechanism for \funcnomulterr. The reduction is given in \Cref{alg:incgadgetlb}. We choose $d$ to be $d = \min \left\{ t^{-}(T) , v^{-}(n) \right\}$, so $n' = v(d) \le n$ and $T' = t(d) \le T$.

The reduction uses the extendable bitwise AND gadget and starts by creating an empty graph with $n'$ vertices, inserting the edges of $H_\init$ in arbitrary order.
Then, we embed the dataset by inserting edge $e_i$ if the corresponding bit $x_i = 1$, and inserting $\bot$ otherwise. This leads to the graph $H_x^1$.
Then iteratively for each query $q^j$, we first query the mechanism for $\calA(H^j_x)$ and store it after evaluation with $\dec_j$ as $\apx h$. Then, we insert the edges in $S^j$ in any arbitrary order, resulting in the graph $Q^j_x$, and query the mechanism again for $\calA(Q^j_x)$ and store it after evaluation $\dec_j$ as $\apx q$. Finally, we estimate the inner product $\inprod{x, q^j}$ as $\apx q - \apx h$, then insert the edges in $T^j$ in any arbitrary order, resulting in the graph $H^{j+1}_x$.
This creates an instance with $n'$ vertices and $\le T'$ timesteps.

Neighboring inputs for \innerprodq reduce to event-level neighboring streams for \funcnomulterr, since changing the value of one bit of $x$, say the $i\textsuperscript{th}$ bit, changes the insertion of edge $e_i$ to the insertion of a $\bot$ or vice-versa for \funcnomulterr.

Fix a value of $j$, and let $\apx q$ and $\apx h$ be the values of the corresponding variables after the $j\textsuperscript{th}$ query is inserted and queried. Let $q$ and $h$ be the true query values, namely $\dec_j(g(Q^j_x))$ and $\dec_j(g(H^{j}_x))$.
By guarantee of the gadget, we have
\[
    q - h = \inprod{x, q^j}
\]
By Lipschitzness of $\dec_j$ and the accuracy of the mechanism, we have
\[
    |q - \apx q| \le \| g(Q^j_x) - \calA(Q^j_x) \| \le \alpha
\quad\text{and}\quad
    |h - \apx h| \le \| g(H^j_x) - \calA(H^j_x) \| \le \alpha
\]
which together give that
\[
    q - \alpha \le \apx q \le q + \alpha
    \quad \text{ and } \quad
    h - \alpha \le \apx h \le h + \alpha
\]
Subtracting these two accuracy guarantees and noting that $\textrm{inprod}^j$ is as set in \cref{alg:incgadgetlb} after the $j\textsuperscript{th}$ iteration is $\textrm{inprod}^j = \apx q - \apx h$, we get that
\[
    \inprod{x, q^j} - 2\alpha \le \textrm{inprod}^j \le \inprod{x, q^j} + 2\alpha
\]
for all $j$, giving a mechanism for \innerprodq with additive error $2\alpha$, where the accuracy guarantee holds with probability at least $2/3$.
Since any $(\eps, \delta)$-DP mechanism for \innerprodq has additive error $\Omega(\sqrt{d})$ by \cref{thm:innerproductlb}, this gives us a lower bound of
\[
    \alpha = \Omega \left( \min \left\{ \sqrt{t^{-}(T)} , \sqrt{v^{-}(n)} \right\} \right).
\]
for \funcnomulterr.
\end{proof}

We now use this to prove lower bounds for natural graph problems by presenting such gadgets.

\begin{restatable}{lemma}{incgadgets}
\label{lem:incgadgets}
There exist extendable bitwise AND gadgets for the following problems and parameters:
\begin{alignat*}{3}
    \inc\matching:\quad & v(d) = O(d^2)  &&\quad\text{ and }\quad &&t(d) = O(d^2), \\
    \inc\deghist:\quad & v(d) = O(d) &&\quad\text{ and }\quad &&t(d) = O(d^2), \\
    \inc\kcore:\quad & v(d) = O(d^2) &&\quad\text{ and }\quad &&t(d) = O(d^4).
\end{alignat*}
\end{restatable}
We prove this lemma in \cref{apx:incgraph}.
Combining \cref{lem:incgadgets} with \cref{lem:incgadgetlb}, we get \cref{thm:incgraphlb}.

\subsection{Lower Bound Gadgets for Incremental Graph Problems}
\label{apx:incgraph}

\begin{restatable}{lemma}{incmatching}
\label{lem:incmatching}
There exist extendable bitwise AND gadgets for \inc\matching with
\[
v(d) = O(d^2)  \quad\text{ and }\quad t(d) = O(d^2).
\]
\end{restatable}
\begin{proof}

\begin{figure}[t]
\includegraphics[width=\textwidth]{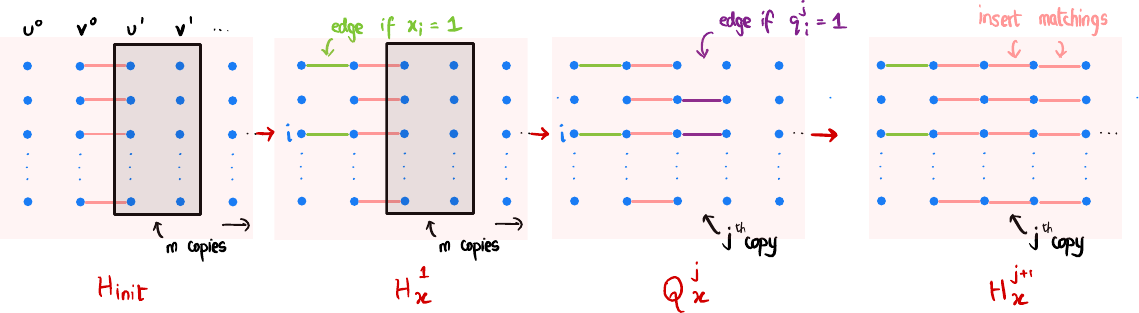}
\caption{Lower bound for \inc\matching. The core idea of the reduction is to note that in $Q^j_x$, there is an augmenting path for the pink matching if and only if both $x_i$ and $q^j_i$ are set to $1$.}
\label{fig:incmatching}
\end{figure}
We present the gadget in \cref{fig:incmatching}. Formally,

\paragraph*{Gadget construction.}

\begin{enumerate}
    \item $H_\init$ has vertex set $( U^j \cup V^j )_{j \in \{ 0 \} \cup [m]} \cup U^{m+1} $ with $|V^j| = |U^j| = d$ for all $j$, and a perfect matching between $V^0$ and $U^1$ given by $\cup_{i \in [d]} (v^0_i, u^1_i)$.
    \item $e_i = (u^0_i, v^1_i)$ for all $i \in [d]$.
    \item $S^j = \cup_{i: q^j_i = 1} (u^j_i, v^{j}_i)$, which consists of a matching of all edges between $U^j$ and $V^j$ whose index corresponds to the index of a query bit that is set to $1$.
    \item $T^j = \left( \cup_{i \in [d]} (v^{j}_i, u^{j+1}_i)  \right) \cup \left( \cup_{i \in [d]} (u^j_i, v^{j}_i)  \setminus S^j \right)$, which consists of a perfect matching between $V^{j}$ and $U^{j+1}$ and additionally edges between $U^j$ and $V^j$ to form a perfect matching between them as well.

\end{enumerate}

\paragraph*{Properties of $H^j_x$ and $Q^j_x$.}

Note that for any $j$, both graphs $H^j_x$ and $Q^j_x$  are the union of (i) vertex-disjoint paths consisting of vertices with the same subscript and (ii) isolated vertices, and for each $i \in [d]$ there exists at most one such path. Denote the path corresponding to index $i$ by $P_i$. Since the size of a matching is not affected by removing isolated vertices, we focus on the paths in both graphs.
For each $j \ge 0$, note that for each path $P_i$ in $H^j_x$,
\begin{itemize}
    \item if $x_i=0$, the length of the path $P_i$ is $2j-1$,
    \item if $x_i=1$, the length of the path $P_i$ is $2j$.
\end{itemize}
In both cases, the size of the maximum matching within the path is $j$. Thus the size of the maximum matching in $H^j_x$ is $j \cdot d$. In $Q^j_x$, each path has the following property:
\begin{itemize}
    \item if $x_i = 0$ and $q^j_i = 0$, then the length of the path is $2j-1$,
    \item if only one of $x_i$ and $q^j_i$ is $1$, then the length of the path is $2j$,
    \item if $x_i = 1$ and $q^j_i = 1$, then the length of the path is $2j+1$.
\end{itemize}
In the first two cases, the size of the maximum matching of all edges on $P_i$ is $j$, while in the last case, the size is $j+1$. Thus, the size of the maximum matching in $Q^j_x$ is $j\cdot d + \inprod{x,q^j}$ as desired.

\paragraph*{Retrieving the inner product.}
Since the size of the matching is $j \cdot d$ in $H^j_x$ and $j \cdot d + \inprod{x,q^j}$ in $Q^j_x$, the difference between them gives the inner desired inner product $\langle x, q^j \rangle$. Thus, the claim follows.

\paragraph*{Parameters of the gadget.}
The number of vertices in the graph is $O(d^2)$ and the number of edges in the final graph is $O(d^2)$.
\end{proof}

\begin{restatable}{lemma}{inckcore}
\label{lem:inckcore}
There exist extendable bitwise AND gadgets for \inc\kcore with
\[
v(d) = O(d^2)  \quad\text{ and }\quad t(d) = O(d^4).
\]
\end{restatable}

\begin{proof}

Define the $k$-core of a graph to be the subset of all vertices of the graph with core number at least $k$. Note that any vertex with degree $< k$ does not belong to the $k$-core, and removing all the vertices of degree $< k$ does not change the identity of all $k'$ cores for $k' \ge k$.

\paragraph*{Gadget construction.}

\begin{enumerate}
    \item $H_\init$ has vertex set $\{ v \} \cup U \cup ( V^j \cup W^j )_{j \in \{ 0 \} \cup [m]} $ with $|U| = |V^j| = |W^j| = d$ for all $j$, and a complete graph $K_{2d(m+1)}$ embedded inside $\cup_{j \in \{ 0 \} \cup [m]}(V^j \cup W^j)$, along with a star $\{ v \} \times (V^0 \cup W^0)$ and a complete bipartite graph $U \times V^0$. Thus the core number of all vertices outside $\{ v \} \cup U$ is at least $2d(m+1)$.
    \item $e_i = (v, u_i)$ for all $i \in [d]$.
    \item $S^j = \cup_{i: q^j_i = 1} ( \{ u_i \} \times (W^{j-1} \cup V^j))$ is a star from $u_i$ to all vertices in $W^{j-1} \cup V^j$ for each query bit $i$ set to $1$.
    \item $T^j = ((U \times (W^{j-1} \cup V^j)) \setminus S^j) \cup (\{v\} \times (V^j \cup W^j) )$ is the remaining edges in the complete bipartite graph along with a star from $v$ to $V^j \cup W^j$.
\end{enumerate}

\paragraph*{Properties of $H^j_x$ and $Q^j_x$.}
Define $S = \cup_{s \in \{ 0 \} \cup [j-1]} V^s \cup W^s$ as a shorthand.
We first discuss $H^j_x$. Note that $v$ has edges only to $S \cup U$ and the degree of $v$ in $H^j_x$ is $2jd + \|x\|_1$, which upper bounds its core number. Since $v$ has edges to all of $S$, and since $S$ is a complete graph, the core number of $v$ is at least $2jd$. Since the degree of all vertices in $U$ is at most $2(j-1)d + d + 1 < 2jd$, they do not belong to any $k$-core  for $k> 2jd$. Thus the core number of $v$ is exactly $2jd$.

Next, we discuss $Q^j_x$. The degree of $v$ is still $2jd + \|x\|_1$, and we will show that the core number of $v$ is $2jd + \inprod{x, q^j}$.
Note that all vertices $u_{i}$ in $U$ with $q^j_i = 0$ have degree $\le 2(j-1)d + d + 1$ and do not contribute to any larger cores. Thus the core number of $v$ is at most $2jd + \inprod{x, q^j}$ since none of its neighbors with $q^j_i = 0$ contribute to its core number.
Note that every vertex $u_i$ in $U$ with $q^j_i = 1$ has edges to all of $S \cup V^j$ in $Q^j_x$, and thus has degree $\ge 2jd + d$. Let us denote this set of vertices of $U$ by $U^j$.
Thus, if $|U^j| \neq 0$, every vertex in $U^j \cup S \cup V^j$ has induced degree $\ge 2jd + d$.
Thus the subgraph $\{ v \} \cup U^j \cup S \cup V^j$ of $Q^j_x$ has minimum degree at least $2jd + \inprod{x, q^j}$
for all vertices, which shows that the core number of $v$ is the same as well.

\paragraph*{Retrieving the inner product.}
Since the core value of $v$ is $2jd$ in $H^j_x$ and $2jd + \inprod{x,q^j}$ in $Q^j_x$, the difference between the two values gives $\inprod{x,q^j}$ and the claim follows.

\paragraph*{Parameters of the gadget.}
The number of vertices in the graph is $O(d^2)$ and the number of edges in the final graph is $O(d^4)$.

\end{proof}

\begin{restatable}{lemma}{incdeghist}
\label{lem:incdeghist}
There exist extendable bitwise AND gadgets for \inc\deghist with
\[
v(d) = O(d)  \quad\text{ and }\quad t(d) = O(d^2).
\]
\end{restatable}

\begin{proof}

\begin{figure}[t]
\includegraphics[width=\textwidth]{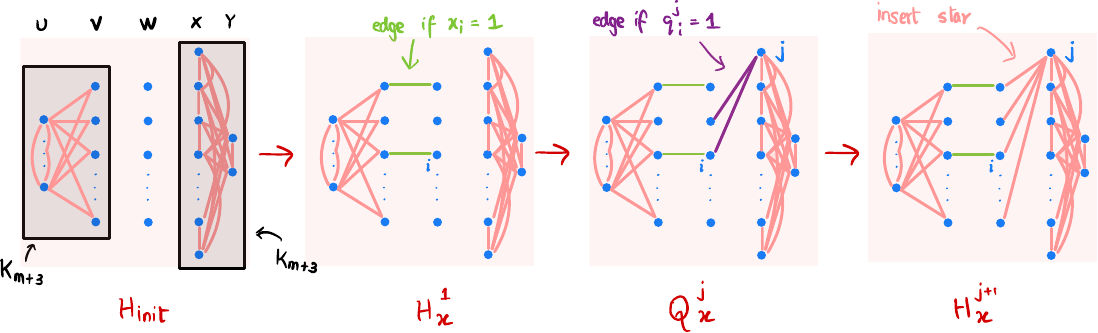}
\caption{Lower bound for \inc\deghist. The core idea of the reduction is to note that in $Q^j_x$, the $i^{\textrm{th}}$ vertex in the middle layer has degree $j+1$ if and only if both $x_i$ and $q^j_i$ are set to $1$.}
\label{fig:incdeghist}
\end{figure}
Recall that we are reducing from the inner product queries problem with parameters $d$ and $m$.
We present the gadget in \cref{fig:incdeghist}.

\paragraph*{Gadget construction.}

\begin{enumerate}
    \item $H_{\init}$ has vertex set $U \cup V \cup W \cup X \cup Y$ with $|U| = m - d+3$, $|V| = d$, $|W| = d$, $|X| = m$, and $|Y| = 3$. The initial graph consists of a complete graph $K_{m+3}$ on $U \cup V$ and of a complete graph $K_{m+3}$ on $X \cup Y$.
    \item $e_i = (v_i, w_i)$ for all $i \in [d]$.
    \item $S^j = \cup_{i: q_i^j = 1} (w_i, x_j)$ is a star at $x_j$  to all vertices in $W$ with the query bit set to $1$.
    \item $T^j = W \times \{ x_j \} \setminus S_j$ is the set of remaining vertices in the star.
    \item $g_j = \Pi_{j+1}$ is the projection onto the $(j+1)\textsuperscript{st}$ coordinate of the degree histogram as the decoding function for the $j\textsuperscript{th}$ query.
\end{enumerate}

\paragraph*{Properties of $H^j_x$ and $Q^j_x$.}
Note that all vertices in $U, V, X,$ and $Y$ have degree at least $m+2$ throughout the stream. We will focus on the degree of vertices in $W$, which remain smaller than $m+1$ at all times. After the $j\textsuperscript{th}$ query is inserted (recall that $1\le j \le m$), the number of vertices with degree exactly $j+1$ in $Q^j_x$ is equal to $\inprod{x, q^j}$. At the same time, the number of vertices with degree $j+1$ in $H^j_x$ is $0$, which means that $g_j$ works as the decoding function as stated.

\paragraph*{Retrieving the inner product.}
Since the number of vertices of degree $j+1$ is $0$ in $H^j_x$ and $\inprod{x,q^j}$ in $Q^j_x$, the claim follows.

\paragraph*{Parameters of the gadget.}
The number of vertices in the graph is $O(d)$ and the number of edges in the final graph is $O(d^2)$.
\end{proof}

\section{Insertions-only Graph Upper Bounds}
\label{sec:incgraphub}
In this section we first use the sparse vector technique (\cref{apx:SVT}) to show an insertions-only upper bound for a set of graph problems including \inc\matching and \inc\kcore, then we use another mechanism that composes multiple continual counting mechanisms to give an upper bound for \inc\deghist (hence for event-level privacy).
\subsection{Upper Bound for Monotone Functions using SVT}
\begin{lemma}
\label{thm:incGraphMonotone}
Let $g:\mathcal G\rightarrow\R$ be a monotone function on graphs with constant sensitivity such that for every incremental stream of length $T$ on graphs with $n$ vertices, the output of $g$ lies in the range $[L(n,T), R(n,T)]$ of length $\ell=R(n,T)-L(n,T)$.
Then, for $\eps>0$ and $\delta\in[0,1)$ there is an $(\eps,\delta)$-DP mechanism $M$ that answers $g$ with additive error $\alpha$,
where for $\delta>0$, $\alpha=\tO({\ell^{1/3}\sqrt{\ln(1/\delta)}}/{\eps})$ and for $\delta=0$, $\alpha=\tO({\sqrt{\ell}}/{\eps})$.
\end{lemma}
\begin{proof}
To this end, we set a parameter $k$ (to be determined later), then we run the SVT mechanism as mentioned in \cref{apx:SVT}, by setting the query to be the value of the function $g$ itself for the graph of the first $i$ time steps ($G_i$) for all $i\in[T]$ and setting $t_i$'s adaptively from the sequence $\tau_j=L(n,T)+k.j$ for $j\in[\frac{\ell}{k}]\cup\{0\}$. Note that $\tau_0=L(n,T)$ and $\tau_{\frac{\ell}{k}}=R(n,T)$. We start by setting $t_1=\tau_0$ and then after each positive output we set the next threshold to be the next threshold on the $\tau$ sequence ($t_{i+1}=t_i+k$ if the answer to the $i$-th query was positive, $t_{i+1}=t_i$ otherwise).

Now, in both cases of $\delta=0$ and $\delta>0$ there exist allocations for $k$ where $k=\tilde\theta(\sqrt{\ell})$ for $\delta=0$ and $k=\tilde\theta(\ell^{1/3})$ for $\delta>0$, such that since the above threshold condition in the statement of \cref{thm:SVT} occurs $\tilde O(\ell^{2/3})$ times for $\delta>0$ and $\tilde O(\sqrt\ell)$ times for $\delta=0$. Thus, applying \cref{thm:SVT} with setting $c$ according to the mentioned upper bounds, will imply the accuracy guarantees in the statement of this theorem. Note that since the sensitivity of $g$ can be greater than 1, we might need to apply \cref{thm:SVT} with scaled privacy parameter; however, this will not change our bounds asymptotically (since we assume constant sensitivity for $g$).
\end{proof}
\begin{corollary}
For $\eps>0$ and $\delta\in[0,1)$ there we have the following upper bound for the additive error of $(\eps,\delta)$-DP mechanisms for \inc\matching:
for $\delta>0$, $\alpha=\tilde O({\min(n,T)^{1/3}\sqrt{\ln(1/\delta)}}/{\eps})$ and for $\delta=0$, $\alpha=\tilde O({\sqrt{\min(n,T)}}/{\eps})$. Also, for $\inc\kcore(v)$, we have the following bounds:  for $\delta>0$, $\alpha=\tilde O({\min(n,\sqrt{T})^{1/3}\sqrt{\ln(1/\delta)}}/{\eps})$ and for $\delta=0$, $\alpha=\tilde O({\sqrt{\min(n,\sqrt{T})}}/{\eps})$
\end{corollary}
\begin{proof}
This follows directly from \cref{thm:incGraphMonotone} since all of the problems are monotone (\inc\cc is decreasing and the other two are increasing) and the range for the answers of these problems are $[0,\min(\frac{n}{2}, T)]$ for \matching, $[\max(n-T,1),n]$ for \inc\cc and $[0,\min(n,\kappa)]$  where $\kappa$ is the largest integer such that $\binom{\kappa}{2} \leq T$ ($\kappa=O(\sqrt{T})$) for $\inc\kcore(v)$.
\end{proof}

\subsection{Upper Bound for \deghist}
In this subsection, we use $n$ separate counters (one for each degree) to calculate the number of vertices with each degree. Then we use the advanced composition theorem (\cref{thm:advancedcomposition}) to find the right privacy parameter for the counters so that the whole mechanism is private. Then, using those parameters, we discuss the accuracy of this mechanism.
\begin{lemma}
\label{thm:incdeghistub}
For any $\eps,\delta\in(0,1)$, there is an $(\eps,\delta)$-differentially private \nomulterrmech that solves \inc\deghist with additive error $\alpha$, where $\alpha=\tO({\sqrt{n \log (1/\delta)}}/{\eps})$.
\end{lemma}
\begin{proof}
For each possible degree $d \in\{0,\ldots,n-1\}$, we build an $\eps'$-DP counter $\hat f_d$ with $\eps'=\frac{\eps}{2\sqrt{2n\ln{1/\delta}}}$. Then after receiving an update which inserts an edge $(v,u)$, for each endpoint we insert -1 on its previous degree's counter, 1 on its new degree's counter, and 0 on the other counters. Advanced composition theorem implies that the whole mechanism is $(\eps,\delta)$-DP. Note that since we are considering the insertions-only case, each vertex can enter each degree once, and exit it once; thus, the sensitivity of each counter is 4 (2 for each endpoint of the edge). Therefore, by applying \cref{theo: BCM} with $\beta=\theta(n^{-1}T^{-1})$, and privacy parameter $\frac{\eps'}{4}$ (for some constant $c$, where $c$ comes from the sensitive of degree counters), there exists an $\eps'$-DP counter which except with probability $O(1/n)$, at all time steps simultaneously, has additive error $\tO({\sqrt{n}}/{\eps})$. Thus, the accuracy guarantee for all of the counters hold with a constant probability.
\end{proof}
\section{Lower Bounds for \textsc{1-Way-Marginals}}
\label{sec:owmlb}
In this section, we will first review the definition of sample complexity and \owmm and then prove two different lower bounds for its sample complexity. The first lower bound is for $\varepsilon$-DP mechanism and uses a more restrictive notion of accuracy, namely $(\eta,\alpha)$-accuracy than the second lower bound that applies for $(\varepsilon,\delta)$-DP mechanisms and uses $(\eta,\alpha,\beta)$-accuracy.

\begin{definition}
    A set of counting queries $Q$ has \emph{sample complexity} $n^*$ for $(\eps,\delta)$-DP and ($\eta,\alpha)$-accuracy, resp.,~$(\eta,\alpha,\beta)$-accuracy  if $n^*$ is the smallest number of samples such that there exists an $(\eps,\delta)$-DP and $(\eta,\alpha)$-accurate, resp.~$(\eta,\alpha,\beta)$-accurate mechanism for $Q$.
\end{definition}

\begin{definition}[\owm{$n,d$}]
In this problem, given a private matrix $Y\in\{0,1\}^{n\times d}$, the mechanism must output an estimation of the average of the $j\textsuperscript{th}$ column ($\frac{1}{n}\sum_iY_{i,j}$) for each $j\in[d]$.
\end{definition}

\subsection{Pure Differential Privacy}
\label{sec:owmpuredp}

Now, we prove the following bound on the sample complexity of approximating (with multiplicative error $\eta$ and additive error $\alpha$) this problem under pure differential privacy. This is an extension of the proofs of Hardt and Talwar~\cite{HT10dpgeometry} and De~\cite{De12dplb} to allow multiplicative approximations.

\begin{definition} [$(\eta,\alpha)$-Accuracy for \owm{$n,d$}]
Let $\eta\geq1$ and $\alpha \in[0,1]$ be parameters. An output sequence $(a_1,\ldots,a_d)$ is an $(\eta,\alpha)$-accurate answer to \owm{$n,d$} on database $D\in(\{0,1\}^d)^n$ if
\[
\frac{1}{\eta} \cdot \frac{1}{n}\sum_{i=1}^nD_{i,j}-\alpha\leq a_j\leq \eta\cdot\frac{1}{n}\sum_{i=1}^nD_{i,j}+\alpha
\]
for all columns $j\in[d]$.

A mechanism $\calM$ is $(\eta,\alpha)$-accurate for \owm{$n,d$} if for every database $D\in(\{0,1\}^d)^n$,
\[
\Pr[\calM(D)\textrm{ is an }(\eta,\alpha)\textrm{-accurate answer to \owm{$n,d$} for }D]\geq 2/3,
\]
\end{definition}

\begin{theorem}
Let $\eta\ge 1$ and $\alpha>0$ such that $\eta \cdot \alpha < \frac{1}{2}$.
For every $\varepsilon$-DP mechanism $\calM\colon\{0,1\}^{n\times d}\rightarrow[0,1]^d$ for \owm{$n,d$} that is $(\eta,\alpha)$-accurate,
it holds that $n=\Omega({d}/{\varepsilon})$.
\end{theorem}
\begin{proof}
We prove this theorem by a packing argument.
For every $d$-length binary string $a\in\{0,1\}^d$, define the database $Y^a\in\{0,1\}^{n\times d}$ to be an $n \times d$ matrix with all of its rows being $a$, i.e, $Y^a_{i,j}=a_j$ for all $i,j \in [n] \times [d]$. For every database $Y^a$, the answer to the \owm{$n,d$} problem on $Y^a$ is the vector $a$ itself.

Towards defining the set of acceptable accurate answers of the mechanism, we define the $0$-accurate and $1$-accurate intervals in $\R$ to be:
\[
I_z=[\eta^{-1}z - \alpha, \eta z + \alpha] \text{ for }z\in\{0,1\}
\textrm{ which implies that }
I_0 = [-\alpha, \alpha] \quad\text{and}\quad I_1 = [\eta^{-1} - \alpha, \eta + \alpha]
\]
In the general case, we define the (asymmetric) accuracy ball $\calB_a$ around $a \in \{0,1\}^d$ as
\[
\calB_a = \left\{ x = (x_1, \ldots, x_d) \in \R^d \,\middle\vert\,
x_j \in I_{a_j},
\,\, \forall j\in[d] \right\}
\]
which is the set of all acceptable answers for $Y^a$ under the accuracy guarantee of the mechanism.
When $ \eta \cdot \alpha < \frac{1}{2}$, it holds that $\alpha<\eta^{-1}-\alpha$, so $I_0$ and $I_1$ are disjoint. Thus, when $\eta \cdot \alpha < \frac{1}{2}$, the sets $\calB_a$ are disjoint for all $a\in\{0,1\}^d$.
Thus, for every $a\in\{0,1\}^d$, we have:
\[
\sum_{a'\in\{0,1\}^d} \Pr[\calM(Y^a) \in \calB_{a'}] \leq 1
\implies
\exists\, a'\in\{0,1\}^d\colon \Pr[\calM(Y^a) \in \calB_{a'}] \leq 2^{-d}
\]
From the accuracy guarantee of the mechanism we have:
\[
\Pr[\calM(Y_{a'})\in\calB_{a'}]\geq \frac{2}{3}
\]
The two inequalities above imply the following:
\[
\frac{\Pr[\calM(Y_{a'})\in\calB_{a'}]}{\Pr[\calM(Y_{a})\in\calB_{a'}]} \geq   \frac{2}{3} \cdot 2^{d}
\]
However, since $\calM$ is $\varepsilon$-DP, we have by group privacy that:
\[
\frac{\Pr[\calM(Y_{a'})\in\calB_{a'}]}{\Pr[\calM(Y_{a})\in\calB_{a'}]} \leq e^{n\varepsilon}
\]
which for $n<\frac{(d+1)\ln{2} - \ln(3)}{\varepsilon}$ is a contradiction.
\end{proof}
Thus, we have the following lower bound on the additive error on $\varepsilon$-DP $\eta$-approximations of \owm{$n,d$}:
\begin{corollary} \label{cor:owmepslb}
There exist constants $c_1,c_2>0$, such that if $n \le c_1 \cdot d/\varepsilon$, then for every $\varepsilon$-DP mechanism $\calM\colon\{0,1\}^{n\times d}\rightarrow[0,1]^d$ that is $(\eta,\alpha)$-accurate
it holds that $\eta \cdot \alpha > c_2$.
\end{corollary}

\subsection{Approximate Differential Privacy}
\label{sec:owmapproxdp}

We show that the lower bound for approximate DP given by \cite{bun2018fpc} can be extended to mechanisms with multiplicative errors. We follow their proof structure, and recall two definitions that will be used in this section. We first introduce the definition of accuracy that we will be using in this subsection, which allows for a $\beta$ fraction of the queries to be answered with unbounded error.

\begin{definition} [$(\eta,\alpha,\beta)$-Accuracy for \owm{$n,d$}]
Let $\eta\geq 1$ and $\alpha,\beta\in[0,1]$ be parameters. An output sequence $(a_1,\ldots,a_d)$ is an $(\eta,\alpha,\beta)$-accurate answer to \owm{$n,d$} on database $D\in(\{0,1\}^d)^n$ if
\[
\frac{1}{\eta} \cdot \frac{1}{n}\sum_{i=1}^nD_{i,j}  - \alpha
\leq  a_j
\leq \eta \cdot \frac{1}{n}\sum_{i=1}^nD_{i,j} + \alpha
\]
for at least a $1-\beta$ fraction of columns $j\in[d]$. A mechanism $\calA$ is $(\eta,\alpha,\beta)$-accurate for \owm{$n,d$} if for every database $D\in(\{0,1\}^d)^n$,
\[
\Pr[\calA(D)\textrm{ is an }(\eta,\alpha,\beta)\textrm{-accurate answer to \owm{$n,d$} for }D]\geq 2/3
\]
\end{definition}
Note that if a mechanism $\calM$ is $(\eta,\alpha,\beta)$-accurate, then for every $\beta'$ such that $\beta\leq\beta'\leq1$, $\calM$ is also $(\eta,\alpha,\beta')$-accurate. Thus, if there is no $(\eta,\alpha,\beta')$-accurate mechanism for \owm{$n,d$}, then there does not exist an $(\eta,\alpha,\beta)$-accurate mechanism for \owm{$n,d$} (for $0\leq\beta\leq\beta'\leq1$). This means proving an impossibility result for $(\eta,\alpha,\beta)$-accurate mechanisms for \owm{$n,d$} will imply the impossibility of $(\eta,\alpha,0)$-accurate mechanisms (which is equivalent to $(\eta,\alpha)$-accuracy).

\begin{definition}[Reidentifiable Distribution] \label{def:reidentifiabledist}
For $n,d \in \mathbb{N}$, let $\calD$ be a distribution on $n$-row databases with $d$ columns, $D \in (\{0,1\}^d)^n$.
The distribution $\calD$ is $\gamma$-reidentifiable from $(\eta, \alpha, \beta)$-accurate answers to \owm{$n,d$} if there exists an adversary $\calB\colon(\{0,1\}^d)^n\times[0,1]^d\rightarrow [n]\cup\{\bot\}$ such that for every mechanism $\calA\colon (\{0,1\}^d)^*\rightarrow[0,1]^d$, the following both hold:

\begin{enumerate}
\item
$
\Pr_{D\leftarrow_R\calD}[(\calB(D,\calA(D))=\bot)\wedge (\textrm{$\calA(D)$ is $(\eta,\alpha,\beta)$-accurate for \owm{$n,d$})}] \leq \gamma.
$
\item
For every $i \in [n]$,
$
\Pr_{D \leftarrow \calD}[{\calB(D,  \calA(D_{-i})) = i}] \leq \gamma.
$
\end{enumerate}
Here, $D_{-i}$ corresponds to $D$ after removing its $i\textsuperscript{th}$ row and the probabilities are taken over the choice of $D$, $i$, and coin flips of $\calA$ and $\calB$. Also, $\calD$ and $\calB$ can share a common state.
\end{definition}
The common state shared by $\calD$ and $\calB$, is an auxiliary string $aux$ that $\calD$ sends to $\calB$ (but not $\calA$) about the realization of $D$ and is supposed to help $\calB$ reidentify the index $i$.

Intuitively, the first condition says that with high probability $1-\gamma$, it does not simultaneously occur that $\calB$ receives an accurate answer for the whole database $D$ from $\calA$ and it does not accuse any index at all. The second condition says that the probability of $\calB$ accusing some index, whose row has been removed from the database used by $\calA$ (which is then a false accusation) is at most $\gamma$.

We show that the existence of a reidentifiable distribution for \owmm implies (under a condition on the parameters) strong lower bounds on the accuracy guarantees of any $(\varepsilon, \delta)$-differentially private mechanism for \owmm for certain parameters.
Note that since $D$ is public, the output of $\calB$ is a post-processing of the output of $\calA$.
\begin{lemma}
\label{lem:ReIdToNoDP}
Let $\alpha,\beta,\gamma\in[0,1]$, $\eps>0$, $\delta \in(0,1]$, $\eta\geq 1$ be parameters and let $d,n \ge 1$ be integers.
If there exists a $\gamma$-reidentifiable distribution $\calD$ from $(\eta, \alpha, \beta)$-accurate answers to \owm{$n,d$}, then there is no $(\eta, \alpha, \beta)$-accurate $(\eps,\delta)$-DP mechanism for \owm{$n,d$} such that
\[
e^{-\eps} \cdot \frac{1}{n} \cdot \left( \frac{2}{3} - \gamma \right) - \delta
> \gamma.
\]
In particular, for $\eps=1$, this holds for any $\delta < 2/(3en) - \gamma - \gamma/(en)$, and so choosing $\gamma = 1/(100n)$ and  $\delta \le 1/(5n)$ suffices.

\end{lemma}

\begin{proof}
Assume to the contrary that there exists an $\calA$ with such properties.
First, by the accuracy guarantee of $\calA$ we have:
\[
\Pr[\calA(D) \textrm{ is \textbf{not} } (\eta,\alpha,\beta)\textrm{-accurate for \owm{$n,d$}}]\leq 1/3
\]
and from the first property of \cref{def:reidentifiabledist},
\[\Pr[\calA(D) \textrm{ is } (\eta,\alpha,\beta)\textrm{-accurate for \owm{$n,d$} }\wedge (\calB(D,\calA(D))=\bot)]\leq\gamma
\]
Taking the complement of this event and a union bound, we get
\[
\Pr[\calA(D) \textrm{ is \textbf{not} } (\eta,\alpha,\beta)\textrm{-accurate for \owm{$n,d$} }]
+\Pr[(\calB(D,\calA(D))\neq\bot)]
\geq 1-\gamma
\]
which when combined with the accuracy guarantee gives
\[
\Pr_{D \leftarrow_R \calD} [\calB(D, \calA(D)) \neq \bot]
\geq \frac{2}{3} - \gamma
\]
By an averaging argument, there exists an $i^*\in[n]$ such that the following holds:
\[
\Pr_{D \leftarrow_R \calD} [\calB(D,\calA(D)) = i^*]
\geq \frac{1}{n} \cdot \left( \frac{2}{3} - \gamma \right)
\]
On the other hand, since $\calB$ is a post-processing of $\calA$, it is $(\eps, \delta)$-DP. So we have:
\[
\Pr_{D \leftarrow_R \calD} [\calB(D,\calA(D_{-i^*})) = i^*]
\geq e^{-\eps} \cdot \Pr_{D \leftarrow_R \calD}[\calB(D,\calA(D)) = i^*] - \delta
\geq e^{-\eps} \cdot \frac{1}{n} \cdot \left( \frac{2}{3} - \gamma \right) - \delta
\]
However, from the second property of \cref{def:reidentifiabledist}, we have:
$$\Pr_{D\leftarrow_R\calD}[\calB(D,\calA(D_{-i^*}))=i^*]\leq \gamma$$
which leads to a contradiction when
\[
e^{-\eps} \cdot \frac{1}{n} \cdot \left( \frac{2}{3} - \gamma \right) - \delta
> \gamma.
\qedhere
\]
\end{proof}

Now, we restate a variant of the definition of fingerprinting codes introduced in \cite{bun2018fpc}. Fingerprinting codes were initially introduced by Boneh and Shaw~\cite{Boneh1998}. Assume there is a set of $n$ users, $\{1, \dots, n\}$. A fingerprinting code is a pair (Gen, Trace) of randomized mechanisms where Gen outputs a binary codebook $C\in(\{0,1\}^d)^n$ containing $n$ codewords of length $d$  such that $c_i$ is the codeword of user $i$ and $\textrm{Trace}\colon (\{0,1\}^d)^n\times\{0,1\}^d\rightarrow [n]$ is supposed to trace back one of the codewords used in a \textit{coalition}'s codeword. In particular, a set $S\subseteq[n]$ can form a coalition to create a certain codeword $c'$ from their own codewords in the codebook ($C_S$), maintaining some constraints (formalized  in the set $F_\beta(C_S)$ of feasible codewords). Trace, given the codebook and the new codeword $c'$, has to trace back one of the codewords $i$ or does not accuse anyone of participating in the coalition. However, the probability of the new codeword maintaining the constraint and Trace not accusing anybody should be low; moreover, Trace should not accuse somebody who has not participated in the coalition (except with a small probability). The definition below is a variant of fingerprinting codes with a weaker (compared to the standard definition of fingerprinting codes)
constraint on feasible codewords ($F_\beta(C_S)$).

\begin{definition}[Error-Robust Fingerprinting Codes~\cite{bun2018fpc}] \label{def:rfpc}
For any $n, d \in \mathbb{N}$, $\gamma, \beta \in [0,1]$, a pair of mechanisms $(\textrm{Gen}, \textrm{Trace})$ is an \emph{$(n,d)$-fingerprinting code with security $\gamma$ robust to a $\beta$ fraction of errors} if Gen outputs a codebook $C \in \{0,1\}^{n \times d}$ such that $c_i$ is the codeword of user $i$ and for every (possibly randomized) adversary $\calA_{FP}$, and every coalition $S \subseteq [n]$
with codewords set $C_S=\cup_{i\in S}\{c_s\}$, if we set $c' \leftarrow \calA_{FP}(C_S)$, then
\begin{enumerate}
\item
$
\Pr[c' \in F_{\beta}(C_S) \wedge \textrm{Trace}(C, c') = \bot] \leq \gamma,
$ \\ where $F_{\beta}(C_S) = \set{c' \in \{0,1\}^{d} \mid \Pr_{j \leftarrow [d]}[\exists i \in S, c'_{j} = c_{ij}] \geq 1 - \beta}$
\item
$
\Pr[\textrm{Trace}(C, c') \in [n] \setminus S] \leq \gamma,
$
\end{enumerate}
where the probability is taken over the coins of Gen, Trace, and $\calA_{FP}$.  The mechanisms Gen and Trace may share a common state.
\end{definition}
The existence of such Gen and Trace is proved in the same paper. (For the standard definition of fingerprinting codes, their existence was shown in \cite{Tardos2008FPC}.)
\begin{theorem}[\citealp{bun2018fpc}]\label{thm:rfpc}
For every $d \in \mathbb{N}$, and $\gamma \in (0,1]$, there exists an $(n, d)$-fingerprinting code with security $\gamma$ robust to a $1/75$ fraction of errors for
$$
n = n(d, \gamma) = \tOmega(\sqrt{d / \log(1/\gamma)}).
$$
\end{theorem}
Note that for any $n\in \mathbb N$, if there exists an $(n, d)$-fingerprinting code with security $\gamma$ robust to a $\beta$ fraction of errors, then there also exists one for
any $n'\in\mathbb N$ where $n'\leq n$.

It then follows that for every $d \in \mathbb{N}$, and $\gamma \in (0,1]$, there exists a function $n'(d,\gamma)=\tTheta(\sqrt{d / \log(1/\gamma)})$ (possibly slower growing than $n (d, \gamma)$) such that for every $n\leq n'(d,\gamma)$, there exists an $(n, d)$-fingerprinting code with security $\gamma$ robust to a $1/75$ fraction of errors. Since we will need the exact terms inside the asymptotic notation in our proofs, we use constants $c_1>0$ and $c_2<0$ such that for large enough $d$, the above-mentioned fingerprinting code exists for every $n\leq c_1\sqrt{d}\log^{c_2}\frac{d}{\gamma}$.

The following lemma will show that the existence of fingerprinting codes implies the existence of reidentifiable distribution for \owm{$n,d$}.
\begin{lemma}
\label{lem:FpcToReId}
Given $n,d\in\mathbb{N}$ and $\beta,\gamma\in[0,1]$, if there exists an $(n,d)$-fingerprinting code with security parameter $\gamma$ robust to a $\beta$ fraction of errors, then there exists a distribution $\calD$ on databases $D\in(\{0,1\}^d)^n$ that is $\gamma$-reidentifiable from $(\eta,\alpha,\beta)$-accurate answers to \owm{$n,d$} if $\eta \cdot \alpha < \frac{1}{2}$.
\end{lemma}
\begin{proof}
First, we define the distribution $\calD$ and adversary $\calB$. Let $\calA$ be the $(\eta,\alpha,\beta)$-accurate mechanism for \owm{$n,d$}. In this construction, $\calD$ will be Gen itself and $\calB$ will be $\textrm{Trace}(D, \textrm{round}_\alpha(\calA(D)))$ where $\textrm{round}_\alpha(\calA(D))$ is the output of the mechanism $\calA(D)$ after doing the following: for each entry of $\calA(D)$, we round it to 0 if it is less than $\alpha$, and to 1 otherwise. Also, the common state of $\calD$ and $\calB$ will be the same as the common state of Gen and Trace.
Now, we will prove the properties of \cref{def:reidentifiabledist}.
Note that an all 0s (resp.~1s) column in $C_S$ corresponds to a bit that must be set to 0 (resp.~1) in $c'$.

Every mechanism $\mathcal A$ for \owm{$n,d$} that is $(\eta,\alpha,\beta)$-accurate
satisfies that for at least $(1-\beta) \cdot d$ many of the columns,
\[
\frac{1}{\eta} \cdot \frac{1}{n}\sum_{i=1}^nD_{i,j}  - \alpha
\leq  a_j
\leq \eta \cdot \frac{1}{n}\sum_{i=1}^nD_{i,j} + \alpha
\]
In particular, for any column that is all $0$s or all $1$s, the accuracy guarantee $\eta \cdot \alpha < \frac{1}{2}$ also implies that $a_j(0)<\alpha<a_j(1)$, where $a_j(i)$ is the output on the all $i$s column.
Thus, by the definition of $F_\beta$ (except for a $\beta$ fraction of the columns), every entry corresponding to an all 0s or all 1s column is decoded by $\textrm{round}_\alpha$ as 0 or 1, respectively.
Thus, by definition of $F_\beta(D)$, we have that $\textrm{round}_\alpha(\calA(D))\in F_{\beta}(D)$.
This gives that the first property of error-robust fingerprinting codes implies the first property of \cref{def:reidentifiabledist}. Finally, we show that the second property of fingerprinting codes (of $\textrm{Trace}$ wrongly returning an index that does not belong to $S$) also implies the second property of reidentifiable distributions. For every $i\in[n]$:
\begin{align*}
    \Pr_{D \leftarrow \calD}[{\calB(D,  \calA(D_{-i})) = i}]=\Pr_{D \leftarrow \textrm{Gen}}[\textrm{Trace}(D, \textrm{round}_\alpha(\calA(D_{-i})))=i]\leq\gamma
\end{align*}
where the equation is from the constructions above and the inequality comes from the second property in \cref{def:rfpc} by applying $S=[n]\setminus{i}$. which completes the proof.
\end{proof}
Now, we are ready to prove the main result of this section.

\begin{restatable}{theorem}{apxowmepsdelta}
\label{thm:apxowmepsdelta}
For $d\in\mathbb{N}$, $n=\tO(\sqrt{d})$, $\alpha\in[0,1]$ and $\eta\geq 1$ such that $\eta \cdot \alpha < \frac{1}{2}$, there exists no $(1,O(1/n))$-DP mechanism for \owm{$n,d$} that is $(\eta,\alpha,1/75)$-accurate.
\end{restatable}
\begin{proof}

For any $d\in\mathbb{N}$ and $\gamma\in(0,1]$, it follows from \cref{thm:rfpc} and \cref{lem:FpcToReId} that there exists a function $n'(d,\gamma)=\tTheta(\sqrt{d/\log{1/\gamma}})$ such that for any $n\leq n'(d,\gamma)$ there exists a distribution $\calD$ on databases $D\in(\{0,1\}^d)^n$ that is $\gamma$-reidentifiable from $(\eta,\alpha,1/75)$-accurate answers to \owm{$n,d$}, whenever $\eta \cdot \alpha < \frac{1}{2}$.
Now, for $\gamma=\frac{c}{d}$ and $n\leq n'(d,\gamma)=\tTheta(\sqrt{d})$, we have that $\gamma=\tO(1/n^2)$. We can choose $c$ such that $\gamma\leq\frac{1}{100n}$.
Thus, we can use \cref{lem:ReIdToNoDP} to conclude that for every $\alpha \in [0,1]$, $\eta \ge 1$ and every $d\in\mathbb{N}$, $n\leq n'(d,\frac{1}{d})=\tTheta(\sqrt{d})$, and $\eta \cdot \alpha < \frac{1}{2}$, there is no
$(1,O(1/n))$-DP mechanism for \owm{$n,d$} that is $(\eta,\alpha,1/75)$-accurate.
\end{proof}

In particular, this also shows that any $(\eta, \alpha)$-approximation mechanism must have $\eta \cdot \alpha \ge 1/2 = \Omega(1)$.

Now, to extend the result above from $(\varepsilon_0,\delta)$-DP for constant $\varepsilon_0$ to $(\varepsilon_0,\delta)$-DP for any  $\eps\leq 1$,
 we use the following lemma which is similar to Lemma 2.6 of \cite{bun2018fpc} implied by \cite{Kasiviswanathan2008privately}. However, since they only showed the lemma for the case of $\eta=1$, we prove it here.

\begin{lemma}
    \label{lem:SampleComplexityAllEps}
 Let $\alpha,\beta\in[0,1]$, $1\leq\eta$, let constant $\eps_0>0$ and $0<\eps\leq\eps_0$.  If \owm{$n,d$} has sample complexity $n^*$ for $(\eta,\alpha,\beta)$-accuracy and $(\eps_0,\delta)$-differential privacy for $\delta=O(1/n)$, then \owm{$kn,d$}
 (for $k=\lfloor{\frac{\eps_0}{\eps}}\rfloor$)
 has sample complexity $\Omega(\frac{n^*}{\eps})$ for $(\eta,\alpha,\beta)$-accuracy and $(\eps,\delta')$-differential privacy where $\delta'(kn)=O(1/(kn))=O(\eps/n)$.
\end{lemma}

\begin{proof}
Let $k=\lfloor{\frac{\eps_0}{\eps}}\rfloor \ge 1$. We prove the contrapositive.
We show that for every $n$, if there exists an $(\eps,\delta')$-DP (for a suitable $\delta'(n)=O(1/n)$ to be chosen later), $(\eta,\alpha,\beta)$-accurate mechanism $\calM'(\cdot)$ for \owm{$kn,d$} with $k \cdot n = \lfloor{\frac{\eps_0}{\eps}}\rfloor n=O(\frac{n}{\eps})$ samples, then there exists an $(\eps_0,\delta)$-DP $(\eta,\alpha,\beta)$-accurate mechanism $\calM(\cdot)$ with $\delta(n)=O(1/n)$ that solves \owm{$n,d$} with $n$ samples.

Now, we assume that $\calM'$ exists and show how to construct $\calM(X)$. Let $X\in(\{0,1\}^d)^n$ be the private input dataset with $n$ samples that $\calM$ receives. $\calM$ simply outputs $\calM'(f(X))$, where $f(X)\in (\{0,1\}^d)^{kn}$ is a dataset constructed by $k$ copies of $X$. Since the correct answers of \owm{$\cdot,d$} are equal for $X$ and $f(X)$, $\calM$ is $(\eta,\alpha,\beta)$-accurate. Now, we prove that $\calM$ is $(\eps_0, \delta)$-DP. For every pair of neighboring datasets $X,X'$ and subset of outputs $R$, we have the following:
    \begin{align*}
    \Pr[\calM(X)\in R]
    &=\Pr[\calM'(f(X))\in R] \\
    &\leq e^{k\eps} \cdot \Pr[\calM'(f(X'))\in R]+(1+e^\eps+\ldots+e^{(k-1)\eps}) \cdot \delta'(kn) \\
    &\leq e^{k\eps} \cdot \Pr[\calM'(f(X'))\in R]+\frac{e^{k\eps}}{e^\eps-1} \cdot \delta'(kn) \\
    &\leq e^{k\eps} \cdot \Pr[\calM'(f(X'))\in R]+\frac{e^{k\eps}}{\eps} \cdot \delta'(kn) \\
    &\leq e^{\eps_0} \cdot \Pr[\calM'(f(X'))\in R]+\frac{e^{\eps_0}}{\eps} \cdot \delta'(kn),
    \end{align*}
where the first inequality holds since $\calM'$ is $(\eps,\delta')$-DP and $f(X)$ and $f(X')$ are different in $k$ rows and the last inequality holds since $k\eps\leq\eps_0$.

To show that $\calM$ is $(\eps_0,\delta)$-DP, for each $n'\in\mathbb{N}$ and $\delta(n)=O(1/n)$, we
set $\delta(n) = \frac{e^{\eps_0}}{\eps} \cdot \delta'(kn)$, which implies that
$\delta'(n') = \frac{\eps}{e^{\eps_0}} \cdot \delta(\frac{n'}{k})$. Since $\delta(n'/k) = O(k/n')$, and since $\eps_0$ is a constant, we get
\[
\delta'(n') = O \left( \eps \cdot \frac{k}{n'} \right) = O\left( \frac{\eps_0}{n'} \right) = O\left( \frac{1}{n'} \right)
\]
as well, as  required. Thus, from the inequalities above we have:
$$\Pr[\calM(X)\in R]\leq e^{\eps_0} \cdot \Pr[\calM'(f(X'))\in R]+\delta(n)=e^{\eps_0} \cdot \Pr[\calM(X))\in R]+\delta(n)$$
Therefore, $\calM$ is $(\eps_0,\delta)$-DP.
\end{proof}
Note that the above arguments also hold for any set of counting queries. Now, by \cref{lem:SampleComplexityAllEps} and \cref{thm:apxowmepsdelta}, we have the following:
\begin{restatable}{theorem}{apxowmGenepsdelta}
\label{thm:apxowmGenepsdelta}
For $\eps\leq 1$, $d\in\mathbb{N}$, $n=\tO(\sqrt{d}/\eps)$, $\eta\geq1$ and $\alpha \in[0,1]$ such that $\eta \cdot \alpha < \frac{1}{2}$, there exists no $(\eps,O(1/n))$-DP mechanism for \owmm that is $(\eta,\alpha,1/75)$-accurate.
\end{restatable}
As $(\eta,\alpha,1/75)$-accuracy implies $(\eta,\alpha)$-accuracy, the corresponding statement also holds for $(\eta,\alpha)$-accuracy.

\section{Item-Level Lower Bounds for Fully Dynamic Graph Mechanisms}
\label{sec:fd-item}

In this section, we will use reductions from \owmm to show strong lower bounds on fully dynamic item-level graph problems.
We first introduce the framework of graph problems for which we show our lower bounds.
\begin{definition}[Marginals Solving Family]
\label{def:MSF}
Let $n_0\in\mathbb N$ and $g$ be a real-valued graph function. We call a family of tuples $\{(H_n=(V_n,E_n),(e_1,\ldots,e_n))\}_{n>n_0}$
a \textit{marginals solving family for function $g$ with weight $w$ and size $(\nu(n),\xi(n))$} (where $\nu\colon\mathbb N\rightarrow\mathbb N$  is an increasing function of $n$, $\xi\colon\mathbb N\rightarrow\mathbb N$ is an increasing polynomial functions of $n$, and $w$ is a positive number) if the following holds for all $n>n_0$:
\begin{itemize}
    \item $H_n$ is a graph with $\nu(n)$ vertices and $\xi(n)$ edges, i.e., $|V_n|=\nu(n), |E_n|=\xi(n)$,
    \item  One of the following cases happen: (Note that the same case should hold for all values of $n>n_0$)
    \begin{itemize}
        \item $e_1,\ldots,e_n\notin E_n$ and for every subset $S\subseteq[n]$, $g((V_n,E_n\cup\{e_i \mid i\in S\}))=w\cdot|S|$, or
        \item $e_1,\ldots,e_n\in E_n$ and for every subset $S\subseteq[n]$, $g((V_n,E_n\setminus\{e_i \mid i\in S\}))=w\cdot|S|$
    \end{itemize}
\end{itemize}
If there exists such a sequence for a problem $g$, we say that $g$ has a marginal solving family with weight $w$ and size $(\nu,\xi)$.
\end{definition}

Recall that inserting $\bot$ is equivalent to a no-op and has no influence on the graph (and thus the value of the function).

Now, we use a reduction from \owmm to prove a lower bound for graph problems that have marginal solving families.

\begin{theorem} \label{thm:itemLB}
Let $\eps, \delta, \alpha \in [0,1]$ and $\eta \ge 1$. Let $g:\mathcal{G}[V]\rightarrow \mathbb R$ be a function that has a marginals solving family with weight $w$ and size $(\nu,\xi)$, where $\mathcal G[V]$ is the set of all graphs with vertex set $V$. Then, for all large enough $T$ and $N$, if there exists an $(\eps, \delta)$-differentially private mechanism that solves $g$ with multiplicative error $\eta$ and additive error $\alpha$, for inputs with at most $T$ time steps on graphs with at most $N$ vertices, then
\[
\eta \cdot \alpha =
\begin{cases}
\tOmega\left( w \cdot \min\left(\frac{\sqrt[3]{T}}{\eps^{2/3}},\nu^{-}(N),\xi^{-}(T)\right)\right)
& \text{when $\delta>0$ and $\delta=o(1/T)$} \\
\tOmega\left( w \cdot \min\left(\sqrt{\frac{T}{\eps}},\nu^{-}(N),\xi^{-}(T)\right)\right)
& \text{when $\delta = 0$}
\end{cases}
\]
where $f^-(x) = \sup \{ y \mid f(y) \le x \}$ is the generalized inverse as earlier.
\end{theorem}

First, we show a reduction from \owmm to graph problems with a marginal solving family, and then complete the proof of this theorem by fixing the right parameters.

\begin{lemma}[Reduction from \owm{$n,d$} to graph problems that have marginals solving family]\label{lem:reduction_from_owm}
Let $n,d\in\mathbb N$ and $Y\in\{\{0,1\}^d\}^n$. Then for all $w\in\mathbb {R}^+$ and functions $\nu$ and $\xi$ that satisfy conditions of \cref{def:MSF} and each function $g:\mathcal G(V)\rightarrow \mathbb R$ with a marginal solving family with weight $w$ and size $(\nu,\xi)$, there exists a transformation from $Y$
to a fully dynamic graph stream on $N$-node graphs, where $T=\xi(n)+2nd$, $N=\nu(n)$, and the graph after the update on time step $t$ is $G_t$ ($\forall t\in[T]$), such that
\begin{itemize}
    \item The transformations of neighboring $Y,Y'\in\{\{0,1\}^d\}^n$ give item-level neighboring streams, and
    \item defining $t_0=\xi(n)$, for all $j\in[d]$ and
    $t_j=\xi(n)+2jn,$
    we have $g(G_{t_j-n})=w\sum_{i=1}^nY_{i,j}$.
\end{itemize}
\end{lemma}

\begin{proof}
 Assume we are given an instance $Y$ of \owm{$n,d$} with $d$ columns and $n$ rows.
First, we show how the graph sequence is built and then we show the two properties mentioned in the lemma. Assume without loss of generality that the marginals solving family $(H_n=(V_n,E_n),(e_1,\ldots,e_n))$ considered for $g$ is in the first case of the definition ($e_1,\ldots,e_n\notin E_n$). Recall that the number of vertices is $\nu(n)$ and the number of edges in the initial graph is $\xi(n)$.

The graph of this instance will be exactly the graph $H_n$ from the marginals solving family. Intuitively, each one of edges $e_i$ is corresponding to the $i\textsuperscript{th}$ row. We are going to first add the edges $E_n$, and then in $d$ rounds we are going to use this graph to answer the column sum for each one of the columns in $Y$ by modeling the bits set in the column by the subset of edges from $\{e_1, \dots, e_n\}$ that are present in the graph.
Formally, in the time steps $1\leq t\leq \xi(n)$ the edges of $E_n$ are inserted. Then, in the $j\textsuperscript{th}$ round for each $j\in[d]$ ($t_{j-1}<t\leq t_j)$, in time steps $t_{j-1}+i$ for each $i\in [n]$, if $Y_{i,j}=1$ we insert $e_i$ and otherwise we insert $\bot$. Then, in the second half of this round (for the next $n$ time steps), at time step $t_{j-1}+n+i$ we undo what we did in time step $t_{j-1}+i$. The reduction is formalized in \cref{alg:itemGadgetlb}. Note that inserting or deleting edge $e$ (for any edge $e$) means inserting $(+,e)$ or $(-,e)$ in the input stream, respectively.

\begin{algorithm}[th]
\DontPrintSemicolon
\caption{Reduction from \owmm to $g$ with a Marginals Solving Family $\{(H_n=(V_n,E_n),(e_1,\ldots,e_n))\}_{n>n_0}$ of size $(\nu(n),\xi(n))$ and weight $w$}
\label{alg:itemGadgetlb}
\KwInput{A private dataset $Y \in (\{ 0,1 \}^d)^n$}
\KwOutput{Marginal sums $y_i=\frac{1}{n}\sum_{i=1}^nY_{i,j}$}
Set $N = \nu(n)$ and $T = \xi(n)+2dn$.\;
Initialize an $(\eps, \delta)$-DP mechanism $\mathcal A$ solving $g$ with $N$ nodes and $T$ time steps with multiplicative error $\eta$ and additive error $\alpha$.\;
Insert edges of $E_n$ into the graph of $\mathcal A$ in any arbitrary order.\;

\For{$j = 1, \ldots, d$}{
    \For{$i=1,\ldots,n$}{
        \If{$Y_{i,j}=1$}{
            Insert $e_i$ to the graph of $\calA$\;
        }
        \Else{Insert $\bot$ to the graph of $\calA$ \;}
    }
    $\apx y_j$ $\leftarrow$ $\mathcal A$ on the current graph divided by $wn$.\;
    \For{$i=1,\ldots,n$}{
        \If{$Y_{i,j}=1$}{
            Delete $e_i$ from the graph of $\calA$\;
        }
        \Else{Insert $\bot$ to the graph of $\calA$ \;}
    }
}
\end{algorithm}

For the first property of the lemma, note that all of the updates in the graph sequence corresponding to the $i\textsuperscript{th}$ row are done to the edge $e_i$. Thus, the graph sequences corresponding to the neighboring databases $Y,Y'$ are item-level edge-neighboring.

    Regarding the second item, note that after time step $t_0$ the graph is $H_n$ and after time step $t_j-n$ the graph is $(V_n,E_n\cup\{e_i \mid Y_{i,j}=1\}$, thus $g(G_{t_j-n})=w\sum_{i=1}^nY_{i,j}$.
\end{proof}

Now we are ready to prove \cref{thm:itemLB}.
\begin{proof}[Proof of \cref{thm:itemLB}]
Given a mechanism that is $(\varepsilon, \delta)$-DP for solving $g$, we will first choose integer values for $n$ and $d$ such that $\nu(n)\leq N$, $\xi(n)+2nd\leq T$ and
(1) for $\delta=0$,
$n \le c_1 d/\varepsilon$, where $c_1$ is the constant from \Cref{cor:owmepslb}, and
(2) for $\delta>0$, $n = \tO(\sqrt d/\varepsilon)$  as needed for \Cref{thm:apxowmGenepsdelta}.
Then we will use the reduction from~\Cref{lem:reduction_from_owm} to construct a fully dynamic graph stream on $\nu(n)$-node graphs and length $\xi(n)+2nd$ such that any $(\eta,\alpha)$-accurate algorithm for the function $g$ on this stream gives an
$(\eta/(n \cdot w),\alpha)$-accurate algorithm for \owm{$n,d$}.
Note that an $(\eta/(n \cdot w),\alpha)$-accurate mechanism is also an
$(\eta/(n \cdot w),\alpha, 0.75)$-accurate mechanism.
Then, since we show that a mechanism to solve $g$ for $N$ and $T$ can solve any instance of \owm{$n,d$} (with the privacy and accuracy guarantees mentioned above), we apply the lower bounds from \Cref{sec:owmlb} to finish the proof.
The details are as follows.

\textbf{Case 1: $\delta>0$:}
We set
\[
d=\left\lfloor\left(T\eps\right)^{2/3}\right\rfloor
\quad\text{and}\quad
n=\min\left(\left\lfloor\frac{\sqrt d}{\eps}p(d)\right\rfloor,\nu^{-}(N),\xi^{-}\left(T-2d\left\lfloor\frac{\sqrt d}{\eps}p\left(d\right)\right\rfloor\right)\right)
\]
where $p(d)$ hides the polylogarithmic factors in \cref{thm:apxowmGenepsdelta} (so \cref{thm:apxowmGenepsdelta} holds for $n\leq{\sqrt{d}}p(d)/\eps$). As discussed in \cref{sec:owmapproxdp}, we have that there exist constants $c_1>0$ and $c_2<0$ such that $p(d)\leq c_1\log^{c_2}d$.

Now, we show that the conditions $n= \tO({\sqrt{d}}/{\eps})$ (according to \cref{thm:apxowmGenepsdelta}), $\nu(n)\leq N$ and $\xi(n)+2nd\leq T$  hold.
The first condition trivially follows from the first argument in the minimum used in construction of $n$ and the second condition follows from the second argument of the minimum and since $\nu$ is increasing.
For the third condition, we know that since $n\leq\xi^{-}\big(T-2d\lfloor{\sqrt d}p(d)/{\eps}\rfloor\big)$, we have $\xi(n)\leq T-2d\lfloor{\sqrt d}p(d)/{\eps}\rfloor$. Thus,
\begin{equation}
\label{eq:delg0}
\xi(n)+2d\left\lfloor{\sqrt d}p(d) / \eps\right\rfloor \leq T
\end{equation}
On the other hand, we have that
\begin{align*}
\xi(n) + 2nd
&\le \xi(n)+2 \cdot \left\lfloor{\sqrt d}p(d) / \eps\right\rfloor \cdot d \tag*{\text{(since $n \le \left\lfloor{\sqrt d}p(d)/\eps\right\rfloor$)}} \\
&\le T \tag*{(by \cref{eq:delg0})}
\end{align*}

Note that $\lfloor{\sqrt d}p(d) / \eps\rfloor=\ttheta(\sqrt{d}/\eps)$. Also, since as mentioned above $p(d)=o(1)$ and $d=\lfloor(T\eps)^{2/3}\rfloor$, $T-2d\lfloor\frac{\sqrt d}{\eps}p(d)\rfloor=\theta(T)$. Thus, $\xi^{-}\big(T-2d\lfloor\frac{\sqrt d}{\eps}p(d)\rfloor\big)=\theta(\xi^-(T))$. Therefore, in the constructions mentioned above,
\[
d=\theta\left((T\eps)^{2/3}\right)
\quad\text{and}\quad
n=\ttheta\left(\min\left({T^{1/3}}/{\eps^{2/3}},\nu^{-}(N),\xi^{-}(T)\right)\right)
\]

Then, \Cref{lem:reduction_from_owm} shows that for $\eps=O(1), \delta=o(1/T)$ any item-level $(\eps,\delta)$-edge-DP $(\eta,\alpha)$-accurate mechanism for $g$ will imply an $(\eps,\delta)$-DP $(\eta,\frac{\alpha}{wn})$-accurate mechanism for \owm{$n,d$} since $\frac{1}{n}\sum_{i=1}^nY_{i,j}=\frac{g(G_{t_j-n})}{nw}$.

Thus, \cref{thm:apxowmGenepsdelta} implies  that for any $(\eps,o(1/T))$-DP mechanism that solves $g$ $(\eta,\alpha)$-accurately, $\alpha=\tOmega(\frac{nw}{\eta})=\tOmega(\frac{1}{\eta}\cdot w\min(\frac{T^{1/3}}{\eps^{2/3}},\nu^{-1}(N),\xi^{-1}(T)))$.

\textbf{Case 2: $\delta=0$:}
This case can be solved similar to the previous case. We set
\[
d=\theta\left(\sqrt{T\eps}\right)
\quad\text{and}\quad
n= \theta\left(\min\left(\sqrt{{T}/{\eps}},\nu^{-1}(N),\xi^{-1}(T)\right)\right)
\]
such that the conditions $\nu(n)\leq N$, $\xi(n)+2nd\leq T$ and $n= O({d}/{\eps})$ (according to \cref{cor:owmepslb}) hold. We show that this is possible similar to the previous case: As $n=O(\sqrt{{T}/{\eps}})$ and $d=O(\sqrt{T\eps})$, it holds that $nd=O(T)$. Also, $n=O(\xi^{-1}(T))$, $n=O(\nu^{-1}(N))$, and $\nu(\cdot)$ being increasing and $\xi(\cdot)$ being an increasing polynomial implies that $\xi(n)=O(T)$ and $\nu(n)=O(N)$. Finally, by choice of $n$ and $d$, $n=O(\frac{d}{\eps})$. Therefore, by choosing the right constant factors for $n$ and $d$,  all the conditions above hold.

The bound in \cref{cor:owmepslb} implies that for any $\eps$-DP mechanism that solves $g$ $(\eta,\alpha)$-accurately, it must hold that
$\alpha= \Omega(\frac{nw}{\eta})=\Omega(\frac{1}{\eta}\cdot w\min(\sqrt{\frac{T}{\eps}},\nu^{-1}(N),\xi^{-1}(T)))$.
\end{proof}

Now, we will use our framework to show lower bounds for a variety of problems using \cref{thm:itemLB}. First, let us mention the definition of 1-edge distinguishing gadgets by Raskhodnikova and Steiner~\cite{RS25graphdp}. This gadgets are defined for additive functions, which are defined as follows:
\begin{definition}[Additive Graph Functions]
    A function $g$ is additive if for every graph $G$ with connected components $C_1,\ldots, C_m$, we have $g(G)=\sum_{i\in[m]}g(C_i)$.
\end{definition}
\begin{definition}[1-Edge Distinguishing Gadgets~\cite{RS25graphdp}]
Let $g$ be a real-valued additive graph function,
$H=(V',E')$ be a graph and $e\in V'\times V'$. Set $n_g=|V'|$ and $m_g=|E'|$. The pair $(H,e)$ is called a 1-edge distinguishing gadget for $g$ of size $(n_g,m_g)$ and weight $w$ if either:
\begin{itemize}
    \item $e\notin E'$ and $g((V',E'\cup\{e\}))=g(H)+w$ or
    \item $e\in E'$ and $g((V',E'\setminus\{e\}))=g(H)+w$
\end{itemize}
If there exists such a pair for a problem $g$, we say that \textit{$g$ has a 1-edge distinguishing gadget}.
\end{definition}
In their theorem, Raskhodnikova and Steiner show the following lower bound for every function $g$ that has a 1-edge distinguishing gadget.
In \cref{cor:itemlbgraph}, among other problems, we show a stronger lower bound
for all functions $g$ that have a 1-edge-distinguishing gadget such that $g(H)=0$, as defined below:
\begin{definition}[0-based 1-Edge Distinguishing Gadgets]
    Let $g$ be a real-valued additive graph function, $H=(V',E')$ be a graph and $e\in V'\times V'$. Set $n_g=|V'|$ and $m_g=|E'|$. The pair $(H,e)$ is called a 0-based 1-edge distinguishing gadget for $g$ of size $(n_g,m_g)$ and weight $w$ if $g(H)=0$ and either:
    \begin{itemize}
        \item $e\notin E'$ and $g((V',E'\cup\{e\}))=w$ or
        \item $e\in E'$ and $g((V',E'\setminus\{e\}))=w$
    \end{itemize}
    If there exists such a pair for a problem $g$, we say that \textit{$g$ has a 0-based 1-edge distinguishing gadget}.
\end{definition}
Here are a few examples of 0-based 1-edge distinguishing functions:
\begin{itemize}
    \item \matching: For this problem setting $H$ to be two isolated vertices and $e$ to be the potential edge between them, gives a 0-based 1-edge distinguishing gadget of size $(2,0)$ and weight 1.
    \item Subgraph Counting: For the problem of counting instances of any subgraph $G$, we can have a 0-based 1-edge distinguishing gadget for it, by setting $H$ to be $G-e$ for any arbitrary edge $e$ in $G$. Then $(G-e, e)$ will be a 0-based 1-edge distinguishing gadget of size $(|V(G)|,|E(G)-1|)$ and weight $1$. When $G = C_3 = K_3$, the cycle graph on $3$ vertices, this gives the well-studied triangle counting problem.
\end{itemize}
Our lower bounds are stronger than that of \cite{RS25graphdp} since our bounds hold even for mechanisms that have multiplicative error.
Now, we are ready to show our lower bounds.

\begin{corollary}
\label{cor:itemlbgraph}
Let $\eps, \delta \in [0,1]$ and $\eta \ge 1$. For every large enough $T$ and $N$, if there is a mechanism that is $(\eps,\delta)$-DP and $(\eta,\alpha)$-accurate for the functions below, then the bound in front of them holds for $\eta$ and $\alpha$:
\begin{itemize}
    \item Every $g$ that has a 0-based 1-edge distinguishing gadget of size $(n_g,m_g)$ and weight $w$: If $0<\delta<o(1/T)$, then $\eta \cdot \alpha =\tOmega(w\cdot\min(\frac{\sqrt[3]{T}}{\eps^{2/3}},\frac{N}{n_g},\frac{T}{m_g}))$, and if $\delta=0$, then $\eta \cdot \alpha =\tOmega(w\cdot\min(\sqrt{\frac{T}{\eps}},\frac{N}{n_g},\frac{T}{m_g}))$.
    \item $f_{\geq\tau}$ and \kcore: If $0<\delta<o(1/T)$, then $\eta \cdot \alpha =\tOmega(\min(\frac{\sqrt[3]{T}}{\eps^{2/3}},{N},\sqrt{T}))$, and if $\delta=0$, then $\eta \cdot \alpha =\tOmega(\min(\sqrt{\frac{T}{\eps}},N))$.
    \item \mincut: If $0<\delta<o(1/T)$, then $\eta \cdot \alpha =\tOmega(\min(\frac{\sqrt[3]{T}}{\eps^{2/3}},{N},\sqrt{T}))$, and if $\delta=0$, then $\eta \cdot \alpha =\tOmega(\min(\sqrt{\frac{T}{\eps}},N))$.
    \item \stmincut: If $0<\delta<o(1/T)$, then $\eta \cdot \alpha =\tOmega(\min(\frac{\sqrt[3]{T}}{\eps^{2/3}},{N},T))$, and if $\delta=0$, then $\eta \cdot \alpha =\tOmega(\min(\sqrt{\frac{T}{\eps}},N,T))$.
    \item \edgecount: If $0<\delta<o(1/T)$, then $\eta \cdot \alpha =\tOmega(\min(\frac{\sqrt[3]{T}}{\eps^{2/3}},N^2))$, and if $\delta=0$, then $\eta \cdot \alpha =\tOmega(\min(\sqrt{\frac{T}{\eps}},N^2))$.
\end{itemize}
\end{corollary}
\begin{proof}
    We will prove the bounds by showing marginals solving families for each one of the problems mentioned above. The lower bounds follow directly from \cref{thm:itemLB}.
    \begin{itemize}
        \item $g$ that has a 0-based 1-edge distinguishing gadget $(H,e)$ of size $(n_g,m_g)$ and weight $w$: We will make $n$ copies of $H$ and call the copies of $e$ on them $e_1,\ldots,e_n$. One can see that this is a marginals solving family for function $g$ with weight $w$ and size $(\theta(nn_g),\theta(nm_g))$.
        \item $f_{\geq\tau}$: Let $H_n=(V_n,E_n)$ be as follows: $V_n=\{v_1,\ldots,v_n\}\cup\{u_1,\ldots,u_n\}\cup\{z_1,\ldots,z_n\}$. Then, we give the vertices $v_1,\ldots,v_n$ a cyclic order (so $v_1$ is right after $v_n$) and for each $i\in[n]$ put an edge between $v_i$ and all of the $\lfloor\frac{\tau-1}{2}\rfloor$ vertices before and all of the $\lfloor\frac{\tau-1}{2}\rfloor$ vertices after it. So now every vertex $v_i$ (for all $i\in[n]$) has degree $\tau-1$ if $\tau$ is odd, and has degree $\tau-2$ if $\tau$ is even. Then, if $\tau$ is even, we add the edge $(v_i,u_i)$ for all $i\in[n]$. Thus, in the graph $H_n$, every vertex $v_i$ (for $i\in[n]$) has degree $\tau-1$ and all other vertices have degree 0 or 1. Now, we set the edges $e_1,\ldots,e_n$ to be $e_i=(v_i,z_i)$ for each $i\in[n]$. It can be seen that for every $\tau\geq 2$, after adding any subset $S$ of edges $\{e_1,\ldots,e_n\}$, there will be exactly $|S|$ vertices with degree $\tau$ and other vertices' degree will be less. For $\tau=1$, it will be $2|S|$ such vertices and others will have degree 0. Therefore, we have a marginals solving family for $f_{\geq\tau}$ of size $(\theta(n),O(n^2))$ (since $\tau<n$) and weight $1$ for $\tau\geq2$ and weight $2$ for $\tau=1$.
        \item \mincut: Let $H_n$ consist of a vertex $v$ and $n$ vertices that form a $K_n$. Also, let the edges $e_1,\ldots,e_n$ be the edges between $v$ and the $n$ vertices of the $K_n$. The minimum cut of the graph after adding each subset of the edges $\{e_1,\ldots,e_n\}$ will be $(\{v\}, V_n\setminus\{v\})$. Thus, $H_n$ is a marginals solving family for this problem with weight $1$ and size $(\theta(n),\theta(n^2))$.
        \item \stmincut: $H_n$ contains 2 distinguished vertices $s$ and $t$, and $n$ vertices such that $t$ has an edge to all of them and $e_1,\ldots,e_n$ will be the edges between these vertices and $s$. After adding any subset of the edges $\{e_1,\ldots,e_n\}$, the minimum $(s,t)$-cut will be $(\{s\},V_n\setminus\{s\})$. So $H_n$ is a marginals solving family for this problem with weight $1$ and size $(\theta(n),\theta(n))$.
        \item $\kcore(v)$: We will take exactly the same family as \stmincut. Here also the answer will be the number of edges incident to $v$ ($e_1,\ldots,e_n$) existing in the graph. Thus, it has a marginals solving family with weight $1$ and size $(\theta(n),\theta(n^2))$.
        \item \edgecount: $H_n$ for this problem will be just an empty graph with $\lceil2\sqrt{n}\rceil$ vertices and $e_1,\ldots,e_n$ any arbitrary $n$ edges of this graph. This will give us a marginals solving family with weight $1$ and size $(\theta(\sqrt{n}),0)$. \qedhere
    \end{itemize}
\end{proof}
\paragraph{Remark. } Note that for the problems \stmincut and \mincut our arguments will still hold if we extend the model to weighted graphs with maximum weight $W$ where neighboring sequences can differ arbitrarily in updates corresponding to the edge between a single pair of vertices. The reductions would have the weight of all of the edges set to $W$. In that case, the lower bounds will be multiplied by $W$.
\section{Bounds on Monotone Symmetric Norm Estimation}
\label{sec:inctopk}

We begin with a polynomial lower bound for \nomulterrmechs for private incremental \topk in \cref{sec:inctopkLB}, then show a polylogarithmic (in $n$) upper bound for the static case in \cref{sec:statictopkub} and a polylogarithmic (in $n$ and $T$) upper bound for the incremental monotone symmetric norm estimation problem in \cref{sec:incsneub}. Since \topk norms are all monotone symmetric norms, the lower bound stated in \cref{sec:inctopkLB} extends to all monotone symmetric norms as well.

\subsection{Insertions-Only \topk Lower Bound}
\label{sec:inctopkLB}
In this subsection, we show a lower bound on estimating all \topk norms under insertions of elements, even for the data structure version.
We reduce from the inner product problem.
\begin{algorithm}[t]
\DontPrintSemicolon
\caption{Reduction from \innerprodq to \inc\topk}
\label{alg:inctopk}
\KwInput{A secret dataset $x \in \{ 0,1 \}^d$, and  $m = \psi d$ queries $q_1, \ldots, q_m$ with $q^j \in \{ 0,1 \}^d$.}
\KwOutput{Inner product answers $\inprod{x, q^1}, \ldots, \inprod{x, q^m}$}
Set $n = d$ and $T = d + 2 \psi d^2$\;
Initialize an $(\eps, \delta)$-DP mechanism $\mathcal A$ for \inc\topk with $n$ elements for $T$ timesteps with additive error $\alpha$.\;
\For{$t = 1, \ldots, n$}{
    \textbf{if} $x_t = 1$ \textbf{then} insert element $t$ \textbf{else} insert $\bot$ \algcomment{insert dataset $x$}
}
\For{$j = 1, \ldots, m$}{
    \For{$t' = 1, \ldots, n$}{
        \textbf{if} $q^j_{t'} = 1$ \textbf{then} insert element $t'$ \textbf{else} insert $\bot$ \algcomment{insert query $j$}
    }
    Set $\textrm{inprod}^j$ as the smallest $k \in \{ 1, \ldots, n \}$ such that $\mathcal A(k) < (j+1)k - \alpha$\;
    \For{$t' = 1, \ldots, n$}{
        \textbf{if} $q^j_{t'} = 0$ \textbf{then} insert element $t'$ \textbf{else} insert $\bot$ \algcomment{neutralize query $j$}
    }
}
\end{algorithm}

\begin{figure}[t]
\includegraphics[width=\textwidth]{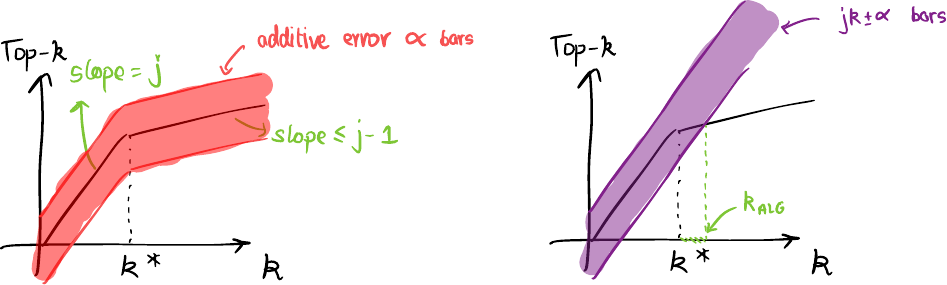}
\caption{Idea for \inc\topk lower bound reductions. This shows the structure of \topk values after the $j-1^{\text{st}}$ query has been inserted. It increases by $j$ until $k^* = \inprod{x, q^{j-1}}$, then reduces to at most $j-1$ later. The left figure shows the accuracy guarantee, while the right figure shows how $k_\alg = \textrm{inprod}^{j-1}$ is chosen. The goal is to accurately estimate $k^*$ by $k_\alg$. The worst case choice of $k_\alg$ occurs at the latest $k'$ when the red-shaded region on the left figure is completely disjoint from the purple-shaded region on the right figure.
}
\label{fig:inctopk}
\end{figure}

\inctopklb*

\begin{proof}
Let $\mathcal{A}$ be a mechanism that solves the data structure version of \inc\topk. We denote by $\mathcal A(k)$ the query value from the mechanism for the value of the $\top{k}$ norm of the current frequency vector $f^t$.
The accuracy guarantee tells us that with probability at least $2/3$, at all times $t \in [T]$ and for all $k \in [n]$,
\[
    | \mathcal A(k) - \|f^t\|_{\top{k}} | \le \alpha
.\]

We condition on the event that the accuracy guarantee holds, and
given an instance of \innerprodq of dimension $d$, we show how to privately answer it with an additive error of at most $2\alpha + 1$ using the mechanism for \inc\topk. The reduction is given in \Cref{alg:inctopk}.

We maintain the invariant that before query $j$ is inserted, element $i$ is present $j$ times if $x_i = 1$, and $j-1$ times if $x_i = 0$.
The reduction starts by inserting element $i$ if $x_i=1$, and inserting $\bot$ otherwise.
For query $q^j$, we perform a sequence of $n$ insertions, by inserting element $i$ if $q^j_i=1$, and inserting $\bot$ otherwise. After these insertions, we have that
\begin{itemize}
    \item if $x_i = 0$ and $q^j_i = 0$, then element $i$ is present $j-1$ times,
    \item if only one of $x_i$ and $q^j_i$ is $1$, then element $i$ is present $j$ times,
    \item if $x_i = 1$ and $q^j_i = 1$, then element $i$ is present $j+1$ times,
\end{itemize}
We use this to estimate the inner product value as explained below, then restore the invariant by inserting element $i$ if $q^j = 0$.
This creates an instance with $n=d$ and $T = d + 2\psi d^2 = d+2dm$ for some constant $\psi>0$.

We observe first that neighboring inputs for \innerprodq reduce to event-level neighboring streams for \inc\topk, since changing the value of one bit of $x$, say the $i\textsuperscript{th}$ bit, changes the insertion of element $i$ to the insertion of a $\bot$ or vice-versa for \inc\topk.

The reduction hinges on the following observation: Consider the frequency vector $f$ after the $(j-1)\textsuperscript{st}$ query was inserted, and let $k^* = \inprod{x, q^{j-1}}$. Then the frequency vector $f$ satisfies
\[
\|f\|_{\top{k}}
\begin{cases}
   = j k & \text{if $k \le k^*$} \\
   \le  j k^* + (j-1) (k-k^*) & k > k^*
\end{cases}
\]
That is, the value of $\|f\|_{\top{k}}$, when plotted against $k$, has slope $j$ for all $k \le k^*$, and it has slope $\le j-1$ for the remaining $k > k^*$. The idea of how this leads to the lower bound is presented in \cref{fig:inctopk}.

Now using $k_\alg = \textrm{inprod}^{j-1}$ (as defined in \Cref{alg:inctopk}) as a shorthand, we first show that $k_\alg > k^*$.
By the accuracy guarantee of $\mathcal A$ and our conditioning it holds for any $k \le k^*$ that
\[
\mathcal A(k) \ge \|f\|_{\top{k}} - \alpha = jk - \alpha
\]
As $k_\alg$ is defined to be the smallest $k'$ such that $\mathcal A(k') < jk'-\alpha$, it follows that $k_\alg > k^*$.

Next, we show that $k_\alg \le k^* + O(\alpha)$. Since the mechanism did not choose $k_\alg - 1$,
\[
\mathcal A(k_\alg - 1) \ge
j \cdot (k_\alg - 1) - \alpha
\]
By the accuracy guarantee of $\mathcal A$,
\[
\mathcal A(k_\alg - 1) \le \|f\|_{\top{ (k_\alg - 1) }} + \alpha \le j k^* + (j-1) \cdot (k_\alg - 1 - k^*) + \alpha
\]
Combining these two inequalities, we get
\[
k_\alg \le k^* + 2 \alpha + 1
\]
which in other words translates to
\[
\textrm{inprod}^{j-1} - \inprod{x, q^{j-1}} \le 2\alpha + 1
\]
for all $j$, giving a mechanism for \innerprodq for any set of $m$ queries  that has, with probability at least $2/3$, an additive error of at most $2\alpha + 1$.
Since any $(\eps, \delta)$-DP mechanism for \innerprodq has additive error $\Omega(\sqrt{d})$ by \cref{thm:innerproductlb}, this gives us a lower bound of
\[
\alpha = \Omega(\sqrt d) = \Omega \left( \min \left\{  \sqrt[4]{T} , \sqrt{n}  \right\} \right)
\]
for \inc\topk.
\end{proof}
Since \topk norms are special cases of monotone symmetric norms, we also have the following:
\begin{corollary}
\label{cor:incSNElb}
Let $\eps \in [0,1]$ and $\delta \in [0,1/3]$.
Any incremental $(\eps, \delta)$-DP \nomulterrmech for the data structure version of \SNE that runs with $n$ elements for $T$ time steps must have additive error $\alpha = \Omega ( \min \{ \sqrt[4]{T} , \sqrt{n} \} )$.
\end{corollary}

\subsection{Static \topk Upper Bound}
\label{sec:statictopkub}

Next, we show that no lower bound that is polynomial in $n$ can exist for the  static \topk estimation problem. Specifically we show how it can be solved using a prefix-sum estimation mechanism as subroutine, thereby giving a mechanism with error polylogarithmic in $n$.

\paragraph*{Prefix sum estimation.}
In the prefix sum estimation problem, the input is a stream $\stream u = u^1, \ldots, u^n$ with $u^i \in \R$ and the output are estimates $a^1, \ldots, a^n$ of the prefix sums. The error is calculated as
\[
\alpha =
\max_{i \in [n]} \left| a^i - \sum_{j = 1}^i u^j \right|
\]
and two streams $\stream u$ and $\stream v$ are said to be neighboring if $\| \stream u - \stream v \|_1 \le 1$.

\paragraph{Offline problems.}
Like in the static setting, all the updates are provided up front in the offline setting. The goal in the offline setting however is to estimate the property after \emph{each} update, while in the static setting, only one estimate sufficed at the end of the stream.

\begin{restatable}{lemma}{statictopkub}
\label{lem:statictopkub}
Any private mechanism for offline prefix-sum estimation can be used to solve the static \topk problem privately with the same additive error guarantees.
\end{restatable}
\begin{proof}
Let $\stream u$ and $\stream v$ be two neighboring inputs for static \topk with frequency vectors $f^{\stream u}$ and $f^{\stream v}$. Let $s^{\stream u}$ and $s^{\stream v}$ be the frequency vectors sorted in decreasing order.
Thus $\|f^{\stream u}\|_{\top{k}} = \sum_{i=1}^k s^{\stream u}_i$ for any $k \in [n]$.

Since $\stream u$ and $\stream v$ are neighboring inputs for \topk, there exists one index $k^*$ such that their frequency vectors agree on all indices except for $k^*$, and it holds that $|f^{\stream u}_{k^*} - f^{\stream v}_{k^*}| = 1$.
Thus the frequency vectors in particular satisfy $\|f^{\stream u} - f^{\stream v}\|_1 \le 1$.
Assume without loss of generality that $f^{\stream u}_{k^*} + 1 = f^{\stream v}_{k^*}$.

Let $\sigma$ be the permutation used to sort $f^{\stream u}$, i.e., $s^{\stream u}_i = f^{\stream u}_{\sigma(i)}$. Note that swapping the indices of different elements with the same frequency do not affect the sorted frequency vector. To this end, consider the permutation $\pi$ that swaps the index $\sigma^{-1}(k^*)$ to be the first index with frequency $f^{\stream u}_{k^*}$. Thus, $i' = \min \{ i \in [n] \mid f^{\stream u}_{\sigma (i)} = f^{\stream u}_{k^*} \}$, and $\pi$ swaps the positions of $i'$ and $\sigma^{-1}(k^*)$. Then $\pi \circ \sigma$ is still a sorting permutation for $f^{\stream u}$. We will show that $\pi \circ \sigma$ is a sorting permutation for $\stream v$ as well.

The sorting property holds for all pairs $i,j \neq \sigma^{-1}(k^*)$ since the streams $\stream u$ and $\stream v$ coincide on all coordinates other than $k^*$. Fix $j = \sigma^{-1}(k^*)$. For all $i$ with $i > \sigma^{-1}(k^*)$, it held that $f^{\stream u}_{\sigma(i)} > f^{\stream u}_{k^*}$ (since $\sigma^{-1}(k^*)$ is the first index in $s^{\stream u}$ with frequency $f^{\stream u}_{k^*}$), and thus it still holds that $f^{\stream v}_{\sigma(i)} \ge f^{\stream v}_{k^*}$.
For all $i$ with $i < \sigma^{-1}(k^*)$, it held that $f^{\stream u}_{\sigma(i)} \le f^{\stream u}_{k^*}$, and thus it still holds that $f^{\stream v}_{\sigma(i)} \le f^{\stream v}_{k^*}$.

Since the $\ell_1$-norm is permutation invariant,
and since the $s$ vectors are obtained by applying $\pi \circ \sigma$ to the corresponding $f$ vectors, we have that $\|s^{\stream u} - s^{\stream v}\|_1 \le 1$ as well.

Thus, since estimating all the prefix-sums of $s$ accurately is the same as estimating all the \topk norms of $f$ accurately, using an $\alpha$-additive error mechanism for prefix sum estimation on the streams $s^{\stream u}$ (with answers $a^k$) gives that for all $k \in [n]$,
\[
  \left| a^k - \|f\|_{\top{k}} \right|
= \left| a^k - \sum_{i=1}^k s_i^{\stream u} \right|
\le \alpha
.\qedhere
\]
\end{proof}

Using the continual counting mechanism of Dwork et al.~\cite{dwork2010differentially} and Chan et al.~\cite{chan2011private} with additive error $O(\log^2 n/\eps)$, we get the following corollary.

\begin{restatable}{corollary}{statictopk}
\label{cor:statictopk}
There exists an $\eps$-DP mechanism that solves static \topk with additive error $O(\log^2 n / \eps)$.
\end{restatable}

\subsection{Incremental Upper Bound for Monotone Symmetric Norm Estimation with Multiplicative Approximation}
\label{sec:incsneub}
We show next that no lower bound polynomial in $T$ and $n$ can exist in the incremental setting when a multiplicative error is allowed.
More specifically, we show that in the incremental setting there exists an $(\eps,\delta)$-differentially private $\eta$-multiplicative approximation mechanism with polylogarithmic in $T$ and $n$ additive error for \topk estimation, and even for the harder problem of monotone symmetric norm estimation.

\probaBoosted*

Towards proving this theorem, we first describe a mechanism whose accuracy guarantee holds for the vector version for a single time step, then extend it to the setting where the accuracy bounds hold simultaneously for all time steps. We defer proofs to \cref{apx:SNE}.

\begin{restatable}{theorem}{main}
\label{thm:main}
Let $\eps>0$ and $\zeta \in (0, 1/2]$.
There exists an $\eps$-DP $(1+\zeta)$-approximate mechanism for the vector version of \inc\SNE
with the additive error upper bounded by
$\tO \lpar \frac{1}{\eps^2 {\zeta}^5} \cdot \log^4(nT)\rpar
\cdot L(e_1)$ for a single fixed time step $t \in [T]$, where $L(e_1)$ is the norm of the first standard basis vector.
\end{restatable}

The high-level intuition of our mechanism, is the following: the continual histogram of \Cref{cor:histogram} are very precise for estimating the size of each large set -- more precisely,
the estimation is correct up to an $(1\pm \zeta)$ factor for each set that has size at least $\calE(n)/\zeta$, where $\calE(n)$ is whp the maximum error of the respective continual $n$-dimensional histogram mechanism $\calH_1$ over all time steps simultaneously and $\zeta >0$.
Thus, we use a first histogram mechanism with privacy parameter $\eps/2$ to estimate each frequency: let $\hat f$ be the outcome of the histogram. $\hat f_i$ is correct up to a factor $1\pm \zeta$ for every element $i$ with $f_i \ge \calE(n)/\zeta$.
Let $C_L = 4 \cdot \calE(n)/\zeta^2$, $C_U = (4+2\zeta) \cdot \calE(n)/\zeta^2$ and let $\tau^f$ be a number chosen uniformly at random in
$[C_L, C_U]$.

Using $\hat f$, we can construct a private vector which captures the elements with ``high" frequency. To do this formally, we define $H^{\hat f} := \{j : \hat f_j > \tau^f\}$ to be the set of elements with estimated frequency more than $\tau^f$ (i.e., the frequencies that are correctly estimated by $\hat f$).
The vector $V^{\hat f}$ is defined as follows: the first $n-|H^{\hat f}|$ coordinates are $0$, and the next ones are the elements of the multiset $\{\hat f_j, j \in H^{\hat f}\}$. Formally, let $j_1, ..., j_{|H^{\hat f}|}$ be the indices in $H^{\hat f}$, we define:
\begin{equation}
    \label{eq:Vhat}
    V^{\hat f} := (\underbrace{0,\ldots,0}_{n-|H^{\hat f}|}, \hat f_{j_1}, \hat f_{j_2}, ..., \hat f_{j_{|H^{\hat f}|}})\in \mathbb{R}^n
\end{equation}

It remains to capture the elements with small frequency -- i.e., estimated frequency less than $\tau^f$. This is delicate, as the estimated frequency of \emph{all} elements might change at every time step, even if the true frequency of only a single element changes -- due to the noise added by the histogram mechanism.
However, since $\hat f_i \leq f_i$, the set  $\{i : \hat f_i > \tau^f\}$ does not overlap with $\{i: f_i \leq \tau^f\}$: our strategy is therefore the following. (1): Capture the contribution of elements with \emph{true} frequency at most $\tau^f$, and (2) show that elements with $\hat f_i \leq \tau^f < f_i$ do not contribute significantly to the norm.
The latter is perhaps our most technical contribution, and builds on the random choice of $\tau^f$. We present it in \Cref{sec:dropped}.
To build a vector that answers (1), we group elements into ``levels" based on their frequency:

\begin{definition} \label{def:level}
    (Levels, definition 3.1 in \cite{Blasiok2017norms}) For $i\in \{1, ..., \lceil\log_{1+\zeta}{\tau^f}\rceil\}$,  the $i$-th level is
    $B_i:=\{j \in [n]:(1+\zeta)^{i-1}\leq f_j < \min((1+\zeta)^i, \tau^f+1)\}$.
    The vector $b\in\mathbb{R}^{\lceil\log_{1+\zeta}{\tau^f}\rceil}$ records the \emph{size} of each level: $b_i:=|B_i|$.  \footnote{Note that in \cite{Blasiok2017norms}, there is no upperbound such as $\tau^f$ on the frequencies of elements in a level and the levels are simply defined as
$B_i:=\{j:(1+\zeta)^{i-1}\leq f_j < (1+\zeta)^i\}$. However, to answer (1), we only need to care about elements with frequency less than $\tau^f$.}
\end{definition}

Importantly, note that the definitions above do not include elements with frequency larger than $\tau^f$.
At each update from the input, the frequency of a single element changes: hence, the size of at most one level increases by $1$, and the size of at most one decreases by $1$.
Thus, we can use a second continual histogram mechanism $\calH_2$, this one for estimating privately the size of each level: The input universe of the histogram mechanism consists of $\lceil\log_{1+\zeta}{\tau^f}\rceil$ many different levels, each new element in the input stream $u$ of the norm estimation mechanism creates up to two entries in the input stream for the histogram mechanism, being  $+1$ when the size of a level increases, $-1$ when it decreases, and 0 when it is unchanged.

Note, however, that the privacy guarantee decreases: two neighboring input streams for the SNE problem can create streams of updates for the level sizes differing at $\lceil\log_{1+\zeta}{\tau^f}\rceil$ entries.
Indeed, suppose the input stream differ for an update concerning element 1: the level size streams will differ any time element 1 changes level, which can happen $\lceil\log_{1+\zeta}{\tau^f}\rceil$ times. Thus, we need to use the privacy parameter $\frac{\eps}{2\lceil\log_{1+\zeta}{\tau^f}\rceil}$ for $\calH_2$, which in turn increases the error by a factor of $2\lceil\log_{1+\zeta}{\tau^f}\rceil$.
Also, the number of columns in $\calH_2$ is $\lceil \log{\tau^f}\rceil$. Therefore, the error bound is $2\lceil\log_{1+\zeta}{\tau^f}\rceil \calE(\lceil\log_{1+\zeta}{\tau^f}\rceil)$.

Let $\hat b$ be the vector produced by this mechanism.
This vector correctly estimates all the level sizes larger than $\tau^b=
2 \lceil\log_{1+\zeta}{\tau^f}\rceil \calE(\lceil\log_{1+\zeta}{\tau^f}\rceil)/\zeta$
, up to a factor $1\pm \zeta$. Let $h^{\hat b}:=\{i:\hat{b}_i\geq \tau^b\}$.
To convert this information into an estimation of the contributions of the ``low" frequency elements, we round all frequencies that are at  level $i$ to $(1+\zeta)^i$, and have a vector with $\hat b_i$ many occurrences of it (instead of $b_i$).
Formally, the estimated level vector $\estV_i$ is:
\begin{equation}
    \label{eq:Ei}
    \estV_i:=(\underbrace{0,\ldots,0}_{\sum_{k\leq i,k\in h^{\hat b}} \hat{b}_k}, \underbrace{(1+\zeta)^i,\ldots,(1+\zeta)^i}_{\hat{b}_i}, 0,\ldots,0)\in \mathbb{R}^n
\end{equation}

This definition ensures that the non-zero elements of two level vectors $\estV_i$ and $\estV_{i'}$ are non-overlapping for $i \neq i'$. Thus, the vector $\sum_{i\in h^{\hat b}}\estV_i$ approximately captures the contribution of all elements that have low frequency {\em and} whose level has large size.

The contribution of the remaining elements to the norm is negligible: indeed, (a) they are in levels with small size, and (b) they have small frequency, hence they do not contribute much to the norm. We drop them, i.e., estimate their frequency by $0$. We show in \Cref{sec:dropped} that with constant probability the resulting error is small.

Our mechanism uses the vector  $\estV = \sum_{i\in h^{\hat b}}\estV_i+V^{\hat{f}}$ as an estimate for $f$ and we show first that, depending on the choice of the continual histogram mechanisms that it uses,  it is either $\eps$-DP or $(\eps, \delta)$-DP. Afterwards we analyze the approximation guarantees.

\begin{algorithm}[t]
\DontPrintSemicolon
\KwIn{Stream $u_1,\ldots, u_t$ of updates}
\KwOut{At each time step, a vector $\estV\in\mathbb{R}^n$}
Choose $\tau^f$ uniformly at random from the range $[4\cdot \calE(n)/\zeta^2, (4+2\zeta) \cdot \calE(n)/\zeta^2]$\;
\For{$i \in \{1, \dots ,\lceil\log_{1+\zeta}{\tau^f}\rceil\}$}{
    $B_i \gets \emptyset$\tcp*[r]{$B_i$ consists of elements with frequency in $\left[(1+\zeta)^{i-1}, \min((1+\zeta)^i, \tau^f+1)\right)$}
}
$\tau^b \leftarrow 2 \lceil\log_{1+\zeta}{\tau^f}\rceil \calE(\lceil\log_{1+\zeta}{\tau^f}\rceil)/\zeta$\;
Initialize two private continual histogram mechanisms, $\calH_1$ for $\hat f$ (with privacy parameter $\frac{\eps}{2}$) and $\calH_2$ for $\hat b$ (with privacy parameter $\frac{\eps}{2\lceil\log_{1+\zeta}{\tau^f}\rceil}$)\;

\ForEach{update $u_t$}{
    $f_{u_t} \gets f_{u_t}+1$\;
    Feed the update into $\calH_1$ to update $\hat{f}$\;
    Feed the changes in $f$ into $\calH_2$ to update $\hat{b}$\;
    Build $V^{\hat{f}}$ according to $\hat{f}$, as in \Cref{eq:Vhat}\;
    Let $h^{\hat b} \gets \{i : \hat b_i \geq \tau^b\}$\;
    Build $\estV_i$ for all $i \in \{1, \dots ,\lceil\log_{1+\zeta}{\tau^f}\rceil\}$ as in \Cref{eq:Ei}\;
    $\estV \gets \sum_{i\in h^{\hat b}}\estV_i+V^{\hat{f}}$\;
    \textbf{Output} $\estV$\;
}
\caption{Output a vector $\estV$ such that $L(\estV)$ is a good approximation of $L(f)$}
\label{alg:main}
\end{algorithm}

We also prove the following results for the case of $(\eps,\delta)$-differential privacy in \Cref{sec:epsdelta}.
\begin{theorem}\label{thm:SNEUBEpsDelta}
    Let $\eps>0$ and $\delta>0$.
    For any $0< \zeta \le 1/2$, there exist an $(\eps, \delta)$-DP mechanism for the vector version of \inc\SNE such that
    at any time step $t$ it holds with probability at least $2/3$ that the multiplicative approximation is $1+\zeta$  and the additive error
    $\tilde O \lpar \frac{1}{\eps^2\zeta^5}\cdot \log^{2.5}(T) \cdot  \log^{0.5}(nT)  \log (1/\delta) \rpar \cdot L(e_1)$.
\end{theorem}
The probability in this result can also be boosted for the data structure version, see \Cref{sec:epsdelta}.

We note that our mechanisms assume the knowledge of $T$, the number of time steps: this is for sake of simplicity, and we can use the technique of \cite{chan2011private} to extend our mechanism to the case where $T$ is unknown at a multiplicative increase by $O((\log T)^{0.1})$ in the additive error bound.\footnote{the constant $0.1$ can be chosen an arbitrarily small $a > 0$, and the error becomes $O(\log T)^a / a)$. See Theorem 4.1 in \cite{chan2011private}.}

\bibliographystyle{gamma}

\bibliography{Ref}

@String{esa = "European Symposium on Algorithms (ESA)"}

@String{stoc = "Symposium on Theory of Computing (STOC)"}

@String{focs = "Foundations of Computer Science (FOCS)"}

@String{tcc = "Theory of Cryptography Conference (TCC)"}

@String{random = "Randomization and Computation (RANDOM)"}

@String{itcs = "Innovations in Theoretical Computer Science (ITCS)"}

@String{neurips = "Neural Information Processing Systems (NeurIPS)"}

@String{pods = "Principles of Database Systems (PODS)"}

@string{soda = "Symposium on Discrete Algorithms (SODA)"}

@String{icml = "International Conference on Machine Learning (ICML)"}

@String{colt = "Conference on Learning Theory (COLT)"}

@inproceedings{jain23countdistinct,
  author       = {Palak Jain and
                  Iden Kalemaj and
                  Sofya Raskhodnikova and
                  Satchit Sivakumar and
                  Adam D. Smith},
  title        = {Counting Distinct Elements in the Turnstile Model with Differential
                  Privacy under Continual Observation},
  booktitle    = neurips,
  year         = {2023},
  url          = {http://papers.nips.cc/paper\_files/paper/2023/hash/0ef1afa0daa888d695dcd5e9513bafa3-Abstract-Conference.html},
  bibsource    = {dblp computer science bibliography, https://dblp.org}
}

@inproceedings{NRS07smoothsensitivity,
 author = {Nissim, Kobbi and Raskhodnikova, Sofya and Smith, Adam D.},
 title = {Smooth sensitivity and sampling in private data analysis},
 year = {2007},
 url = {https://doi.org/10.1145/1250790.1250803},
 doi = {10.1145/1250790.1250803},
 bibsource = {dblp computer science bibliography, https://dblp.org},
 booktitle = stoc
}

@inproceedings{nguyen21icml,
  author       = {Dung Nguyen and
                  Anil Vullikanti},
  title        = {Differentially Private Densest Subgraph Detection},
  booktitle    = icml,
  year         = {2021},
  url          = {http://proceedings.mlr.press/v139/nguyen21i.html},
  bibsource    = {dblp computer science bibliography, https://dblp.org}
}

@inproceedings{DLR22kcore,
 author = {Dhulipala, Laxman and Liu, Quanquan C. and Raskhodnikova, Sofya and Shi, Jessica and Shun, Julian and Yu, Shangdi},
 title = {Differential Privacy from Locally Adjustable Graph Algorithms: k-Core
Decomposition, Low Out-Degree Ordering, and Densest Subgraphs},
 year = {2022},
 url = {https://doi.org/10.1109/FOCS54457.2022.00077},
 doi = {10.1109/FOCS54457.2022.00077},
 bibsource = {dblp computer science bibliography, https://dblp.org},
 booktitle = focs
}

@inproceedings{farhadi22aistats,
  author       = {Alireza Farhadi and
                  MohammadTaghi Hajiaghayi and
                  Elaine Shi},
  title        = {Differentially Private Densest Subgraph},
  booktitle    = {Artificial Intelligence and Statistics (AISTATS)},
  year         = {2022},
  url          = {https://proceedings.mlr.press/v151/farhadi22a.html},
  bibsource    = {dblp computer science bibliography, https://dblp.org}
}

@inproceedings{dali2023neurips,
  author       = {Mina Dalirrooyfard and
                  Slobodan Mitrovic and
                  Yuriy Nevmyvaka},
  title        = {Nearly Tight Bounds For Differentially Private Multiway Cut},
  booktitle    = neurips,
  year         = {2023},
  url          = {http://papers.nips.cc/paper\_files/paper/2023/hash/4e8f257e054abd24c550d55e57cec274-Abstract-Conference.html},
  bibsource    = {dblp computer science bibliography, https://dblp.org}
}

@inproceedings{Fan22neurips,
  author       = {Chenglin Fan and
                  Ping Li and
                  Xiaoyun Li},
  title        = {Private Graph All-Pairwise-Shortest-Path Distance Release with Improved
                  Error Rate},
  booktitle    = neurips,
  year         = {2022},
  url          = {http://papers.nips.cc/paper\_files/paper/2022/hash/71b17f00017da0d73823ccf7fbce2d4f-Abstract-Conference.html},
  bibsource    = {dblp computer science bibliography, https://dblp.org}
}

@inproceedings{bodwin2024icalp,
  author       = {Greg Bodwin and
                  Chengyuan Deng and
                  Jie Gao and
                  Gary Hoppenworth and
                  Jalaj Upadhyay and
                  Chen Wang},
  title        = {The Discrepancy of Shortest Paths},
  booktitle    = {International Colloquium on Automata, Languages, and Programming (ICALP)},
  year         = {2024},
  url          = {https://doi.org/10.4230/LIPIcs.ICALP.2024.27},
  doi          = {10.4230/LIPICS.ICALP.2024.27},
  bibsource    = {dblp computer science bibliography, https://dblp.org}
}

@inproceedings{chen2023soda,
  author       = {Justin Y. Chen and
                  Badih Ghazi and
                  Ravi Kumar and
                  Pasin Manurangsi and
                  Shyam Narayanan and
                  Jelani Nelson and
                  Yinzhan Xu},
  title        = {Differentially Private All-Pairs Shortest Path Distances: Improved
                  Algorithms and Lower Bounds},
  booktitle    = soda,
  year         = {2023},
  url          = {https://doi.org/10.1137/1.9781611977554.ch184},
  doi          = {10.1137/1.9781611977554.CH184},
  bibsource    = {dblp computer science bibliography, https://dblp.org}
}

@inproceedings{liu24soda,
  author       = {Jingcheng Liu and
                  Jalaj Upadhyay and
                  Zongrui Zou},
  title        = {Optimal Bounds on Private Graph Approximation},
  booktitle    = soda,
  year         = {2024},
  url          = {https://doi.org/10.1137/1.9781611977912.39},
  doi          = {10.1137/1.9781611977912.39},
  bibsource    = {dblp computer science bibliography, https://dblp.org}
}

@inproceedings{DN03pods,
  author       = {Irit Dinur and
                  Kobbi Nissim},
  title        = {Revealing information while preserving privacy},
  booktitle    = pods,
  year         = {2003},
  url          = {https://doi.org/10.1145/773153.773173},
  doi          = {10.1145/773153.773173},
  bibsource    = {dblp computer science bibliography, https://dblp.org}
}

@inproceedings{DBLP:conf/nips/DenisovMRST22,
  author       = {Sergey Denisov and
                  H. Brendan McMahan and
                  John Rush and
                  Adam D. Smith and
                  Abhradeep Guha Thakurta},
  title        = {Improved Differential Privacy for {SGD} via Optimal Private Linear
                  Operators on Adaptive Streams},
  booktitle    = neurips,
  year         = {2022},
  url          = {http://papers.nips.cc/paper\_files/paper/2022/hash/271ec4d1a9ff5e6b81a6e21d38b1ba96-Abstract-Conference.html},
  bibsource    = {dblp computer science bibliography, https://dblp.org}
}

@inproceedings{WangPS22,
  author       = {Lun Wang and
                  Iosif Pinelis and
                  Dawn Song},
  title        = {Differentially Private Fractional Frequency Moments Estimation with
                  Polylogarithmic Space},
  booktitle    = {International Conference on Learning Representations (ICLR)},
  year         = {2022},
  url          = {https://openreview.net/forum?id=7I8LPkcx8V},
  bibsource    = {dblp computer science bibliography, https://dblp.org}
}

@inproceedings{Smith0T20,
  author       = {Adam D. Smith and
                  Shuang Song and
                  Abhradeep Thakurta},
  title        = {The Flajolet-Martin Sketch Itself Preserves Differential Privacy:
                  Private Counting with Minimal Space},
  booktitle    = neurips,
  year         = {2020},
  url          = {https://proceedings.neurips.cc/paper/2020/hash/e3019767b1b23f82883c9850356b71d6-Abstract.html},
  bibsource    = {dblp computer science bibliography, https://dblp.org}
}

@inproceedings{BlockiBDS12,
  author       = {Jeremiah Blocki and
                  Avrim Blum and
                  Anupam Datta and
                  Or Sheffet},
  title        = {The Johnson-Lindenstrauss Transform Itself Preserves Differential
                  Privacy},
  booktitle    = focs,
  year         = {2012},
  url          = {https://doi.org/10.1109/FOCS.2012.67},
  doi          = {10.1109/FOCS.2012.67},
  bibsource    = {dblp computer science bibliography, https://dblp.org}
}

@inproceedings{MirMNW11,
  author       = {Darakhshan J. Mir and
                  S. Muthukrishnan and
                  Aleksandar Nikolov and
                  Rebecca N. Wright},
  title        = {Pan-private algorithms via statistics on sketches},
  booktitle    = pods,
  year         = {2011},
  url          = {https://doi.org/10.1145/1989284.1989290},
  doi          = {10.1145/1989284.1989290},
  bibsource    = {dblp computer science bibliography, https://dblp.org}
}

@inproceedings{BravermanMWZ23,
  author       = {Vladimir Braverman and
                  Joel Manning and
                  Zhiwei Steven Wu and
                  Samson Zhou},
  title        = {Private Data Stream Analysis for Universal Symmetric Norm Estimation},
  booktitle    = random,
  year         = {2023},
  url          = {https://doi.org/10.4230/LIPIcs.APPROX/RANDOM.2023.45},
  doi          = {10.4230/LIPICS.APPROX/RANDOM.2023.45},
  bibsource    = {dblp computer science bibliography, https://dblp.org}
}

@inproceedings{epasto2023differentially,
  author       = {Alessandro Epasto and
                  Jieming Mao and
                  Andres Mu{\~{n}}oz Medina and
                  Vahab Mirrokni and
                  Sergei Vassilvitskii and
                  Peilin Zhong},
  title        = {Differentially Private Continual Releases of Streaming Frequency Moment
                  Estimations},
  booktitle    = itcs,
  year         = {2023},
  url          = {https://doi.org/10.4230/LIPIcs.ITCS.2023.48},
  doi          = {10.4230/LIPICS.ITCS.2023.48},
  bibsource    = {dblp computer science bibliography, https://dblp.org}}

@inproceedings{DBLP:conf/colt/PeterTU24,
  author       = {Naty Peter and
                  Eliad Tsfadia and
                  Jonathan R. Ullman},
  title        = {Smooth Lower Bounds for Differentially Private Algorithms via Padding-and-Permuting
                  Fingerprinting Codes},
  booktitle    = colt,
  year         = {2024},
  url          = {https://proceedings.mlr.press/v247/peter24a.html},
  bibsource    = {dblp computer science bibliography, https://dblp.org}
}

@article{bun2018fpc,
  author       = {Mark Bun and
                  Jonathan R. Ullman and
                  Salil P. Vadhan},
  title        = {Fingerprinting Codes and the Price of Approximate Differential Privacy},
  journal      = {{SIAM} J. Comput.},
  volume       = {47},
  number       = {5},
  pages        = {1888--1938},
  year         = {2018},
  url          = {https://doi.org/10.1137/15M1033587},
  doi          = {10.1137/15M1033587},
  bibsource    = {dblp computer science bibliography, https://dblp.org}
}

@inproceedings{bassily2014private,
  author       = {Raef Bassily and
                  Adam D. Smith and
                  Abhradeep Thakurta},
  title        = {Private Empirical Risk Minimization: Efficient Algorithms and Tight
                  Error Bounds},
  booktitle    = focs,
  year         = {2014},
  url          = {https://doi.org/10.1109/FOCS.2014.56},
  doi          = {10.1109/FOCS.2014.56},
  bibsource    = {dblp computer science bibliography, https://dblp.org}
}

@inproceedings{chan2012differentially,
  author       = {T.{-}H. Hubert Chan and
                  Mingfei Li and
                  Elaine Shi and
                  Wenchang Xu},
  title        = {Differentially Private Continual Monitoring of Heavy Hitters from
                  Distributed Streams},
  booktitle    = {Privacy Enhancing Technologies (PETS)},
  year         = {2012},
  url          = {https://doi.org/10.1007/978-3-642-31680-7\_8},
  doi          = {10.1007/978-3-642-31680-7\_8},
  bibsource    = {dblp computer science bibliography, https://dblp.org}
}

@inproceedings{hardt2010multiplicative,
  author       = {Moritz Hardt and
                  Guy N. Rothblum},
  title        = {A Multiplicative Weights Mechanism for Privacy-Preserving Data Analysis},
  booktitle    = focs,
  year         = {2010},
  url          = {https://doi.org/10.1109/FOCS.2010.85},
  doi          = {10.1109/FOCS.2010.85},
}

@article{dwork2014algorithmic,
  author       = {Cynthia Dwork and
                  Aaron Roth},
  title        = {The Algorithmic Foundations of Differential Privacy},
  journal      = {Found. Trends Theor. Comput. Sci.},
  volume       = {9},
  number       = {3-4},
  pages        = {211--407},
  year         = {2014},
  url          = {https://doi.org/10.1561/0400000042},
  doi          = {10.1561/0400000042},
  bibsource    = {dblp computer science bibliography, https://dblp.org}
}

@article{SLM18graphdp,
 author = {Song, Shuang and Little, Susan and Mehta, Sanjay and Vinterbo, Staal Amund and Chaudhuri, Kamalika},
 title = {Differentially Private Continual Release of Graph Statistics},
 journal = {CoRR},
 year = {2018},
 volume = {abs/1809.02575},
 url = {http://arxiv.org/abs/1809.02575},
 bibsource = {dblp computer science bibliography, https://dblp.org},
 eprint = {1809.02575},
 eprinttype = {arXiv}
}

@inproceedings{dwork2006calibrating,
 author = {Dwork, Cynthia and McSherry, Frank and Nissim, Kobbi and Smith, Adam D.},
 title = {Calibrating Noise to Sensitivity in Private Data Analysis},
 year = {2006},
 url = {https://doi.org/10.1007/11681878_14},
 doi = {10.1007/11681878_14},
 bibsource = {dblp computer science bibliography, https://dblp.org},
 booktitle = tcc,
}

@inproceedings{gupta2010differentially,
 author = {Gupta, Anupam and Ligett, Katrina and McSherry, Frank and Roth, Aaron and Talwar, Kunal},
 title = {Differentially Private Combinatorial Optimization},
 year = {2010},
 url = {https://doi.org/10.1137/1.9781611973075.90},
 doi = {10.1137/1.9781611973075.90},
 bibsource = {dblp computer science bibliography, https://dblp.org},
 booktitle = soda
}

@inproceedings{gupta2012iterative,
  author       = {Anupam Gupta and
                  Aaron Roth and
                  Jonathan R. Ullman},
  title        = {Iterative Constructions and Private Data Release},
  booktitle    = tcc,
  year         = {2012},
  url          = {https://doi.org/10.1007/978-3-642-28914-9\_19},
  doi          = {10.1007/978-3-642-28914-9\_19},
  bibsource    = {dblp computer science bibliography, https://dblp.org}
}

@article{chan2011private,
	Author = {T.{-}H. Hubert Chan and Elaine Shi and Dawn Song},
	Bibsource = {dblp computer science bibliography, https://dblp.org},
	Doi = {10.1145/2043621.2043626},
	Journal = {{ACM} Trans. Inf. Syst. Secur.},
	Number = {3},
	Pages = {26:1--26:24},
	Title = {Private and Continual Release of Statistics},
	Url = {http://doi.acm.org/10.1145/2043621.2043626},
	Volume = {14},
	Year = {2011}}

@inproceedings{dwork2010differentially,
	Author = {Cynthia Dwork and Moni Naor and Toniann Pitassi and Guy N. Rothblum},
	Booktitle = stoc,
	Doi = {10.1145/1806689.1806787},
	Title = {Differential privacy under continual observation},
	Url = {http://doi.acm.org/10.1145/1806689.1806787},
	Year = {2010}}

@inproceedings{dwork2010pan,
  author       = {Cynthia Dwork and
                  Moni Naor and
                  Toniann Pitassi and
                  Guy N. Rothblum and
                  Sergey Yekhanin},
  title        = {Pan-Private Streaming Algorithms},
  booktitle    = itcs,
  year         = {2010},
  url          = {http://conference.iiis.tsinghua.edu.cn/ICS2010/content/papers/6.html},
  bibsource    = {dblp computer science bibliography, https://dblp.org}
}

@inproceedings{sealfon2016shortest,
  author       = {Adam Sealfon},
  title        = {Shortest Paths and Distances with Differential Privacy},
  booktitle    =  pods,
  year         = {2016},
  url          = {https://doi.org/10.1145/2902251.2902291},
  doi          = {10.1145/2902251.2902291},
  bibsource    = {dblp computer science bibliography, https://dblp.org}
}

@inproceedings{jain2021price,
  author       = {Palak Jain and
                  Sofya Raskhodnikova and
                  Satchit Sivakumar and
                  Adam D. Smith},
  title        = {The Price of Differential Privacy under Continual Observation},
  booktitle    = icml,
  year         = {2023},
  url          = {https://proceedings.mlr.press/v202/jain23b.html},
  bibsource    = {dblp computer science bibliography, https://dblp.org}
}

@inproceedings{eliavs2020differentially,
  author       = {Marek Eli{\'{a}}s and
                  Michael Kapralov and
                  Janardhan Kulkarni and
                  Yin Tat Lee},
  title        = {Differentially Private Release of Synthetic Graphs},
  booktitle    = soda,
  year         = {2020},
  url          = {https://doi.org/10.1137/1.9781611975994.34},
  doi          = {10.1137/1.9781611975994.34},
  bibsource    = {dblp computer science bibliography, https://dblp.org}
}

@inproceedings{Blasiok2017norms,
  author       = {Jaroslaw Blasiok and
                  Vladimir Braverman and
                  Stephen R. Chestnut and
                  Robert Krauthgamer and
                  Lin F. Yang},
  title        = {Streaming symmetric norms via measure concentration},
  booktitle    = stoc,
  year         = {2017},
  url          = {https://doi.org/10.1145/3055399.3055424},
  doi          = {10.1145/3055399.3055424},
  bibsource    = {dblp computer science bibliography, https://dblp.org}
}

@InProceedings{fichtenberger2023differentially,
  title = 	 {Constant Matters: Fine-grained Error Bound on Differentially Private Continual Observation},
  author =       {Fichtenberger, Hendrik and Henzinger, Monika and Upadhyay, Jalaj},
  booktitle = 	 icml,
  year = 	 {2023},
  url = 	 {https://proceedings.mlr.press/v202/fichtenberger23a.html},
}

@InProceedings{sheffet17differentially,
  title = 	 {Differentially Private Ordinary Least Squares},
  author =       {Or Sheffet},
  booktitle =  icml,
  year = 	 {2017},
  url = 	 {https://proceedings.mlr.press/v70/sheffet17a.html},
}

@article{choi2020differentially,
  author       = {Seung Geol Choi and
                  Dana Dachman{-}Soled and
                  Mukul Kulkarni and
                  Arkady Yerukhimovich},
  title        = {Differentially-Private Multi-Party Sketching for Large-Scale Statistics},
  journal      = {Proc. Priv. Enhancing Technol.},
  volume       = {2020},
  number       = {3},
  pages        = {153--174},
  year         = {2020},
  url          = {https://doi.org/10.2478/popets-2020-0047},
  doi          = {10.2478/POPETS-2020-0047},
  bibsource    = {dblp computer science bibliography, https://dblp.org}
}

@inproceedings{bu2021differentially,
title={Fast and Memory Efficient Differentially Private-{SGD} via {JL} Projections},
author={Zhiqi Bu and Sivakanth Gopi and Janardhan Kulkarni and Yin Tat Lee and Judy Hanwen Shen and Uthaipon Tantipongpipat},
booktitle=neurips,
year={2021},
url={https://openreview.net/forum?id=WwZbupAKWo}
}

@InProceedings{blocki2023differentially,
  author =	{Blocki, Jeremiah and Grigorescu, Elena and Mukherjee, Tamalika and Zhou, Samson},
  title =	{{How to Make Your Approximation Algorithm Private: A Black-Box Differentially-Private Transformation for Tunable Approximation Algorithms of Functions with Low Sensitivity}},
  booktitle =	random,
  year =	{2023},
  URL =		{https://drops.dagstuhl.de/entities/document/10.4230/LIPIcs.APPROX/RANDOM.2023.59},
  URN =		{urn:nbn:de:0030-drops-188849},
  doi =		{10.4230/LIPIcs.APPROX/RANDOM.2023.59}
}

@inproceedings{FHO21graphdp,
 author = {Fichtenberger, Hendrik and Henzinger, Monika and Ost, Lara},
 title = {Differentially Private Algorithms for Graphs Under Continual Observation},
 year = {2021},
 url = {https://doi.org/10.4230/LIPIcs.ESA.2021.42},
 doi = {10.4230/LIPIcs.ESA.2021.42},
 booktitle = esa,
}

@inbook{bhatia2013matrix,
author="Bhatia, Rajendra",
title="Symmetric Norms",
booktitle="Matrix Analysis",
year="1997",
publisher="Springer New York",
pages="84--111",
isbn="978-1-4612-0653-8",
doi="10.1007/978-1-4612-0653-8_4",
url="https://doi.org/10.1007/978-1-4612-0653-8_4"
}

@inproceedings{HSS24random,
  author       = {Monika Henzinger and
                  A. R. Sricharan and
                  Teresa Anna Steiner},
  title        = {Private Counting of Distinct Elements in the Turnstile Model and Extensions},
  booktitle    = random,
  year         = {2024},
  url          = {https://doi.org/10.4230/LIPIcs.APPROX/RANDOM.2024.40},
  doi          = {10.4230/LIPICS.APPROX/RANDOM.2024.40},
}

@inproceedings{RS25graphdp,
  author       = {Sofya Raskhodnikova and
                  Teresa Anna Steiner},
  title        = {Fully Dynamic Algorithms for Graph Databases with Edge Differential Privacy},
  booktitle    = {Principles of Database Systems (PODS)},
  year         = {2025},
  url          = {https://arxiv.org/abs/2409.17623},
}

@inproceedings{HT10dpgeometry,
 author = {Hardt, Moritz and Talwar, Kunal},
 title = {On the geometry of differential privacy},
 year = {2010},
 url = {https://doi.org/10.1145/1806689.1806786},
 doi = {10.1145/1806689.1806786},
 bibsource = {dblp computer science bibliography, https://dblp.org},
 booktitle = stoc
}

@inproceedings{De12dplb,
 author = {De, Anindya},
 title = {Lower Bounds in Differential Privacy},
 year = {2012},
 url = {https://doi.org/10.1007/978-3-642-28914-9_18},
 doi = {10.1007/978-3-642-28914-9_18},
 bibsource = {dblp computer science bibliography, https://dblp.org},
 booktitle = tcc
}

@article{Boneh1998,
  title={Collusion-Secure Fingerprinting for Digital Data},
  author={Dan Boneh and James Shaw},
  journal={IEEE Trans. Inf. Theory},
  year={1998},
  volume={44},
  pages={1897-1905},
  url={https://doi.org/10.1109/18.705568}
}

@inproceedings{Dwork2009complexity,
author = {Dwork, Cynthia and Naor, Moni and Reingold, Omer and Rothblum, Guy N. and Vadhan, Salil},
title = {On the complexity of differentially private data release: efficient algorithms and hardness results},
year = {2009},
isbn = {9781605585062},
url = {https://doi.org/10.1145/1536414.1536467},
doi = {10.1145/1536414.1536467},
booktitle = stoc,
}

@inproceedings{Roth2010median,
author = {Roth, Aaron and Roughgarden, Tim},
title = {Interactive privacy via the median mechanism},
year = {2010},
url = {https://doi.org/10.1145/1806689.1806794},
doi = {10.1145/1806689.1806794},
booktitle = stoc,
}

@article{lyu2017svt,
author = {Lyu, Min and Su, Dong and Li, Ninghui},
title = {Understanding the sparse vector technique for differential privacy},
year = {2017},
issue_date = {February 2017},
publisher = {VLDB Endowment},
volume = {10},
number = {6},
issn = {2150-8097},
url = {https://doi.org/10.14778/3055330.3055331},
doi = {10.14778/3055330.3055331},
journal = {Proc. VLDB Endow.},
month = feb,
pages = {637–648},
numpages = {12}
}

@inproceedings{ELM24sublinear,
  author       = {Alessandro Epasto and
                  Quanquan C. Liu and
                  Tamalika Mukherjee and
                  Felix Zhou},
  title        = {Sublinear Space Graph Algorithms in the Continual Release Model},
  year         = {2025},
  url          = {https://doi.org/10.48550/arXiv.2407.17619},
  doi          = {10.48550/ARXIV.2407.17619},
  booktitle    = random,
}

@article{Kasiviswanathan2008privately,
 author = {Kasiviswanathan, Shiva Prasad and Lee, Homin K. and Nissim, Kobbi and Raskhodnikova, Sofya and Smith, Adam D.},
 title = {What Can We Learn Privately?},
 journal = {SIAM J. Comput.},
 year = {2011},
 number = {3},
 volume = {40},
 pages = {793--826},
 url = {https://doi.org/10.1137/090756090},
 doi = {10.1137/090756090},
 bibsource = {dblp computer science bibliography, https://dblp.org}
}

@article{Tardos2008FPC,
  title={Optimal probabilistic fingerprint codes},
  author={G{\'a}bor Tardos},
  journal={J. ACM},
  year={2008},
  volume={55},
  pages={10:1-10:24},
  url={https://api.semanticscholar.org/CorpusID:2175998}
}

@inproceedings{DMT07dplb,
 author = {Dwork, Cynthia and McSherry, Frank and Talwar, Kunal},
 title = {The price of privacy and the limits of LP decoding},
 year = {2007},
 url = {https://doi.org/10.1145/1250790.1250804},
 doi = {10.1145/1250790.1250804},
 bibsource = {dblp computer science bibliography, https://dblp.org},
 booktitle = stoc
}

\appendix
\section{Preliminaries on Differential Privacy}
\label{apx:dpPre}
\subsection{Post-processing and Composition}
\label{apx:post-composition}
Here, we mention two key properties of differential privacy (post-processing and composition) that are used in this paper.
\begin{theorem}[Post-processing~\cite{dwork2014algorithmic}]
    Let $\calM$ be an $(\eps,\delta)$-DP mechanism with range $\calR$. For any arbitrary randomized mapping $f\colon\calR\rightarrow\calR'$, $f\circ\calM$ is also $(\eps,\delta)$-DP.
\end{theorem}
Now, we mention two composition theorems for differential privacy.
\begin{theorem}[Basic Composition~\cite{dwork2014algorithmic}]
Let $\calM_1,\ldots,\calM_k$ be differentially private mechanism on the same domain such that $\calM_i$ is $(\eps_i,\delta_i)$-DP (for all $i\in[k]$). Then $$\calM_{[k]}(x)=(\calM_1(x),\ldots,\calM_k(x))$$ is $(\sum_{i=1}^k\eps_i,\sum_{i=1}^k\delta_i)$-DP.
\end{theorem}

\begin{theorem}[Advanced Composition~\cite{dwork2014algorithmic}]
\label{thm:advancedcomposition}
Given target privacy parameters $0<\eps'<1$ and $\delta'>0$, To ensure $(\eps', k\delta+\delta')$ cumulative privacy loss over $k$ mechanisms it suffices that each mechanism is $(\eps,\delta)$-DP, where $\eps=\frac{\eps'}{2\sqrt{2k\ln(1/\delta')}}$.
\end{theorem}

\section{Monotone Symmetric Norm Estimation Upper Bound} \label{apx:SNE}
In this section, we prove \cref{thm:main}, followed by \cref{thm:probaBoosted}. We then prove the corresponding upper bound for the case of $(\eps,\delta)$-DP, and show a more refined analysis for \topk and $\ell_p$ norms.
We provide some preliminaries in the next three subsections.
For clarity, we provide a table of the variable definitions used in the proof in \cref{tab:ref}.
\subsection{Continual Counting Problem}
\label{sec:bincnt}
The Continual Counting Problem denotes the following problem. Given a stream $u=(u_1,\ldots,u_T)$ where $u_t\in\{-1,0,1\}$ for all $t\in T$, output a value $y_t$ for all time steps $t$ that is a good estimate on  $\sum_{t'\leq t}u_{t'}$.
The binary tree mechanism was introduced in \cite{dwork2010differentially,chan2011private} to solve this problem privately.
\begin{theorem} \label{theo: BCM}
    There exists an $\eps$-DP mechanism such that given an input stream $u=(u_1,\ldots,u_T)$ where $u_t\in\{-1,0,1\}$ for all $t\in [T]$, at any time step $t$ outputs $y_t$ such that with probability $1-\beta$ and some constant $c$ (depending on $\eps$):
    $$|y_t-\sum_{t'\leq t}u_{t'}|\in c\log(T) \sqrt{\log(1/\beta)} \lpar \sqrt{\log (T)} + \sqrt{\log(1/\beta)}\rpar/\eps.$$
\end{theorem}

With this neighboring definition, running $d$ binary tree mechanisms~\cite{dwork2010differentially,chan2011private} for the Continuous Counting Problem concurrently, one per column, gives the following result which we prove for completeness.

\begin{lemma}\label{lem:parallelCounting}
    There exists an $\eps$-differentially private mechanism for continual $n$-dimensional  histogram such that for any individual time step, with probability $1-\beta$, the maximum error over all $n$ columns is at most $c \log T \sqrt{\log(n/\beta)} \lpar \sqrt{\log (T)} + \sqrt{\log(n/\beta)}\rpar /{\eps}$, for some constant $c>0$ (depending on $\eps$),
    and over all $T$ time steps and all $n$ columns simultaneously with probability $1-1/T^d$, for any constant $d>0$,  it is
    $O(\log (nT) \log T)/\eps)$,  i.e.,
    \[\max_{j, t} \lnor y_j^t - \sum_{t' \leq t: x_{t'} = j} u_{t'}\rnor \leq \frac{c (d+1) \log(nT) \log T}{\eps},\]
\end{lemma}

\begin{proof}
    Consider the following procedure $\calA$ for $\eps$-differentially private histograms under continual observation. During initialization $\calA$ starts $n$
    $\eps$-DP continual counting mechanisms $\calA_1, ..., \calA_{n}$ be $n$ that use independent randomness and then starts processing the input stream $u$. At each time step $t$
    given the input $u_t = (j, s_t)$ for some $j \in [n]$ the mechanism creates $n$ streams of input, one for each column of the histogram, and feeds $s_t$ to $\calA_j$ and feeds 0 to all other counting mechanisms. In this way $\calA$ creates $n$ input streams. Let us denote by $\calI_{i}$ the stream for column $i$.

    We show first that if each $\calA_i$ is $\eps$-DP, then $\calA$ is $\eps$-DP. For a dataset $\calI$ and any potential solution $s = (s_{1}, ..., s_{n})$ we have $\Pr[\calA(\calI) = s] = \prod_{i=1}^n\Pr[\calA_{i}(\calI_{i}) = s_{i}]$ as the random choices of the $\calA_{i}$ are independent.
    Now consider two neighboring streams $u, u'$. Note that, by assumption, only one of the streams $\calI_i$ differs from $\calI'_i$, and only at a single time step. Thus the ratio of probability is $$\frac{\Pr[\calA(\calI) = s]}{\Pr[\calA(\calI') = s]}= \prod_{i=1}^n  \frac{\Pr[\calA_{i}(\calI_{i}) = s_{i}]}{\Pr[\calA_{i}(\calI'_{i}) = s_{i} ]}.$$
    The ratio inside the product is equal to 1 for all $\calI_{i}$ where $\calI_{i} = \calI'_{i}$, and since the $\calA_{i}$ are $\eps$-DP, the ratio is at most $\exp(\eps)$ for the remaining one.
    Therefore, $$\frac{\Pr[\calA(\calI) = s]}{\Pr[\calA(\calI') = s]}\leq \exp(\eps),$$ which shows that $\calA$ is $\eps$-differentially private.

    We use the continual counting mechanism from \Cref{theo: BCM} with parameters $\beta' = \beta/(nT^{f+1})$ as mechanism $\calA_i$ for each $i\in [n]$.
    Thus, at any time step $t$ and each $i \in [n]$ the additive error is at most
    $c \eps^{-1} \log(T) \sqrt{\log(nT^{f+1}/\beta)}\lpar \sqrt{\log(T)} + \sqrt{\log(nT^{f+1}/\beta)}\rpar = c(f+1) \eps^{-1} \log (T)\log(nT/\beta)$ with probability $1-\beta/(nT)$.
    Thus with probability $1-\beta/T^f$ simultaneously  over all time steps and all $i\in [n]$, the additive error is at most $c(f+1) \eps^{-1} \log (T)\log(nT/\beta)$.
\end{proof}

Throughout this section, we will use $\calE(n)$ to denote the error bound  of a private $n$-dimensional histogram mechanism.
Subtracting $\calE(n)$ from the output frequency of the previous lemma gives Lemma~\ref{cor:histogram}.

\subsection{Differentially Private Histogram} \label{apx:DPHist}
Differentially private histogram refers to the problem of privately estimating the column sum of an $n$-dimensional table with entries from $\{-1, 0, 1\}$, while at every time step a new $n$-dimensional row is added to the table containing at most one non-zero entry. Formally, we are given an input stream $u=(u_1,\ldots,u_T)$ where for every $t\in[T]$, $u_t=(x_t,s_t)$ where $x_t\in[n]$ is the updated column and $s_t \in \{-1, 1\}$. The goal is to, at every time step $t$, output the values $(y^t_1, \ldots,y^t_n)$ such that for every $j\in [n]$, $y^t_j$ is a good estimation of the column sum $\sum_{t' \leq t: x_{t'}=j}s_{t'}$.
The quantity $\sum_{t' \leq t: x_{t'}=j}s_{t'}$ is the \emph{frequency} of element $j$.
Two input streams $u$ and $u'$ are {\em neighboring} if there is at most one time step $t^*$ such that $u_{t^* }\ne u'_{t^*}$.

Throughout the paper, we will use $\calE(n)$
to denote the error bound of a private $n$-dimensional histogram mechanism under continual observation.
As shown in \Cref{lem:parallelCounting} in \Cref{sec:bincnt} for $\eps$-differential privacy $\calE(n) = O(\log(nT)\log T/\eps) = O(\log^2 (nT)/\eps)$ with probability $1-1/T^d$ for any constant $d$. Thus, the lemma below follows.
\begin{lemma} \label{cor:histogram}
 Let $\eps >0$.
    There exists an $\eps$-differentially private mechanism for continual $n$-dimensional histogram such that given an input stream $u=(u_1,\ldots,u_T)$ where for every $t\in[T]$, $u_t=(x_t,s_t)$ where $x_t\in[n]$ and $s_t\in\{-1,0, 1\}$, at every time step outputs $y^t_j$ for every $j\in[n]$ such that with constant probability over all time steps $t\in[T]$  and all elements $j\in [n]$ simultaneously, the following holds:
    $$\sum_{t': x_{t'}=j}s_{t'}-\calE(n) \leq y^t_j\leq\sum_{t': x_{t'}=j}s_{t'}.$$
\end{lemma}

Typical examples of monotone symmetric norms include the $\ell_p$ norms (defined as $\ell_p(x) := (|x_1|^p+\ldots |x_n|^p)^{\frac{1}{p}}$), the \top{k} norms  (the sum of the $k$ largest absolute values of entries of $x$), and weighted top-$k$ norms.
The following fact is a direct consequence of the triangle inequality and the symmetry property of the norms.
\begin{fact} \label{fact:L1}
For every monotone symmetric norm $L$ and $x\in\mathbb{R}^n$, $L(x)\leq \ell_1(x) \cdot L(e_1)$, where $e_1\in\mathbb{R}^n$ is a standard basis vector (i.e., $e_1=(1,0,\ldots,0)$).
\end{fact}

\subsection{\texorpdfstring{The Mechanism for \cref{thm:main}}{The Mechanism for SNE Upper Bound}}

\subsubsection{Definitions and high-level description}\label{sec:descAlg}

\begin{lemma}\label{lem:priv}
    Let $\eps',  \eps>0$.
    If  the continual histogram mechanisms used for $\calH_1$ and $\calH_2$ are $\eps'$-differentially private then with the choice of $\eps'$ as stated in the mechanism,
    the mechanism is $\eps$-differentially private.
\end{lemma}
\begin{proof}
    From \Cref{cor:histogram} and the choice of the $\eps$ parameter for $\calH_1$ we know that $\hat{f}=(\hat{f}_1,\ldots,\hat{f}_n)$ is $\eps/2$-DP. Furthermore, $\tau^f$ is chosen oblivious to the data:
    therefore, the vector $V^{\hat f}$ (which corresponds to all $\hat f_i$ larger than $\tau^f$) is $\eps/2$-DP as well.

    The value of $\tau^b$ only depends on $\tau^f$ and not on the input data.
    Again, \Cref{cor:histogram}  and the choice of the $\eps$ parameter for $\calH_2$ ensures that $\hat{b}$ is $\eps/2$-DP and thus, (from the post-processing lemma), $h^{\hat b}$ is $\eps/2$-DP. Since $\sum_{i\in h^{\hat b}}\estV_i$ is fully determined by $h^{\hat b}$ and the ${\hat b}_i$ values, the post-processing lemma ensures that $\sum_{i\in h^{\hat b}}\estV_i$ is $\eps/2$-DP.
    Thus, using sequential composition of $\eps$-differentially private  mechanisms, $\estV = \sum_{i\in h^{\hat b}}\estV_i + V^{\hat{f}}$ is $\eps$-DP.
\end{proof}
Next we analyze the accuracy of the answer.
Note that both $\calH_1$ and $\calH_2$ have a small probability of failure, which we deal with as follows.
\begin{definition}
    Event $E^t$ is the event that, in time step $t$, it holds that $\forall j\in[n], f_j-\calE(n) \leq \hat{f}_j\leq f_j$, and that $\forall i \in \{1, \dots ,\lceil\log_{1+\zeta}{\tau^f}\rceil\}, b_i-2 \lceil\log_{1+\zeta}{\tau^f}\rceil \calE(\lceil\log_{1+\zeta}{\tau^f}\rceil)/\zeta \leq \hat{b}_i\leq b_i$.
\end{definition}

A direct consequence of \Cref{cor:histogram} and the definition of the event $E^t$ is that, with probability $1-2\beta$, $E^t$ holds for all time steps $t$ simultaneously, and when $E^t$ holds,  $\hat f$ and $\hat b$ are good approximations for $f$ and $b$. Formally:
\begin{lemma}\label{lem:fHatApx}
    With probability $1-2\beta$ over all $t\in [T]$ simultaneously,  event $E^t$ holds. When $E^t$ holds, then

    (1) for any $\zeta \in {\mathbb R}^+$, if $\hat{f}_j>\tau^f$ then $\frac{1}{1+\zeta}f_j\leq \hat{f}_j\leq f_j$, and

    (2) for any $\zeta\in {\mathbb R}^+$, if $\hat{b}_i>\tau^b$ then $\frac{1}{1+\zeta}b_i\leq \hat{b}_i\leq b_i$.
\end{lemma}

Fix $\zeta > 0$ to be our targeted precision, and $\tau^f \in
[4 \cdot \calE(n)/\zeta^2, (4+\zeta) \cdot \calE(n)/\zeta^2]$.
Recall that $B_i:=\{j \in [n]:(1+\zeta)^{i-1}\leq f_j < \min((1+\zeta)^i, \tau^f+1)\}$ with $i \in \{1, ..., \lceil\log_{1+\zeta}{\tau^f}\rceil\}$.
Let $H^{\hat f}:=\{j:\hat{f}_j>\tau^f\}$ be the set of \emph{high frequency} elements,
let $h^{\hat b}:=\{i:\hat{b}_i\geq \tau^b\}$ be the set of all of the \emph{heavy}  levels, and $H^{\hat{b}}:=\cup_{i\in h^{\hat b}}B_i$ be the set of all elements that are in a heavy level.
Note that the definition of $h^{\hat b}$ and $H^{\hat f}$ are based on $\hat b$ and $\hat f$: therefore, those are private. However, $H^{\hat{b}}$ is not private: even so $h^{\hat b}$ is private, for any $i \in h^{\hat b}$, $B_i$ is not. Also recall that $H^{\hat{b}} \cap H^{\hat f} = \emptyset$, as discussed in \Cref{sec:descAlg}.
The elements in $H^{\hat f}\cup H^{\hat{b}}$ are called the \emph{important} elements.

Let $\mathcal{R}$ be the set of remaining elements, i.e., that are not in a heavy level and that do not have high frequency (i.e. $\mathcal{R}:=[n]\setminus (H^{\hat f}\cup H^{\hat{b}})$), called the \emph{dropped} elements. They are dropped by our mechanism, in the sense that in the final output, their frequencies are replaced by zero.
They are in one of the two cases: (1) Their real frequency is greater than $\tau^f$ (let this set be $\mathcal{R}_H$) and (2) their real frequency is smaller than $\tau^f$ (let this set be $\mathcal{R}_L$). Formally, $\mathcal{R}_H=\{j|j\in\mathcal{R}, f_j > \tau^f\}$ and $\mathcal{R}_L=\{j|j\in\mathcal{R}, f_j \le \tau^f\}=\mathcal{R}\setminus \mathcal{R}_H$.
See Table~\ref{tab:ref} for a summary of all variable names.

\subsubsection{Analysis the contribution of dropped elements to the norm}\label{sec:dropped}

The next two lemmas show that the contribution of the elements of $\mathcal{R}_H$ and $\mathcal{R}_L$  to the total norm is negligible. The first one states that the norm of $f^{|\mathcal{R}_L}$ -- namely, the frequency vector of only elements that are not in heavy level, and have small frequency -- is tiny:
\begin{lemma} \label{lem:RLSmall}
    Conditioned on event $E^t$, it holds that for every monotone symmetric norm $L$,
    $$L(f^{|\mathcal{R}_L})\leq \tau^f\tau^b\frac{(1+\zeta)^3}{\zeta}\cdot L(e_1)$$
\end{lemma}
\begin{proof}
To show this lemma, we will essentially show that there are few non-zero entries in $f^{|\mathcal{R}_L}_i$ and then note that, by the definition of $\mathcal{R}_L$, all of them have frequency at most $\tau^f$. Using Fact~\ref{fact:L1} we will conclude.

Recall that $\mathcal{R}_L$ contains all elements $j$ with (a)  $f_j \le \tau^f$ and (b) that belong to a level $i$ with $\hat b_i < \tau^b$.
Note that (a) implies $\hat f_j \le \tau^f$, as $\hat f_j \le f_j$.
Thus for each level $B_i$ either all of its elements are in $\mathcal{R}_L$ or none of them are, i.e., either $B_i\subset \mathcal{R}_L$ or $B_i\cap\mathcal{R}_L=\emptyset$. Thus, for each element $j \in \mathcal{R}_L$, the level $B_i$ to which $j$ belongs does not change its cardinality,  when the set of elements is restricted to the elements in $\mathcal{R}_L$.
Let $B^{\mathcal{R}_L}$ be the set of the levels of the elements in $\mathcal{R}_L$, which is equivalent to say that it is the set of levels computed for the frequency vector $f^{|\mathcal{R}_L}$. As for every $i\in B^{\mathcal{R}_L}$, $\hat{b}_i < \tau^b$, \Cref{cor:histogram} implies that $b_i\leq(1+\zeta)\tau^b$.

Recall that the upper bound on the frequency of the elements in $B_i$ is at most $(1+\zeta)^i$.
Therefore, we have the following:
\begin{align*}
    \ell_1(f^{|\mathcal{R}_L}) &\leq \sum_{i \in \mathcal{R}_L} |B_i| (1+\zeta)^i
    \leq \sum_{i=1}^{\lceil\log_{1+\zeta}\tau^f\rceil}(1+\zeta)\tau^b\cdot(1+\zeta)^i \\
    & \leq (1+\zeta)\tau^b\cdot\frac{(1+\zeta)^{\lceil\log_{1+\zeta}\tau^f\rceil+1}-1}{(1+\zeta)-1}\\
    &\leq \tau^f\tau^b\frac{(1+\zeta)^3}{\zeta}
\end{align*}
Therefore, Fact~\ref{fact:L1} implies that $L(f^{|\mathcal{R}_L})\leq \tau^f\tau^b\frac{(1+\zeta)^3}{\zeta}\cdot L(e_1)$.
\end{proof}

As $\mathcal{R}_H$ may consist of many elements, we cannot show that $f^{\mathcal{R}_H}$ has a tiny norm. However, the random choice of $\tau^f$ ensures that each element is in $\mathcal{R}_H$ with only a small probability: therefore, we can show that the norm of $f^{|[n] \setminus \mathcal{R}_H}$ is very similar to the norm of $f$, as each element is preserved with large probability.

\begin{lemma} \label{lem:RHSmall}
Let $\zeta \in (0,1/2]$. Conditioned on the event $E^t$, with probability at least $3/4$ over the choice of $\tau^f$, it holds that for any monotone symmetric norm $L$, $(1-4\zeta) L(f) \leq L(f^{|[n] \setminus \mathcal{R}_H}) \leq L(f) $.
\end{lemma}
\begin{proof}
The right-hand-side inequality is a direct consequence of the monotonicity of the norm.

In the rest of the proof we show the left-hand-side.
    The elements in $\calR_H$ are exactly the ones with true frequency $f_j > \tau^f$ and estimated frequency $\hat f_j \leq \tau^f$. Thus they are all contained in the set of \emph{potentially dropped} elements $\calD = \lbrace j: f_j \in [\frac{4}{\zeta^2}\cdot \calE(n), \frac{(4+4\zeta + \zeta^2)}{\zeta^2}\cdot \calE(n)] \rbrace$: since $f_j - \calE(n) \leq \hat f_j \leq f_j$ (by $E^t$), $\calR_H$ is a subset of $\lbrace j: f_j \in [\tau^f, \tau^f + \calE(n)] \rbrace$.
    The choice of $\tau^f \in [\frac{4}{\zeta^2}\cdot \calE(n), \frac{(4+4\zeta)}{\zeta^2}\cdot \calE(n)]$ ensures that $\calR_H \subseteq \calD$.

    We proceed as follows: (1) We show that, with probability at least $1/2$, $\calR_H$ is only an $\zeta$ fraction of the elements in $\calD$. (2) We show that when this happens, for any monotone symmetric norm, all elements in $\calD$ contribute roughly the same amount to the norm, \emph{and} that this implies that replacing frequencies of $\calR_H$ by $0$ does not change the norm significantly.

    (1) From
    \Cref{cor:histogram} we have that for any element $|f_j - \hat f_j| \leq \calE(n)$. Recall that elements in $\calR_H$ are exactly the ones with  $\hat f_j \leq \tau^f < f_j$. Furthermore, $\tau^f$ is chosen uniformly at random in an interval of length $\calE(n)/\zeta$.
    Thus, for any $j \in \calD$, the probability that $j \in \calR_H$ equals the probability that $\tau^f$ belongs to the interval $[ \hat f_j, f_j ]$. This probability is at most $\zeta/4$, as the interval $[\hat f_j, f_j]$ has size at most $\calE(n)$ and $\tau^f$ is chosen in an interval of size $4\calE(n)/\zeta$, i.e.,~$4/\zeta$ times bigger.
    Thus, the probability that $j \in  \calD$ belongs to $\calR_H$ is at most ${\zeta/4}$.
    Therefore, in expectation over the choice of $\tau^f$, it holds that $\E_{\tau^f}[|\calR_H|] \leq {\zeta}|\calD|/4$, and  Markov's inequality ensures that, with probability at least $3/4$ over the choice of $\tau^f$, $|\calR_H| \leq \zeta |\calD|$.
We condition on this event for the rest of the proof.

    (2) Let $f'$ be the frequency vector where
    each frequency in $\calD$ is replaced by $\frac{(4+4\zeta + \zeta^2)}{\zeta^2}\cdot \calE(n)$ and the vector $f'' = {f'}^{|[n] \setminus \calR_H}$ -- i.e., $f'$ where we zero out the coordinates in $\calR_H$. Formally,
    \[f'_j = \begin{cases}
        f_j \text{ if } j \notin \calD\\
        \frac{(4+4\zeta+ \zeta^2)}{\zeta^2}\cdot \calE(n) \text{ if } j \in \calD
    \end{cases}
   f''_j = \begin{cases}
        f_j \text{ if } j \notin \calD\\
        \frac{(4+4\zeta+ \zeta^2)}{\zeta^2}\cdot \calE(n) \text{ if } j \in \calD \setminus \calR_H\\
        0 \text{ otherwise.}
       \end{cases}\]
    Note that for any $j$, it holds that $f_j \leq f'_j$; and additionally,
    \begin{equation}
    \label{eq:boundfpp}
        f''_j \leq (1+2\zeta)f^{|[n] \setminus \mathcal{R}_H}_j
    \end{equation}
    Indeed, for $j \notin \calD$, we directly have $f_j = f''_j$, and for $j \in \calD \setminus \calR_H$, it holds that $f_j \geq \frac{4}{\zeta^2} \cdot \calE(n)$, which implies $f''_j = \frac{(4+4\zeta + \zeta^2)}{\zeta^2}\cdot \calE(n) \leq \frac{(4+4\zeta + \zeta^2)}{4}\cdot f_j \le (1+2\zeta)f_j$ as $\zeta \le 1/2$.

Next order the values in $\calD$ in increasing order and consider the permutation $\sigma:\mathbb{R}^n \rightarrow \mathbb{R}^n$ that performs a cyclic over the indices of $\calD$, (i.e., that shifts each index of $\calD$ to the next one, and the last to the first), and that is identity on all indices in $[n]\setminus \calD$.
By slight abuse of notation we use $\sigma(f)$ to denote the resulting shifted vector when $\sigma$ is applied to the indices of $f$.
By construction of $f'$ and  $\sigma$, for $j\in [n]\setminus \calD$, it holds that $\forall k \in \mathbb{N}, \sigma^k(f'')_j = f''_j = f'_j$; and for $j \in \calD$, $\sigma^k(f'')_j =  \frac{(4+4\zeta + \zeta^2)}{\zeta^2}\cdot \calE(n)$ if the $k$-th index of $\calD$ after $j$ is in $\calD \setminus \calR_H$. For any $j \in \calD$, there are $|\calD| - |\calR_H|$ such indices (as there are $|\calD| - |\calR_H|$ values in $\calD \setminus \calR_H$), and therefore:
    \begin{align}
    \label{eq:rot1}
        \sum_{k = 1}^{|\calD|} \sigma^k(f'') &= |\calD| {f'}^{|[n] \setminus \calD} + (|\calD| - |\calR_H|) {f'}^{|\calD}
        \geq (1-\zeta) |\calD| f',
    \end{align}
    where the inequality is coordinate wise, and the inequality holds since we conditioned on $|\calR_H| \leq \zeta |\calD|$.

Let $L$ be a monotone symmetric norm. Since it is monotone symmetric, we have for any $k \in \mathbb{N}$ that $L(\sigma^k(f')) = L(f')$ and since it is monotone $L(f) \le L(f')$. Using the definition of norms, we get the following where the second inequality follows from \eqref{eq:rot1}, the third from the fact that $L$ is monotone symmetric, and the fourth from \eqref{eq:boundfpp}:
    \begin{align*}
        (1-\zeta)|\calD|L(f) &\leq (1-\zeta)|\calD| L(f')
        = L\lpar (1-\zeta)|\calD|f'\rpar &\\
        &\leq L\lpar  \sum_{k = 1}^{|\calD|} \sigma^k(f'') \rpar
        \leq  \sum_{k = 1}^{|\calD|} L\lpar \sigma^k(f'') \rpar = |\calD| L(f'')
        \leq (1+2\zeta)|\calD|L(f^{|[n] \setminus \mathcal{R}_H})
    \end{align*}
    Rearranging the terms and using $\frac{1-\zeta}{1+2\zeta} \geq 1-4\zeta$ concludes the lemma.
\end{proof}

The following is implied by the previous lemma and Lemma~\ref{lem:fHatApx}:

\begin{corollary}
    \label{lem:RSmall}
Conditioned on event $E_t$ it holds with probability at least $2/3$ over the choice of $\tau^f$ that, for any monotone symmetric norm $L$,
    $$(1-4\zeta)L(f) - L(f^{|\calR_L})\leq L(f^{| H^{\hat f}\cup H^{\hat{b}})}) \leq L(f)$$
\end{corollary}
\begin{proof}
The upper-bound is a direct consequence of monotonicity.

For the lower bound, we first
observe that $f^{|[n]\setminus\calR_H}=f^{|[n]\setminus\calR}+f^{|\calR_L}$. Using the definition of norms it follows that $L(f^{|[n]\setminus\calR_H}) \leq L(f^{|[n]\setminus\calR}) + L(f^{|\calR_L})$. Conditioned on $E_t$, \Cref{lem:RHSmall} shows that $L(f^{|[n]\setminus\calR_H})\geq(1-4\zeta)L(f)$ with probability at least $3/4$. Thus, by adding the two equations we have that the lower bound holds with probability at least $3/4$.
\end{proof}

\subsubsection{Estimating the norm for the important elements}
It remains to show how to estimate privately the norm of $f^{|(H^{\hat f}\cup H^{\hat{b}})}$.
For this, we simplify the frequency vector even further, rounding all coordinates of heavy levels and rounding the number of elements in each level. To formalize this idea, we use the idea of level vectors, defined as follows. Recall that $h^{\hat b}=\{i:\hat{b}_i\geq \frac{c \log^{0.5}(nT) \log T}{\zeta\cdot\eps}\}$.
\begin{definition}[Level Vectors and Estimated Level Vectors, see Definition 3.3 in \cite{Blasiok2017norms}]\label{def:levelVect}
    For any level $i\in h^{\hat b}$, the level vector $V_i$ is
    $$V_i:=(\underbrace{0,\ldots,0}_{\sum_{k< i,k\in h^{\hat b}} b_k}, \underbrace{(1+\zeta)^i,\ldots,(1+\zeta)^i}_{b_i}, 0,\ldots,0)\in \mathbb{R}^n$$
    i.e., it is a $n$-dimensional vector consisting of $\sum_{k\leq i,k\in h^{\hat b}} b_k$ many zeros, followed by $b_i$ many ones, followed by all zeros.
The estimated private level vector $\estV_i$ is:
$$\estV_i:=(\underbrace{0,\ldots,0}_{\sum_{k< i,k\in h^{\hat b}} \hat{b}_k}, \underbrace{(1+\zeta)^i,\ldots,(1+\zeta)^i}_{\hat{b}_i}, 0,\ldots,0)\in \mathbb{R}^n$$
Also, we define the following auxiliary vector $\auxV_i$ (where the number of zeros depends on $b$ instead of $\hat b$) as follows:
$$\auxV_i:=(\underbrace{0,\ldots,0}_{\sum_{k< i,k\in h^{\hat b}} b_k}, \underbrace{(1+\zeta)^i,\ldots,(1+\zeta)^i}_{\hat{b}_i}, 0,\ldots,0)\in \mathbb{R}^n$$
\end{definition}
Note that (conditioned on $E^t$) these vectors are indeed in $\R^n$, since $\sum{b_i}\leq{n}$, and \Cref{lem:fHatApx} ensures $\hat{b}_i\leq b_i$, so the non-zero indices in the definitions above will not exceed $n$. These vectors are used to represent the frequency of the elements in the corresponding levels, i.e.,
$V_i$ (respectively $\auxV_i$ and $\estV_i$) represents the rounded frequencies based on $b_i$ (respectively estimated $\hat b_i$) of elements in $B_i$.
We showed in \Cref{lem:priv}  that $\estV_i$ is private, but $\auxV_i$ is not (because the initial number of zeros is defined with the true level sizes $b_k$). We introduced $\auxV_i$ only to simplify the proof of utility.

The vectors $\estV_i$ are helpful for representing items that have low frequency; we now introduce vectors that will serve as proxy for the high frequency elements.
Recall that $H^{\hat f} = \{j: \hat{f_j} > \tau^f \}$ is the set of indices of high frequency elements.
Let $\{j_1,\ldots,j_{|H^{\hat f}|}\}$ be the elements of $H^{\hat f}$. We next define two vectors representing the frequencies ($V^f$) and the estimated frequencies ($V^{\hat f}$) of the elements in $H^{\hat f}$.
Vectors $V^f$ and $V^{\hat{f}}$ are $n$-dimensional vectors defined as follows:
$$V^f:=(\underbrace{0,\ldots,0}_{n-{|H^{\hat f}|}}, f_{j_1},\ldots,f_{j_{|H^{\hat f}|}}),\ \ \
V^{\hat{f}}:=(\underbrace{0,\ldots,0}_{n-{|H^{\hat f}|}}, \hat{f}_{j_1},\ldots,\hat{f}_{j_{|H^{\hat f}|}})$$

Then, we can write a simplified version of the frequency vector $f$, where frequencies are sorted and the small ones are rounded: $V':=\sum_{i\in h^{\hat b}}V_i + V^f$. Note that the non-zero coordinates of $\sum_{i\in h^{\hat b}}V_i$ and $V^f$ are disjoint: as explained before, $H^{\hat{b}}$ and $H^{\hat f}$ are disjoint so $n-|H^{\hat{f}}|\geq |H^{\hat{b}}|=\sum_{i\in h^{\hat{b}}}b_i$.

This vector $V'$ is clearly non-private, but has some structure that we can use to show that the outcome of the mechanism, $E$, is a good solution to the simultaneous norm estimation (SNE) problem. Indeed, we can break the proof into three iterative parts as follows:
\begin{enumerate}
    \item using properties of monotone symmetric norms, \cref{lem:Vprime} shows that $V'$ is a good solution to the SNE problem, i.e., the norm of $V'$ is almost the norm of $f$;
    \item then, we show in \cref{lem:auxVprime} that replacing $V^f$ by $V^{\hat{f}}$ does not loose too much in accuracy, namely that $\auxV:=\sum_{i\in h^{\hat b}}\auxV_i+V^{\hat{f}}$ is still a good solution to the SNE problem
    \item finally, we show in \cref{lem: accuracy} that it is also possible to replace the non-private $\auxV_i$ by the private $E_i$:
    $\sum_{i\in h^{\hat b}}\estV_i+V^{\hat{f}}$ is still a good solution to the SNE problem.
\end{enumerate}

\begin{lemma}
    \label{lem:Vprime}
    Conditioned on the event $E^t$, let $V'=\sum_{i\in h^{\hat b}}V_i + V^f$. Then, with probability $3/4$ over the choice of $\tau^f$, for any monotone symmetric norm $L$,
    $$(1-3\zeta)L(f)-L(f^{|\mathcal{R}_L})\leq L(V')\leq (1+\zeta) L(f)$$
\end{lemma}
\begin{proof}
Since for every element in $j\in H^{\hat{f}}$, a unique coordinate of $V'$ is equal to $f_j$ and for every level $i\in h^b$, $b_i$ coordinates of $V'$ are equal to $(1+\zeta)^i$ and since $H^{\hat f}$ and $H^{\hat{b}}$ are disjoint, there exists a one to one mapping $\sigma : [n]\mapsto [n]$ such that $f^{|(H^{\hat{b}}\cup H^{\hat f})}_i\leq V'_{\sigma (i)}\leq (1+\zeta) f^{|(H^{\hat{b}}\cup H^{\hat f})}_i$.
Therefore, the monotonicity of monotone symmetric norms implies that
\begin{equation}\label{eq:vprime}
    L(f^{|(H^{\hat{b}}\cup H^{\hat f})})\leq L(V') \leq (1+\zeta) L(f^{|(H^{\hat{b}}\cup H^{\hat f})})
\end{equation}

We first show the first inequality of the lemma: \Cref{lem:RSmall} shows that conditioned on $E_t$ it holds with probability at least $3/4$ that $(1-3\zeta)L(f)-L(f^{|\mathcal{R}_L})\leq L(f^{|H^{\hat{b}}\cup H^{\hat f}})$. Combined with the left-hand-side of \eqref{eq:vprime} this concludes the proof of the first inequality.

For the second inequality, simply apply the inequality $L(f^{|(H^{\hat{b}}\cup H^{\hat f})})\leq L(f)$  into the right-hand-side of \eqref{eq:vprime}.
\end{proof}

As mentioned earlier, the vector $V'$ is non-private for several reasons: in particular, in $V_i$, the number of non-zero elements is $b_i$, which is not private; and the high frequencies are exactly released in $V^f$.
To make a step towards privacy, we define  $\auxV$ as follows:
$$\auxV:=\sum_{i\in h^{\hat b}}\auxV_i+V^{\hat{f}}$$
The next lemma shows that $L(\auxV)$ is a good approximation of $L(V')$.

\begin{lemma}
    \label{lem:auxVprime}
    Conditioned on the event $E^t$, for any monotone symmetric norm $L$, it holds that $\frac{1}{1+\zeta}L(V')\leq L(\auxV)\leq L(V')$
\end{lemma}

\begin{proof}
Let us define permutations $\pi^k$ for $k\in \mathbb{N}$ such that $\pi^k$ consists of $k$ cyclic shifts on indices of the non-zero elements of $V_i$
for each $i\in h^{\hat b}$ and indices in range $(\sum_{k\in h^{\hat b}}b_k,n]$  remain in the same position.
We apply each permutation to $W$ instead of to $V'$.
So if $h^b=\{{i_1},\ldots,{i_m}\}$ such that ${i_1}\leq {i_2}\leq\ldots$, we have $(\pi^1(W))_1=W_{b_{i_1}}$, $(\pi^1(W))_2=W_1$, $(\pi^1(W))_{b_{i_1}+1}=W_{b_{i_1}+b_{i_2}}$ and $(\pi^1(W))_{b_{i_1}+2}=W_{b_{i_1}+1}$.

For any $i \in h^{\hat b}$ recall that $\hat b_i \le b_i$. Thus, summing over
$b_i$ (consecutive) cyclic shifts of $W_i$ ``rotates'' $b_i$ entries in $W_i$ in such a way that the sum for every non-zero index
$j \in [1+ \sum_{k< i,k\in h^{\hat b}} b_k,\sum_{k\le i,k\in h^{\hat b}} b_k]$
equals $\hat{b}_i \cdot (1+\zeta)^i $ as the cyclic shift rotates a group of indices of $W_i$ consisting of $\hat{b}_i$ non-zero entries in $W_i$ and $b_i - \hat{b}_i$ zero entries.

Let $M=\Pi_{i\in h^{\hat b}}b_i$. After every $b_i$ permutations all of the indices corresponding to the non-zero components of $V_i$ complete a full rotation. Thus, we have that
$\sum_{k\in [M]}\pi^k(\auxV)=\sum_{i\in h^{\hat b}}\frac{M}{b_i}\hat{b}_i V_i + MV^{\hat{f}}.$
On the other hand, \Cref{lem:fHatApx} ensures that $V^{\hat{f}}\geq \frac{1}{1+\zeta}V^f$ and for all $i$, $\frac{\hat{b}_i}{b_i}\geq \frac{1}{1+\zeta}$. Therefore:
\begin{align*}
    \frac{M}{1+\zeta}L(V') &= L\lpar \frac{M}{1+\zeta}V'\rpar = L\lpar M\lpar \sum_{i\in h^{\hat b}}\frac{1}{1+\zeta}V_i+\frac{1}{1+\zeta}V^f\rpar \rpar\\
    &\leq L\lpar M\lpar \sum_{i\in h^{\hat b}}\frac{\hat{b}_i}{b_i}V_i+V^{\hat{f}}\rpar \rpar=L\lpar \sum_{k\in [M]}\pi^k(\auxV)\rpar
    \leq \sum_{k\in [M]}L\lpar\pi^k(\auxV)\rpar =ML(\auxV),
\end{align*}
which implies $\frac{1}{1+\zeta}L(V') \leq L(\auxV)$. The first inequality holds because of the monotonicity of the norms, and the last equality because the norm is monotone symmetric.
Additionally, since $\auxV\leq V'$ and from monotonicity of monotone symmetric norms, we have that $L(\auxV)\leq L(V')$.
\end{proof}

The $\auxV_i$ are still disclosing the true $b_k$ values because of the location of the non-zero elements. Thus we finally replace them by $E_i$, which are based on the (private) $\hat{b}_k$ values instead of the $b_k$ values, i.e.,~the final output  is  $\estV:=\sum_{i\in h^{\hat b}}\estV_i+V^{\hat{f}}$.
\begin{lemma}
\label{lem: accuracy}
     Conditioned on the event $E^t$, it holds with probability at least $3/4$ over the choice of $\tau^f$ that, for any monotone symmetric norm $L$,
     $$\frac{1-3\zeta}{1+\zeta} L(f)-\frac{1}{1+\zeta}\cdot L(f^{|\mathcal{R}_L})\leq L(\estV) \leq  (1+\zeta) L(f)$$
\end{lemma}
\begin{proof}
\Cref{lem:Vprime} and \Cref{lem:auxVprime} imply
 $   \frac{1-3\zeta}{1+\zeta} L(f)-\frac{1}{1+\zeta}\cdot L(f^{|\mathcal{R}_L})\leq L(\auxV) \leq  (1+\zeta) L(f).$

    Also, for every $i$, $\estV_i$ and $\auxV_i$ contain $\hat b_i$ non-zero coordinates which all are equal to $(1+\zeta)^i$. As conditioning on $E^t$ guarantees that for every $i$, $\hat{b}_i\leq b_i$, it follows that the non-zero coordinates of $\sum_{i\in h^{\hat b}}\estV_i$ and $V^{\hat{f}}$ are disjoint. Therefore,
    there exists a permutation $\sigma$ such that, for all $j \in [n]$, $\estV_j = \auxV_{\sigma(j)}$
Thus $L(\estV)=L(\auxV)$ and combined with the above it follows that the proof is completed.
\end{proof}

\begin{proof}[Proof of \Cref{thm:main}]
The privacy guarantee follows from \Cref{lem:priv}.
    Next, we turn on to the accuracy.
    Recall that event $E_t$ holds with probability at least $1-2\beta$. Setting $\beta = 1/20$ implies that $E_t$ holds with probability at least $9/10$. We will condition on this event in the following.
    Conditioned on $E_t$, \Cref{lem: accuracy} holds with probability at least $3/4$ and all the other lemmas hold with probability 1.
    By replacing $L(f^{|\calR_L})$ from \Cref{lem:RLSmall} in \Cref{lem: accuracy} we have:
    $$\frac{1-3\zeta}{1+\zeta} L(f)-\frac{1}{1+\zeta}\cdot \tau^f\tau^b\frac{(1+\zeta)^3}{\zeta}\cdot L(e_1)\leq L(\estV) \leq  (1+\zeta) L(f)$$
Let    $A:= \frac{1}{1+\zeta}\cdot \tau^f\tau^b\frac{(1+\zeta)^3}{\zeta}$.
The following lemma bounds $A$.
\begin{lemma}\label{claim:A}
 Let    $A:= \frac{1}{1+\zeta}\cdot \tau^f\tau^b\frac{(1+\zeta)^3}{\zeta}$. Then for the choices for $\tau^f$ and $\tau^b$ for $\eps$-differentially mechanisms it holds that
     $A = \tO\lpar \frac{1}{\eps^2 {\zeta'}^5} \cdot \log(T)^3\log(nT)\rpar$.
\end{lemma}
    \begin{proof}
    By inserting the values for $\tau^f$ and $\tau^b$ according to the definitions we have:
    $$A = \frac{(1+\zeta)^2}{\zeta}\cdot O\lpar \frac{\calE(n)}{\zeta^2} \cdot \frac{\lceil\log_{1+\zeta}{\tau^f}\rceil \calE\lpar \lceil\log_{1+\zeta}{\tau^f}\rceil\rpar}{\zeta} \rpar$$
    As $0<\zeta<1$,  $\lceil\log_{1+\zeta}{\tau^f}\rceil = O\lpar1+ (\log \tau^f)/\zeta\rpar  = O\lpar \zeta^{-1}(1 + \log (\calE(n)/\zeta)\rpar$.

     As shown in \Cref{lem:parallelCounting} for $\eps$-differential privacy $\calE(n) = O(\log(nT)\log T/\eps) = O(\log^2 (nT)/\eps)$ with probability $1-1/T^d$ for any constant $d$  and, thus,
     with probability at least $9/10 \cdot 3/4 \cdot (1-1/T^d) \ge 2/3$ for large enough $T$. Thus

     $$\lceil\log_{1+\zeta}{\tau^f}\rceil = O\lpar \frac{1}{\zeta} \cdot \log \frac{\log (nT)}{\eps \zeta}\rpar,$$
      and $\calE\lpar \lceil\log_{1+\zeta}{\tau^f}\rceil \rpar = O \lpar \frac{\log T}{\eps} \cdot \log \lpar \frac{T}{\zeta} \cdot \log \frac{\log (nT)}{\eps \zeta}\rpar \rpar = O \lpar \frac{\log(T)^2 \log(1/(\eps \zeta))\log \log \log(n)}{\eps}\rpar.$

    This gives
    \begin{align*}
        A &= \frac{1}{\zeta}\cdot O\lpar \frac{\log(nT)\log T}{\eps \zeta^2} \cdot \frac{\log \frac{\log (nT)}{\eps \zeta} \cdot \log(T)^2 \log(1/(\eps\zeta))\log \log \log(n)}{\eps\zeta^2 } \rpar\\
        &= \frac{\log^2 1/(\eps\zeta)}{\eps^2\zeta^5}\cdot O\lpar\log(T)^3 \cdot  \log(nT) \cdot \log \log (nT) \cdot \log \log \log(n) \rpar,
    \end{align*}
        where we over-estimated some polynomials to simplify bounds.
    \end{proof}

 Now we define $\zeta'$ such that the multiplicative error to be $\zeta'=\frac{4\zeta}{1-3\zeta}$. As $\zeta \le 1/4$ we get that $\zeta' = \Theta(\zeta) = O(1)$ and it follows that
     $$\frac{1}{1+\zeta'} L(f)-A \cdot L(e_1)\leq L(\estV) \leq  (1+\Theta(\zeta')) L(f)$$

\end{proof}

\subsection{\texorpdfstring{Proof of \cref{thm:probaBoosted}}{Proof of Boosting Probability}}
To boost the success probability to $1-\beta$, we simply take $\log(T/\beta)$ copies of the mechanism from \cref{thm:main}. The guarantee we have is that, at any time step, each copy is correct with probability $2/3$. Therefore, using standard concentration inequality we get that, at any time step, with probability $1-\frac{\beta}{T}$ at least a $3/5$-fraction of the copies are correct -- i.e.,  those correct copies computes a vector that is a good proxy of the frequency vector, for any monotone symmetric norm. Thus, given a monotone symmetric norm, the median of the estimates that each copy computes for this norm is a good approximation of the true norm with probability $1-\frac{\beta}{T}$.

Thus, with a union-bound, the following mechanism is a good solution for the data structure version of \inc\SNE: run $m= \log(T/\beta)$ copies of the mechanism from \cref{thm:main}; to evaluate norm $L$ at time $t$, let $E_1, ..., E_m$ be the output of each copies at time $t$; output the median of $L(E_1), ..., L(E_m)$.

To ensure $\eps$-differential privacy, each copy of \cref{thm:main} is run with privacy parameter $\eps / \log(T/\beta)$, which concludes the proof of \cref{thm:probaBoosted}.

\subsection{Mechanism for Approximate Differential Privacy}
\label{sec:epsdelta}

Corollary B.1 in \cite{fichtenberger2023differentially} with $b=1$ and $u=n$ shows the corresponding result for $(\eps,\delta)$-differential privacy histograms with slightly better error bound. To create an $(\eps,\delta)$-differentially private SNE mechanism we simply use as histogram mechanisms $\calH_1$ and $\calH_2$, each with parameters $\eps/2$ and $\delta/2$ in our $\eps$-differentially private mechanism for SNE in the following mechanism:
\begin{lemma}
\label{cor:pre_histogram} Let $\eps>0 , 1>\delta >0$.
    There exists an $(\eps,\delta)$-differentially private mechanism for continual $n$-dimensional histogram such that for all time steps simultaneously, with probability $1-\beta$, the error for all of the $n$ elements is at most $\frac{c \log^{0.5}(nT) \log T}{\eps}$, for some constant $c$, i.e.,
    \[\max_{j, t} \lnor y_j^t - \sum_{t' \leq t: x_{t'} = j} u_{t'}\rnor \leq \frac{c \log^{0.5}(nT/\beta) \log T \sqrt{\log(1/\delta)}}{\eps}.\]
\end{lemma}
We use $\calE(n)$ to denote the error bound of a private $n$-dimensional histogram mechanism.
Subtracting $\calE(n)$ from the output frequency according to \Cref{cor:pre_histogram}, we have the following corollary:
\begin{corollary} \label{cor:histogram2}
 Let $\eps >0, \delta > 0$.
    There exists an $(\eps,\delta)$-differentially private mechanism for continual $n$-dimensional histogram such that given an input stream $u=(u_1,\ldots,u_T)$ where for every $t\in[T]$, $u_t=(x_t,s_t)$ where $x_t\in[n]$ and $s_t\in\{-1,0, 1\}$, at every time step outputs $y^t_j$ for every $j\in[n]$ such that with probability at least $2/3$ over all time steps $t\in[T]$ simultaneously and all elements $j\in [n]$ simultaneously, the following holds:
    $$\sum_{t': x_{t'}=j}s_{t'}-\calE(n) \leq y^t_j\leq\sum_{t': x_{t'}=j}s_{t'}.$$
\end{corollary}
The resulting mechanism fulfills $(\eps, \delta)$-differential privacy:
\begin{lemma}\label{lem:priv2}
    Let $\eps',  \eps, \delta >0$.
    If  the continual histogram mechanisms used for $\calH_1$ and $\calH_2$ are $(\eps', \delta'/2)$-differentially private then with the choice of $\eps'$ as stated in the mechanism,
    the mechanism is $(\eps, \delta)$-differentially private.
\end{lemma}
\begin{proof}
    From \Cref{cor:histogram2} and the choice of the parameters  for $\calH_1$ we know that $\hat{f}=(\hat{f}_1,\ldots,\hat{f}_n)$ is $(\eps/2, \delta/2)$-DP. Furthermore, $\tau^f$ is chosen oblivious to the data:
    therefore, the vector $V^{\hat f}$ (which corresponds to all $\hat f_i$ larger than $\tau^f$) is $(\eps/2, \delta/2)$-DP as well.

    The value of $\tau^b$ only depends on $\tau^f$ and not on the input data.
    Again, \Cref{cor:histogram2}  and the choice of the $\eps$ parameter for $\calH_2$ ensures that $\hat{b}$ is $(\eps/2, \delta/2)$-DP and thus, (from the post-processing lemma), $h^{\hat b}$ is $(\eps/2, \delta/2)$-DP. Since $\sum_{i\in h^{\hat b}}\estV_i$ is fully determined by $h^{\hat b}$ and the ${\hat b}_i$ values, the post-processing lemma ensures that $\sum_{i\in h^{\hat b}}\estV_i$ is $(\eps/2, \delta/2)$-DP.
    Thus, using sequential composition of $\eps$-differentially private  mechanisms, $\estV = \sum_{i\in h^{\hat b}}\estV_i + V^{\hat{f}}$ is $\eps$-DP.

    Using the fact that the sequential composition of two $(\eps/2, \delta/2)$-differentially private  mechanisms is $(\eps,\delta)$-differentially private (Corollary B.2 in ~\cite{dwork2014algorithmic})
    it follows that the computation of $\estV = \sum_{i\in h^{\hat b}}\estV_i + V^{\hat{f}}$ is $(\eps,\delta)$-differentially private given that  the continual histogram mechanisms used for $\calH_1$ and $\calH_2$ are $(\eps', \delta'/2)$-DP  with the choice of $\eps'$ as stated in the mechanism.
\end{proof}
Now, we can prove \cref{thm:SNEUBEpsDelta}.
\begin{proof} [proof of \cref{thm:SNEUBEpsDelta}]
Privacy is proven in \Cref{lem:priv2}. We proceed to the proof of accuracy. This can be proven analogous to the proof of \Cref{thm:main}, except that we use \Cref{cor:histogram2}, which shows that for $(\eps,\delta)$-differential privacy with $\delta>0$, $\calE(n) = O(\log^{0.5}(nT) \log T  \log^{0.5}(1/\delta)/\eps)$ with probability at least $2/3$ and, thus,
     $$\lceil\log_{1+\zeta}{\tau^f}\rceil = O\lpar \frac{1}{\zeta} \cdot \log \frac{\log (nT)\log(1/\delta)}{\eps \zeta}\rpar,$$
      and
      $\calE\lpar \lceil\log_{1+\zeta}{\tau^f}\rceil \rpar =
      O \lpar \frac{\log T \log^{0.5} (1/\delta)}{\eps} \cdot \log^{0.5}
      \lpar \frac{T}{\zeta } \cdot \log \frac{\log (nT) \log(1/\delta)}{\eps \zeta} \rpar \rpar $,
      which is  simplified to
      $$O \lpar \frac{\log(T)^{1.5}\log^{0.5}(1/\delta) \log^{0.5}(1/(\eps \zeta))\log^{0.5} \log \log(n) \log^{0.5} \log \log (1/\delta)}{\eps}\rpar$$ $$= \tO \lpar \frac{\log(T)^{1.5}\log^{0.5}(1/\delta) \log^{0.5}(1/(\eps \zeta))\log^{0.5} \log \log(n)}{\eps}\rpar.$$

    This gives
    \begin{align*}
        A &=  \frac{(1+\zeta)^2}{\zeta}\cdot O\lpar \frac{\calE(n)}{\zeta^2} \cdot \frac{\lceil\log_{1+\zeta}{\tau^f}\rceil \calE\lpar \lceil\log_{1+\zeta}{\tau^f}\rceil\rpar}{\zeta} \rpar\\
        &=\tO\lpar \frac{\log^{0.5}(nT)\log T \log^{0.5}(\frac{1}{\delta})}{\eps \zeta^3} \cdot \frac{\log \frac{\log (nT) \log(1/\delta)}{\eps \zeta} \cdot \log^{1.5}(T) \log^{0.5}(\frac{1}{\delta})\log^{0.5}(\frac{1}{\eps \zeta})\log^{0.5} \log \log(n)}{\eps\zeta^2 } \rpar\\
        &= \tO \lpar \frac{1}{\eps^2\zeta^5}\cdot \log^{2.5}(T) \cdot  \log^{0.5}(nT)  \log (1/\delta)   \rpar,
    \end{align*}
        where we over-estimated some polynomials to simplify bounds.

 Recall that
 $$\frac{1-3\zeta}{1+\zeta} L(f)- A \cdot L(e_1)\leq L(\estV) \leq  (1+\zeta) L(f).$$
 Now we define $\zeta'$ such that the multiplicative error to be $\zeta'=\frac{4\zeta}{1-3\zeta}$. As $\zeta \le 1/4$ we get that $\zeta' = \Theta(\zeta) = O(1)$ we have for $(\eps,\delta)$-differential privacy for $\delta>0$
     $$\frac{1}{1+\zeta'} L(f)-\tO \lpar \frac{1}{\eps^2\zeta^5}\cdot \log^{2.5}(T) \cdot  \log^{0.5}(nT)  \log (1/\delta)  \rpar \cdot L(e_1)$$
     $$\leq L(\estV) \leq  (1+\Theta(\zeta')) L(f)$$
\end{proof}
Now, analogous to the case of $\delta=0$, we can have the following result for the data structure version of the problem.
\begin{theorem}\label{thm:probaBoosted_epsdelta}
    For any $\zeta > 1, \beta \in (0, 1)$, $\eps>0$ and $\delta>0$, there exist an $(\eps,\delta)$-DP mechanism for the data structure version of \inc\SNE such that, with probability $1-\beta$, it holds
    at any time step $t$ that the multiplicative approximation is $\zeta \in \R^+$ and the additive error is \[
    \tO \lpar \frac{1}{\eps^2\zeta^5}\cdot \log^{2.5}(T) \cdot  \log^{0.5}(nT)  \log (1/\delta) \log^2(T/\beta) \rpar \cdot L(e_1)
    .\]
\end{theorem}

\subsection{\texorpdfstring{Bounds for {\topk} and $\ell_p$}{Bounds for top-k and lp}}
In this section we will show better bounds on the additive error of the special cases of {\topk} and $\ell_p$ norms.
\begin{lemma}
    For every $k\in\mathbb{N}$,
    $$\topk(f^{|\mathcal{R}_L})\leq \min(\tau^f\tau^b\frac{(1+\zeta)^3}{\zeta}, k(1+\zeta)\tau^f)$$
\end{lemma}
\begin{proof}
    The desired inequality is implied from the bound proven in \cref{lem:RLSmall} and the fact that all of the coordinates are smaller than $(1+\zeta)\tau^f$.
\end{proof}
\begin{lemma}
    For every $p \geq 1$,
    $$\ell_p(f^{|\calR_L})\leq {(\tau^b)}^{\frac{1}{p}}\cdot\frac{(1+\zeta)^{2+\frac{1}{p}}\tau^f}{\zeta}$$
\end{lemma}
\begin{proof}
    With similar arguments as in the proof of \cref{lem:RLSmall}, we have:
    \begin{align*}
        \ell_p(f^{|\mathcal{R}_L}) &\leq (\sum_{i \in \mathcal{R}_L}|B_i| (1+\zeta)^{ip})^{\frac{1}{p}}\\
        & \leq {((1+\zeta)\tau^b)}^{\frac{1}{p}}\cdot\frac{(1+\zeta)^{\lceil\log_{1+\zeta}\tau^f\rceil +1}-1}{(1+\zeta)^p-1}\\
        &\leq {((1+\zeta)\tau^b)}^{\frac{1}{p}}\cdot\frac{(1+\zeta)^2\tau^f}{\zeta}\\
        &\leq {(\tau^b)}^{\frac{1}{p}}\cdot\frac{(1+\zeta)^{2+\frac{1}{p}}\tau^f}{\zeta} \qedhere
    \end{align*}
\end{proof}
Now, we prove the following theorem about our main theoretical results for norms $L_p$ and $\topk$.
\begin{theorem}\label{lem:Lpmain}
    Let $\eps>0$.
    For any $0< \zeta \le 1/2$, for \Cref{alg:main} for any $p\geq 1$
    at any time step $t \in [T]$ it holds with probability at least $2/3$ that for $\ell_p$ the multiplicative approximation is $1+\zeta$  and the additive error is
    $\tO\lpar \frac{\log(T)^{1+\frac{2}{p}} \cdot  \log(nT)}{\eps^{1+\frac{1}{p}}\zeta^{3+\frac{2}{p}}} \rpar$. Moreover, with probability $1-\beta$ (for any $\beta\in(0,1)$), it holds at all time steps that the multiplicative approximation is $1+\zeta$ and the additive error is
    $\tO\lpar \frac{\log(T)^{1+\frac{2}{p}} \cdot  \log(nT)\cdot\log^2(T/\beta)}{\eps^{1+\frac{1}{p}}\zeta^{3+\frac{2}{p}}}\rpar$

\end{theorem}
\begin{proof}
The proof is analogous to the proof of \Cref{thm:main}. Let $A:=\frac{1}{1+\zeta}\cdot{(\tau^b)}^{\frac{1}{p}}\cdot\frac{(1+\zeta)^{2+\frac{1}{p}}\tau^f}{\zeta}$ be the upper-bound to the additive error in this case. Then similar to the proof of \Cref{claim:A} we have:
 \begin{align*}
        A&=\frac{(1+\zeta)^{1+\frac{1}{p}}}{\zeta}\cdot{(\tau^b)}^{\frac{1}{p}}\tau^f\\
        &= \frac{1}{\zeta}\cdot O\lpar \frac{\log(nT)\log T}{\eps \zeta^2} \cdot
        \frac{\log^{\frac{1}{p}} \frac{\log (nT)}{\eps \zeta} \cdot \log(T)^{\frac{2}{p}} \log^{\frac{1}{p}}(1/(\eps\zeta))\log^{\frac{1}{p}} \log \log(n)}
        {\eps^{\frac{1}{p}}\zeta^{\frac{2}{p}} } \rpar\\
        &= \frac{\log^{\frac{2}{p}} 1/(\eps\zeta)}{\eps^{1+\frac{1}{p}}\zeta^{3+\frac{2}{p}}}\cdot O\lpar\log(T)^{1+\frac{2}{p}} \cdot  \log(nT) \cdot \log^{\frac{1}{p}} \log (nT) \cdot \log^{\frac{1}{p}} \log \log(n) \rpar\\
        &=\tO\lpar \frac{\log(T)^{1+\frac{2}{p}} \cdot  \log(nT)}{\eps^{1+\frac{1}{p}}\zeta^{3+\frac{2}{p}}} \rpar.
    \end{align*}
    Which completes the proof.
\end{proof}

Similarly, for $\topk$ we have:
\begin{theorem}\label{lem:TopKmain}
    Let $\eps>0$.
    For any $0< \zeta \le 1/2$, for \Cref{alg:main} for any $k\geq \mathbb{N}$
    at any time step $t \in [T]$ it holds with probability at least $2/3$ that for {\topk} the multiplicative approximation is $1+\zeta$  and the additive error is
    $\tO\lpar \min \lpar\frac{\log(T)^3\log(nT)}{\eps^2 {\zeta}^5},k\cdot\frac{\log{T}\log{nT}}{\eps\zeta^2}\rpar\rpar$. Moreover, with probability $1-\beta$ (for any $\beta\in(0,1)$), it holds at all time steps that the multiplicative approximation is $1+\zeta$ and the additive error is
    $\tO\lpar \min\lpar\frac{\log(T)^3\log(nT)\log^2(T/\beta)}{\eps^2 {\zeta}^5},k\cdot\frac{\log{T}\log{nT}\log^2(T/\beta)}{\eps\zeta^2}\rpar\rpar$.
\end{theorem}
Also, for $\delta>0$ we have the following results.
\begin{theorem}\label{lem:LpEpsDelta}
    Let $\eps,\delta>0$.
    For any $0< \zeta \le 1/2$, for our $(\eps,\delta)$-DP mechanism for any $p\geq 1$
    at any time step $t \in [T]$ it holds with probability at least $2/3$ that for $\ell_p$ the multiplicative approximation is $1+\zeta$  and the additive error is
    $\tO \lpar \frac{1}{\eps^{1+\frac{1}{p}}\zeta^{3+\frac{2}{p}}}\cdot \log^{1+\frac{1.5}{p}}(T) \cdot  \log^{0.5}(nT)  \log^{0.5+\frac{0.5}{p}} (1/\delta)\rpar$. Moreover, with probability $1-\beta$ (for any $\beta\in(0,1)$), it holds at all time steps that the multiplicative approximation is $1+\zeta$ and the additive error is
    $\tO \lpar \frac{1}{\eps^{1+\frac{1}{p}}\zeta^{3+\frac{2}{p}}}\cdot \log^{1+\frac{1.5}{p}}(T) \cdot  \log^{0.5}(nT)  \log^{0.5+\frac{0.5}{p}}(1/\delta) \log^2(T/\beta)   \rpar$
\end{theorem}
\begin{proof}
Similar to the proofs above and proof of \Cref{thm:SNEUBEpsDelta}, here we have:
\begin{align*}
        A &=  \frac{(1+\zeta)^2}{\zeta}\cdot O\lpar \frac{\calE(n)}{\zeta^2} \cdot \frac{{\lceil\log_{1+\zeta}{\tau^f}\rceil}^{\frac{1}{p}} {\calE\lpar \lceil\log_{1+\zeta}{\tau^f}\rceil\rpar}^{\frac{1}{p}}}{\zeta^{\frac{1}{p}}} \rpar\\
        &=\tO\lpar \frac{\log^{0.5}(nT)\log T \log^{0.5}(\frac{1}{\delta})}{\eps \zeta^3} \cdot \frac{\log^{\frac{1}{p}} \frac{\log (nT) \log(1/\delta)}{\eps \zeta} \cdot \log^{\frac{1.5}{p}}(T) \log^{\frac{0.5}{p}}(\frac{1}{\delta})\log^{\frac{0.5}{p}}(\frac{1}{\eps \zeta})\log^{\frac{0.5}{p}} \log \log(n)}{\eps^{\frac{1}{p}}\zeta^{\frac{2}{p}} } \rpar\\
        &= \tO \lpar \frac{1}{\eps^{1+\frac{1}{p}}\zeta^{3+\frac{2}{p}}}\cdot \log^{1+\frac{1.5}{p}}(T) \cdot  \log^{0.5}(nT)  \log^{0.5+\frac{0.5}{p}} (1/\delta)   \rpar
    \end{align*}
\end{proof}

\begin{theorem}\label{lem:TopkEpsDelta}
    Let $\eps,\delta>0$.
    For any $0< \zeta \le 1/2$, for our $(\eps,\delta)$-DP mechanism for any $k\in\mathbb{N}$
    at any time step $t \in [T]$ it holds with probability at least $2/3$ that the multiplicative approximation is $1+\zeta$  and the additive error is
    $\tO\lpar \min(\frac{1}{\eps^2\zeta^5}\cdot \log^{2.5}(T) \cdot  \log^{0.5}(nT)  \log (1/\delta),k\cdot\frac{\log^{0.5}(nT)\log{T}\log(1/\delta)}{\eps\zeta^2})\rpar$. Moreover, with probability $1-\beta$ (for any $\beta\in(0,1)$), it holds at all time steps that the multiplicative approximation is $1+\zeta$ and the additive error is
    $$\tO\lpar \min(\frac{1}{\eps^2\zeta^5}\cdot \log^{2.5}(T) \cdot  \log^{0.5}(nT)  \log (1/\delta)\cdot \log^2(T/\beta),k\cdot\frac{\log^{0.5}(nT)\log{T}\log(1/\delta)}{\eps\zeta^2}\cdot\log^2(T/\beta))\rpar$$
\end{theorem}

\begin{table}[ht]
\centering
\begin{tabular}{|l|l|}
\hline
\multicolumn{1}{|l|}{Name} & Definition                                                                                                                                  \\
\hline
$f$                      & Frequency vector                                                                                                                            \\
\hline
$\hat f$                 & Noisy frequency vector (private histogram)                                                                                                  \\
\hline
$\calE(n)$                  & Maximum Error of continual $n$-dimensional histogram mechanism whp       \\
\hline
$b$                      & Histogram of the levels                                                                                                                     \\
\hline
$\hat b$                 & Private histogram of the levels                                                                                                             \\
\hline
$C_U$                      & $(4+2\zeta) \cdot \calE(n)/\zeta^2$\\
\hline
$\tau^f$                &Random variable from $[4 \cdot \calE(n)/\zeta^2, (4+\zeta) \cdot \calE(n)/\zeta^2]$
\\
\hline
$B_i$                   &$\{j \in [n]:(1+\zeta)^{i-1}\leq f_j < \min((1+\zeta)^i, \tau^f+1)\}$
\\
\hline
$\tau^b$                &$2 \lceil\log_{1+\zeta}{\tau^f}\rceil \calE(\lceil\log_{1+\zeta}{\tau^f}\rceil)/\zeta$
\\
\hline
$h^{\hat{b}}$            & $\{i:\hat{b}_i\geq \tau^b\}$                                                                   \\
\hline
$H^{\hat b}$             & $\cup_{i\in h^{\hat b}}B_i$                                                                                                                 \\
\hline
$H^{\hat f}$             & $\{j:\hat{f}_j>\tau^f\}$                                                                                                                      \\
\hline
$\calR$                  & $[n]\setminus (H^{\hat f}\cup H^{\hat{b}})$                                                                                                 \\
\hline
$\calR_H$                & $\{j|j\in\mathcal{R}, f_j > \tau^f\}$                                                                                                         \\
\hline
$\calR_L$                & $\{j|j\in\mathcal{R}, f_j \le \tau^f\}$                                                                                                      \\
\hline
$V_i$                    & $(\underbrace{0,\ldots,0}_{\sum_{k< i,k\in h^{\hat b}} b_k}, \underbrace{(1+\zeta)^i,\ldots,(1+\zeta)^i}_{b_i}, 0,\ldots,0)$              \\
\hline
$\estV_i$                & $(\underbrace{0,\ldots,0}_{\sum_{k< i,k\in h^{\hat b}} \hat{b}_k}, \underbrace{(1+\zeta)^i,\ldots,(1+\zeta)^i}_{\hat{b}_i}, 0,\ldots,0)$  \\
\hline
$\auxV_i$                & $(\underbrace{0,\ldots,0}_{\sum_{k< i,k\in h^{\hat b}} b_k}, \underbrace{(1+\zeta)^i,\ldots,(1+\zeta)^i}_{\hat{b}_i}, 0,\ldots,0)$        \\
\hline
$V'$                     & $\sum_{i\in h^{\hat b}}V_i + V^f$                                                                                                           \\
\hline
$\estV$                  & $\sum_{i\in h^{\hat b}}\estV_i+V^{\hat{f}}$                                                                                                 \\
\hline
$\auxV$                  & $\sum_{i\in h^{\hat b}}\auxV_i+V^{\hat{f}}$                                                                                                 \\
\hline
\end{tabular}
\caption{List of all frequently used variables and their definitions.}
\label{tab:ref}
\end{table}
\section{Sparse Vector Technique}
\label{apx:SVT}
The sparse vector technique (SVT), introduced by \cite{Dwork2009complexity} and improved by \cite{hardt2010multiplicative, Roth2010median, lyu2017svt}, can be used to solve the following problem. Given a sequence of real-valued queries $q_1,\ldots,q_T$ on a private database $D$, the output should answer whether or not the query's output is larger than its threshold $t_i$ where the number of queries that are \emph{above threshold} is bounded (at most $c$ times $q_i(D)>t_i$ so the mechanism can halt after $c$ times). Note that queries and their thresholds might be chosen adaptively. Here, the variant of SVT that we use is \cref{alg:svt}, which is the SVT mechanism from \cite{dwork2014algorithmic}. In the original mechanism in \cite{dwork2014algorithmic}, the thresholds all have the same value ($t_1=t_2=\ldots=t_T$); however, in \cref{alg:svt}, using the argument discussed in \cite{lyu2017svt}, we can assume different thresholds since these cases are both equivalent to having all threshold equal to 0 (by taking queries to be $q'_i(D)=q_i(D)-t_i$). This mechanism has been improved later in other works such as \cite{lyu2017svt}, however, since the accuracy of the version in \cite{dwork2014algorithmic} is asymptotically enough for our guarantees, we stick to their version.
\begin{algorithm}[ht]
\DontPrintSemicolon
\caption{Sparse Vector Technique (SVT)~\cite{dwork2014algorithmic}}
\label{alg:svt}
\KwInput{Dataset $D$, queries $q_1, \ldots, q_T$ each with sensitivity 1, thresholds $t_1, \ldots, t_T$, bound $c$, and privacy parameters $\epsilon$ and $\delta$}
\KwOutput{Responses to queries: ``positive'' if above threshold, ``negative'' otherwise}
\If{$\delta=0$}{Let $\sigma=2c / \epsilon$}\;
\Else{Let $\sigma=\frac{\sqrt{32c\ln{1/\delta}}}{\eps}$}
Sample threshold noise $\rho \sim \textrm{Lap}(\sigma)$\;
Initialize counter $cnt = 0$\;
\For{$i = 1, \ldots, T$}{
    Sample query noise $\nu_i \sim \textrm{Lap}(2\sigma)$\;
    \If {$q_i(D) + \nu_i \ge t_i + \rho$ \algcomment{check noisy threshold}}
    {
        Output ``positive'' \;
        $cnt \gets cnt + 1$\;
        $\rho \sim \textrm{Lap}(\sigma)$\;
        \If {$cnt \ge c$} {\textbf{break}\;}
    }
    \textbf{else} Output ``negative''\;
}
\end{algorithm}

Regarding the privacy and accuracy of \cref{alg:svt}, we have the following:
\begin{theorem}[\citealp{dwork2014algorithmic}]
\label{thm:SVT}
For any $\eps>0, \delta\in[0,1)$, \cref{alg:svt} is $(\eps,\delta)$-DP. Also, for any sequence of $T$ queries $q_1,\ldots,q_T$ such that $|\{i\colon q_i(D)\geq t_i-\alpha\}|\leq c$, if $\delta>0$, \cref{alg:svt} has additive error $\alpha$ with probability at least $1-\beta$ for:
$$\alpha=\frac{(\ln{T}+\ln{\frac{2c}{\beta}})\sqrt{512c\ln\frac{1}{\delta}}}{\eps}$$
and if $\delta=0$, \cref{alg:svt} has additive error $\alpha$ with probability at least $1-\beta$ for:
$$\alpha=\frac{8c(\ln{T}+\ln{\frac{2c}{\beta}})}{\eps}$$
where an SVT mechanism is said to have additive error $\alpha$ with probability at least $1-\beta$ if the following holds: with probability at least $1-\beta$, the mechanism does not halt before $q_T$ and for all queries $q_i$, and if the output is positive, $q_i(D)\geq t_i-\alpha$ and otherwise $q_i(D)\leq t_i+\alpha$.
\end{theorem} \end{document}